\documentclass[11pt,a4paper]{article}

\pdfoutput=1 

\usepackage{jheppub}
\usepackage{amssymb,amsmath,amsthm,graphicx}
\usepackage{graphics, color}
\usepackage{subfigure}
\usepackage{latexsym}
\usepackage{bm}
\usepackage{epsfig}
\usepackage{textcomp}
\usepackage[utf8]{inputenc}
\allowdisplaybreaks[1]

\begin{document}
\title{Localized states of BFSS super quantum mechanics}
\author[1]{\'Oscar J.~C.~Dias,}
\author[2]{Jorge~E.~Santos,}

\affiliation[1]{STAG Research Centre and Mathematical Sciences, University of Southampton, University Road, Southampton SO17 1BJ, UK.}
\affiliation[2]{DAMTP, Centre for Mathematical Sciences, University of Cambridge, Wilberforce Road, Cambridge CB3 0WA, UK}
\emailAdd{ojcd1r13@soton.ac.uk}
\emailAdd{jss55@cam.ac.uk}

\newcommand{\blue}{\color{blue}}
\newcommand{\red}{\color{red}}

\abstract{
We analyze the recently discovered localized and non-uniform phases of the 
Banks-Fischler-Shenker-Susskind (BFSS) matrix quantum mechanics. 
Building on \cite{Dias:2024vsc}, we provide first-principles derivations of 
their properties and extend the results with new analytic and numerical 
insights. 
We show that strongly coupled BFSS dynamics emerge from a specific 
Carrollian transformation of 11-dimensional supergravity, which we 
justify in detail. 
In this framework, the uniform BFSS phase corresponds to a black string 
in a $pp$-wave background. 
We demonstrate that this background is unstable to a Gregory-Laflamme 
instability and, for the first time, compute the associated growth rate. 
The instability gives rise to non-uniform and localized phases that 
dominate the microcanonical ensemble in certain low-energy regimes, 
with the localized phase also prevailing in the canonical ensemble at 
low temperatures. 
We identify the corresponding first- and second-order phase transitions 
and derive analytic formulas for the thermodynamics of the localized phase, 
accurate to better than $0.3\%$ against numerical results.
}
\maketitle


\section{Introduction \label{sec:Intro}}

Maldacena’s original duality~\cite{Maldacena:1997re} states that, in its low-energy limit, ten-dimensional type IIB supergravity on AdS$_5 \times S^5$ is equivalent to $(3+1)$-dimensional $\mathcal{N}=4$ supersymmetric Yang–Mills (SYM) theory with gauge group $SU(N)$, in the limit of large $N$ and strong ’t~Hooft coupling. Despite overwhelming evidence in favor of this duality, no rigorous proof exists, as it relates theories at strong and weak coupling. In particular, the gravitational description is valid precisely when $\mathcal{N}=4$ SYM is strongly coupled~\cite{Maldacena:1997re, Witten:1998qj, GubKle98, Aharony:1999ti}, a regime that remains analytically intractable in its full generality.  

Numerical approaches also face major challenges. Standard lattice Monte Carlo methods are notoriously difficult to apply to $\mathcal{N}=4$ SYM, even though the theory is highly symmetric and conformal. The difficulties arise both from its four-dimensional nature (which makes computations demanding) and from the presence of many fermionic degrees of freedom in the supersymmetric ultraviolet (UV) regime, which introduce sign problems in path integral formulations~\cite{Loh:1990zz}. Nevertheless, significant progress has been made in recent years toward overcoming these obstacles~\cite{Bergner:2021goh, Catterall:2023tmr}.  

More concretely, the AdS$_5$/CFT$_4$ correspondence identifies the near-horizon geometry of a stack of D3-branes - AdS$_5 \times S^5$ - with a conformal quantum field theory (CFT), namely $(3+1)$-dimensional $SU(N)$ $\mathcal{N}=4$ SYM~\cite{Maldacena:1997re, Witten:1998qj, GubKle98, Aharony:1999ti}.

Interestingly for our purposes, soon after the original AdS$_5$/CFT$_4$ correspondence was established~\cite{Maldacena:1997re,Witten:1998qj,GubKle98,Aharony:1999ti}, Itzhaki et al.~\cite{Itzhaki:1998dd} (see also~\cite{Polchinski:1999br,Boonstra:1998mp,Kanitscheider:2008kd,Lin:2025iir}) conjectured a similar duality between the near-horizon limit of non-conformal D$p$-brane backgrounds and $(p+1)$-dimensional non-conformal quantum field theories, for $p=0,1,2,4,\dots,8$.\footnote{\label{Foot:nonCFT}For all D$p$-branes except $p=3$, the dilaton field is non-constant and the near-horizon geometry of the extremal configuration is not AdS$_{p+2}\times S^{8-p}$. However, by performing a conformal transformation to the so-called dual frame, the geometry takes the AdS$_{p+2}\times S^{8-p}$ form~\cite{Boonstra:1998mp,Kanitscheider:2008kd,Lin:2025iir}. In this frame, standard holographic techniques can still be applied, even though the dual QFT is not strictly conformally invariant for $p\neq 3$. The varying dilaton reflects the fact that the Yang--Mills coupling is dimensionful, which induces a scale-dependent effective coupling and running with energy. Despite this, the dual QFT exhibits a generalized conformal structure, allowing one to identify the bulk radial coordinate with the energy scale of the field theory and to perform holographic calculations analogous to those in standard AdS/CFT.}  

In more detail, at large $N$, a stack of $N$ coincident (non-)extremal D$p$-branes is described by the corresponding $p$-brane solution of type II supergravity. Taking a decoupling limit yields the near-horizon geometry of these branes, which is dual to the $(p+1)$-dimensional quantum field theory (QFT) living on their worldvolume. This field theory is obtained by dimensionally reducing ten-dimensional $\mathcal{N}=1$ SYM to $(p+1)$ dimensions, leading to a non-conformal $SU(N)$ SYM$_{p+1}$ theory.

In this work we focus on the case $p=0$ within the above network of holographic dualities (in its low-energy limit). Here, the decoupling limit of D0-branes is dual to a non-conformal $(1+0)$-dimensional $SU(N)$ SYM theory: the Banks--Fischler--Shenker--Susskind (BFSS) model~\cite{Claudson:1984th,deWit:1988wri,Douglas:1996yp,Banks:1996vh}, also known as maximally supersymmetric quantum mechanics or the matrix model (see reviews~\cite{Polchinski:1999br,Lin:2025iir}). This case is particularly timely, since in recent years novel techniques have been developed to probe the gauge-theory side of the duality at strong coupling, including lattice simulations, bootstrap methods, and machine learning approaches~\cite{Hanada:2007ti, Catterall:2007fp, Anagnostopoulos:2007fw, Catterall:2008yz, Hanada:2008gy, Hanada:2008ez, Catterall:2009xn, Hanada:2009ne, Hanada:2013rga, Filev:2015hia, Hanada:2016pwv, Balthazar:2016utu, Han:2019wue, Bergner:2021goh, Rinaldi:2021jbg, Koch:2021yeb, Pateloudis:2022ijr, Hanada:2023rlk, Mathaba:2023non, Bodendorfer:2024egw} (see also~\cite{Lin:2025iir}).  

Our goal is to provide new insights into the gravitational side of the duality, relevant both for understanding BFSS theory in its strongly coupled regime and for enabling precision tests of holography. This work complements~\cite{Dias:2024vsc}, in that it presents important first-principle derivations, analyses, and discussions omitted in the Letter, while also introducing new results that shed light on the physical properties of the D0-brane/BFSS dual system and strengthen the consistency of its holographic interpretation. In particular, the derivations presented in Section~\ref{sec:MapEinsteinBFSS}, together with the results reported in Sections~\ref{sec:GL} and~\ref{sec:Results-Grav}, are entirely new and were not included in~\cite{Dias:2024vsc}.
 
We now provide an executive summary of the key aspects and findings of this work as we outline its structure.  

In Section~\ref{sec:decoupling}, we begin by reviewing the D0-brane/BFSS correspondence, which conjectures a duality between the decoupling (near-horizon) limit of a stack of $N$ (non-)extremal D0-branes~\eqref{NHnonExtDp} and the $(1+0)$-dimensional SYM theory~\eqref{S-BFSS} living on the D0-brane worldvolume~\cite{Itzhaki:1998dd,Boonstra:1998mp,Kanitscheider:2008kd}. Using a standard Kaluza--Klein dimensional reduction/oxidation procedure (reviewed in Section~\ref{sec:IIAto11sugra}), this type IIA solution~\eqref{NHnonExtDp} can be equivalently described within 11-dimensional supergravity by the configuration~\eqref{UniformD0uplifted}--\eqref{UniformD0uplifted:aux}, which we refer to as the \textit{uniform D0 phase}. In this uplifted description, the D0-branes are uniformly smeared along the M-theory circle and reside in a specific $pp$-wave background~\eqref{asympMetricBFSSdual}, which plays a central role in our analysis. The dual field theory state is the \textit{uniform BFSS thermal state}.  

Unlike the well-known D3-brane/SYM$_4$ duality~\cite{Maldacena:1997re,Witten:1998qj,GubKle98,Aharony:1999ti}, where the bulk geometries are asymptotically AdS$_5 \times S^5$, any supergravity solution dual to a BFSS thermal state must asymptote instead to the $pp$-wave background~\eqref{asympMetricBFSSdual}. This asymptotic structure fixes the number $N$ of D0-branes, the size of the M-theory circle, and the time-translation generator of the dual theory.  
A crucial result, demonstrated explicitly in Section~\ref{sec:GL}, is that the uniform D0 phase is unstable to the Gregory--Laflamme instability. This instability occurs at low temperatures that can only be probed within the 11-dimensional supergravity description (and not in type IIA), underscoring the need to uplift the solution, as discussed in Fig.~\ref{Fig:validitySUGRA}. The presence of this instability strongly suggests the existence of at least one novel, stable supergravity solution (and dual BFSS phase) into which the unstable uniform phase can decay.  

Motivated by the findings of Section~\ref{sec:GL}, Section~\ref{sec:MapEinsteinBFSS} develops a strategy to search for such new solutions. The first key observation is that the 11-dimensional uniform D0 phase with $pp$-wave asymptotics~\eqref{asympMetricBFSSdual} can be obtained from the standard black string solution of Einstein gravity with Kaluza--Klein asymptotics ${\mathbb R}^{(1,9)} \times S^1_{\tiny L}$ via a particular Carrolian transformation~\eqref{Carrollian:transf0}.\footnote{\label{foot:Carrol} We use the term \textit{Carrolian} for the transformation~\eqref{Carrollian:transf0} (or~\eqref{SG:CarrollianTransf}) because it resembles a boost in which the speed of light is taken to zero. This ``Carrolian boost'' freezes spatial position while time evolves, in line with the notion that space is absolute but time is relative.}  

\begin{figure}[tb]
    \centering    \includegraphics[width=0.85\textwidth]{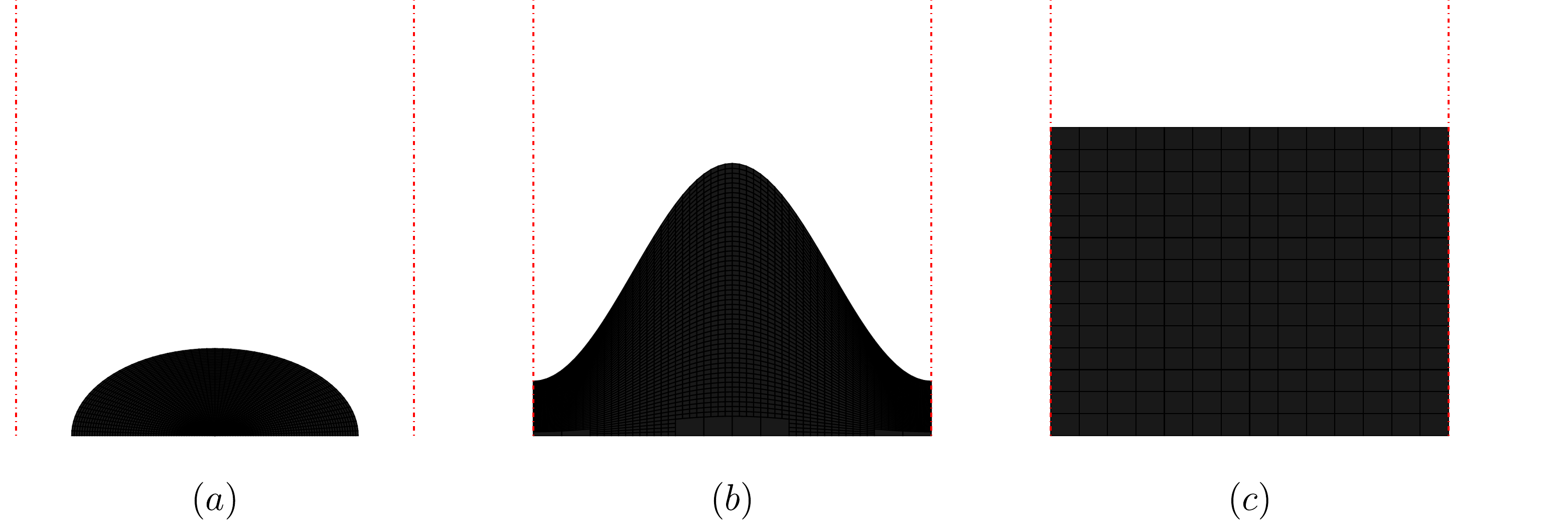}
    \caption{A schematic drawing of the transverse horizon radius for different black hole phases with standard Kaluza-Klein asymptotics: (a) the localized black hole phase, (b) the non-uniform string phase, and (c) the uniform string phase. The boundaries at the horizontal axis describe the two periodically identified boundaries of the circle $S^1_{\hbox{\tiny $L$}}$.}
    \label{fig:sketch}
\end{figure}

The second key observation is that the uniform black string of Einstein gravity is itself Gregory--Laflamme unstable ---  the prototypical and historically first example of this instability~\cite{Gregory:1993vy}. As a consequence, it is now well established that the phase diagram of Einstein gravity with Kaluza--Klein asymptotics ${\mathbb R}^{(1,9)} \times S^1_{\tiny L}$ contains additional families of solutions: namely, \textit{non-uniform black strings} and \textit{localized black holes}~\cite{Gubser:2001ac,Harmark:2002tr,Kol:2002xz,Wiseman:2002zc,Kol:2003ja,Harmark:2003dg,Harmark:2003yz,Kudoh:2003ki,Sorkin:2003ka,Kol:2003if,Sorkin:2004qq,Gorbonos:2004uc,Kudoh:2004hs,Gorbonos:2005px,Dias:2007hg,Harmark:2007md,Wiseman:2011by,Figueras:2012xj,Kalisch:2016fkm,Dias:2017uyv,Dias:2017coo,Kalisch:2017bin,Ammon:2018sin,Cardona:2018shd} (see reviews~\cite{Kol:2004ww,Harmark:2005pp,Horowitz:2011cq}). Figure~\ref{fig:sketch} illustrates how these three black objects fit into the phase structure on the circle $S^1_{\tiny L}$.

Altogether, these elements define a clear four-step strategy to construct the \textit{non-uniform} and \textit{localized D0 phases} and their dual \textit{non-uniform} and \textit{localized BFSS states}.  

\textbf{Step 1} (Sections~\ref{sec:SetupNonUnif} and~\ref{sec:localized}) constructs one-parameter families of asymptotically ${\mathbb R}^{(1,9)}\times S^1_{\tiny L}$ non-uniform black strings and localized black holes in Einstein gravity. These solutions automatically satisfy 11-dimensional supergravity with a vanishing 3-form gauge potential.  

\textbf{Step 2} (Section~\ref{sec:vacuumBHs}) reads the asymptotic Kaluza--Klein quantities of these solutions to obtain their energy, tension, and horizon thermodynamic quantities.  

\textbf{Step 3} (Section~\ref{sec:CarrolMap}) applies a unique Carrolian transformation~\eqref{SG:CarrollianTransf} to each Einstein solution, generating 11-dimensional supergravity solutions with the required $pp$-wave asymptotics~\eqref{asympMetricBFSSdual}. These define the \textit{non-uniform} and \textit{localized D0 phases}, in which the total number of D0-branes is fixed regardless of their distribution along the M-theory circle. Remarkably, the Carrolian map depends on the energy and tension along the M-circle. This transformation also maps the thermodynamics of the Kaluza--Klein solutions directly to that of the $pp$-wave supergravity solutions via the Carrollian thermodynamic map~\eqref{SG:ETzPCarrollian}--\eqref{SG:TSCarrollian}.  

\textbf{Step 4} (Section~\ref{sec:BFSSthermo}) translates the $pp$-wave supergravity thermodynamics to the dual BFSS theory, allowing to obtain the (micro-)canonical phase diagrams of the non-uniform, localized, and uniform BFSS phases. The t'Hooft coupling $\lambda = g_{\rm YM}^2 N$ allows the construction of dimensionless thermodynamic densities from the supergravity results~\eqref{QFTthermoMap}. Using perturbative results for localized Einstein black holes~\cite{Harmark:2003yz} together with the Carrollian map, Section~\ref{sec:ThermoUnifPertBFSS} provides analytical formulas for the BFSS localized phase, which are accurate to better than $0.3\%$ when compared to exact numerics.  

In summary, the workflow reduces to two essential tasks: (i) construct the (non-)uniform and localized solutions of asymptotically ${\mathbb R}^{(1,9)}\times S^1_{\tiny L}$ Einstein gravity and determine their thermodynamics, and (ii) apply the Carrollian thermodynamic map~\eqref{QFTthermoMap} to obtain the thermodynamics of the corresponding BFSS thermal states.  

Section~\ref{sec:Results} presents and discusses our results. In Section~\ref{sec:Results-Grav}, we focus on the (micro-)canonical phase diagrams of the vacuum Kaluza--Klein gravitational solutions. Section~\ref{sec:Results-BFSS} presents the dual BFSS phase diagrams. Our analysis shows that the non-uniform and localized BFSS phases dominate the microcanonical ensemble at low energies, while the localized phase also dominates the canonical ensemble at low temperatures. Above a critical energy or temperature, a first-order transition occurs to the uniform BFSS phase. These supergravity results provide valuable data for the strongly coupled BFSS regime and can be tested using lattice simulations, bootstrap methods, machine learning, and other numerical techniques~\cite{Hanada:2007ti, Catterall:2007fp, Anagnostopoulos:2007fw, Catterall:2008yz, Hanada:2008gy, Hanada:2008ez, Catterall:2009xn, Hanada:2009ne, Hanada:2013rga, Filev:2015hia, Hanada:2016pwv, Balthazar:2016utu, Han:2019wue, Bergner:2021goh, Rinaldi:2021jbg, Koch:2021yeb, Pateloudis:2022ijr, Hanada:2023rlk, Mathaba:2023non, Bodendorfer:2024egw} (see review~\cite{Lin:2025iir}).

\section{The D0-brane/BFSS correspondence \label{sec:SQM}}
The original AdS$_5$/CFT$_4$ duality establishes a correspondence between the near-horizon limit of (conformal) D3-branes $-$ which gives rise to the AdS$_5\times S^5$ geometry $-$ and a CFT, namely $(3+1)$-dimensional $SU(N)$ SYM theory~\cite{Maldacena:1997re,Witten:1998qj,GubKle98,Aharony:1999ti}.  

Soon after~\cite{Maldacena:1997re,Witten:1998qj,GubKle98,Aharony:1999ti}, Itzhaki et al.~\cite{Itzhaki:1998dd} (see also~\cite{Boonstra:1998mp,Kanitscheider:2008kd}) conjectured a similar correspondence between the near-horizon limit of non-conformal D$p$-brane backgrounds and $(p+1)$-dimensional non-conformal quantum field theories for $p=0,1,2,4,\dots,8$ (see footnote~\ref{Foot:nonCFT}). At large $N$, a stack of $N$ coincident non-extremal D$p$-branes is described by the corresponding $p$-brane solution of type II supergravity. A decoupling limit of this solution yields the near-horizon geometry, which is dual to the $(p+1)$-dimensional SYM theory living on the D$p$-brane worldvolume. This theory arises from the dimensional reduction of ten-dimensional $\mathcal{N}=1$ SYM to $(p+1)$ dimensions, producing a non-conformal $SU(N)$ SYM$_{p+1}$ theory.  

In this work, we focus exclusively on the $p=0$ case.\footnote{Detailed discussions of the generic $p$ case and associated dualities can be found in~\cite{Itzhaki:1998dd,Polchinski:1999br,Boonstra:1998mp,Kanitscheider:2008kd,Dias:2017uyv}.} In Section~\ref{sec:decoupling} we review D0-branes and the decoupling limit that gives rise to the duality with $(1+0)$-dimensional $SU(N)$ SYM, also known as matrix quantum mechanics, supersymmetric quantum mechanics, or the BFSS theory~\cite{Claudson:1984th,deWit:1988wri,Douglas:1996yp,Banks:1996vh} (see reviews~\cite{Polchinski:1999br,Lin:2025iir}).  

In this setup, the IIA supergravity description is valid only within a certain range of temperatures. To explore the system at temperatures below the IIA validity bound, we uplift the solution to 11-dimensional supergravity. The details and motivation for this oxidation procedure are described in Section~\ref{sec:IIAto11sugra}.

\subsection{D$0$-branes, their decoupling limit, and the dual SYM$_{1+0}$ (BFSS) theory \label{sec:decoupling}}
We briefly review the duality~\cite{Itzhaki:1998dd} between $(0+1)$ - dimensional $SU(N)$ SYM theory - also known as supersymmetric quantum mechanics, matrix quantum mechanics, or the BFSS theory~\cite{Claudson:1984th,deWit:1988wri,Douglas:1996yp,Banks:1996vh}—and supergravity (see also~\cite{Boonstra:1998mp,Kanitscheider:2008kd,Dias:2017uyv} and the recent reviews~\cite{Polchinski:1999br,Lin:2025iir}).

We begin by considering $N$ parallel D$0$-branes in type IIA string theory separated by a distance $r$.  In general, there are modes that propagate along the worldvolume of the brane and modes that propagate in the bulk.  Ref. \cite{Itzhaki:1998dd} identified a low energy limit where the branes become coincident, the two families of modes decouple, and the worldvolume theory on the branes reduces to the $(1+0)$-dimensional $SU(N)$ SYM, more familiarly known as BFSS theory \cite{Claudson:1984th,deWit:1988wri,Douglas:1996yp,Banks:1996vh}. BFSS theory can be obtained starting from the 10-dimensional SYM theory 
and performing a dimensional reduction to one dimension. This yields a gauge theory that includes finite $N \times N$ traceless Hermitian bosonic degrees of freedom $X^j$ and fermionic degrees of freedom $\Psi^{\alpha}$, transforming as spinors of $SO(9)$ with (low energy) action \cite{Claudson:1984th,deWit:1988wri,Douglas:1996yp,Banks:1996vh}
\begin{equation} \label{S-BFSS} 
S_{\hbox{\tiny BFSS}} =\frac{N}{2\lambda}\int {\rm d}t\;{\rm Tr}\Big\{(D_t X^j)^2+\Psi^{\alpha}D_t \Psi^{\alpha}
+\frac{1}{2}\left[X^j,X^k\right]^2+ \mathrm{i} \Psi^{\alpha}\gamma^j_{\alpha \beta}\left[\Psi^{\beta},X^j\right]\Big\}\,,
\end{equation}
where $D_t=\partial_t - \mathrm{i} [A, ]$  is the covariant derivative associated with the 1-form gauge potential $A$ and summation over spatial indices
$j,k=1, \cdots , 9$ and spinor indices  $\alpha,\beta=1,\cdots , 16$ is implicit.
 In \eqref{S-BFSS}, $\lambda\equiv g_{\rm YM}^2N$ is the t'Hooft coupling and $g_{\rm YM}$ is the gauge coupling. $\lambda$ has units of energy cubed so, at finite temperature $\mathcal{T}$, the system's thermodynamics depends on the dimensionless parameters $\tau \equiv \mathcal{T}/\lambda^{1/3}$ and $N$.

  By identifying the low energy sector of the Dirac-Born-Infeld action (describing open string excitations on  D$0$-branes) with the BFSS action \eqref{S-BFSS}, one can relate the SYM coupling constant $g_{\mathrm YM}$ to the string length $\ell_s=\sqrt{\alpha^\prime}$ and string coupling $g_s$ via
\begin{equation}\label{gYM}
g_{\rm YM}^2\equiv (2 \pi )^{-2} g_s \ell_s^{-3}\,.
\end{equation}
The decoupling limit of \cite{Itzhaki:1998dd}, valid in the t'Hooft large $N$ limit and for strong t'Hooft coupling $\lambda=g_{\rm YM}^2 N$, sends $\ell_s\to0$ while keeping $g_{\rm YM}$ fixed.  This limit suppresses higher-order $\alpha^\prime$ corrections  and sends the gravitational Newton's constant $G_{10}$ to zero. Indeed, recall that identifying the low-energy action of type IIA (closed) string theory  and the action of type IIA supergravity yields $16 \pi  G_{10}\equiv (2 \pi )^7 g_s^2 \ell_s^8$. This limit is taken at fixed energy $U=\frac{r}{\ell_s^2}$ (which measures the mass of the stretched strings between branes) and fixed charge $K_0$, so that it brings all the $N$ branes together ($r\to 0$ as $\ell_s\to 0$) while keeping the Higgs expectation values corresponding to the brane separations fixed (that is, this limit retains finite energy excitations of the D0-branes).\footnote{Another way to see this is to place one of the D$0$-branes at a position $r$. This configuration breaks the symmetries $U(N)\to U(N-1)\times U(1)$, giving an expectation value (with dimensions of energy) to some of the fields that scales as $r$ \cite{Itzhaki:1998dd}.} 
At finite energies $U$, the effective dimensionless SYM coupling is given by $g_{\mathrm{eff}}^2\approx g_{\mathrm{YM}}^2NU^{-3}$. Perturbative SYM is valid for large $U$, and the theory is UV free. 

On the other hand, a stack of $N$ coincident D$0$-branes can also be described within classical type IIA supergravity, provided that curvature scales remain small compared to the string scale (to suppress $\alpha'$ corrections) and the effective dimensionless string coupling is sufficiently small (to suppress string loop effects).  This classical theory contains a non-vanishing graviton $g$, dilaton $\phi$ and Ramond-Ramond (RR) $A_{(1)}$ fields with action (here in the string frame)
\begin{equation}
\label{action}
S_{\hbox{\tiny IIA}}^{(s)} = \frac{1}{(2 \pi )^7 \ell_s^8} \int \mathrm d^{10} x \sqrt{-g} \Big[ e^{-2\phi } \Big(R
+4 \partial_\mu \phi \partial^\mu \phi \Big) - \frac{1}{2} \, \frac{1}{2!}  (\mathrm dA_{(1)})^2 \Big].
\end{equation} 
For completeness,  the corresponding equations of motion \eqref{IIeomStringFr} and their map to the Einstein frame are given in Appendix \ref{sec:EOM}.
A stack of $N$ coincident non-extremal D$0$-branes (at large $N$) is described within IIA supergravity by 0-branes (see e.g. \cite{Itzhaki:1998dd,Harmark:1999xt,Harmark:2005jk,Kanitscheider:2008kd}):
\begin{eqnarray}\label{nonExtDp}
&& {\mathrm ds}^2= -f H^{-\frac{1}{2}} {\mathrm d}t^{\,2} +H^{\frac{1}{2}}\left(\frac{{\mathrm d}r^2}{f}+r^2 {\mathrm d}\Omega_8^2\right),  \nonumber \\
&& e^{\phi}=g_s H^{\frac{3}{4}},\qquad \qquad \qquad 
 A_{(1)}= g_s^{-1}\coth \beta \left(H^{-1}-1\right) {\mathrm d}t,\nonumber \\
&& \hbox{where} \quad f=1-\frac{r_0^{7}}{r^{7}}, \qquad \quad  H=1+\frac{r_0^{7}}{r^{7}}\sinh^2\beta, 
\end{eqnarray}
where the dimensionless string coupling is given by $g_s=e^{\phi_\infty}$, $r_0$ is the horizon location,  $\mathrm{d}\Omega_8^2$ is the line element of a unit radius sphere $S^8$ and $\beta$ is a parameter that sources the gauge potential.
The mass, charge,  temperature, entropy, and chemical potential of these solutions are, respectively,
\begin{eqnarray}
\label{nonExtD$p$Thermo}
&& M_0 = V_0 \frac{\Omega_8}{16\pi G_{10}} r_0^{7}
\Big[ 8 + 7\sinh^2 \beta \Big], \nonumber \\
&&  Q_0 =V_0 \frac{\Omega_8}{16\pi G_{10}} 7 \,r_0^{7}  \sinh \beta \cosh \beta,
\nonumber \\
&& T_0 = \frac{7}{4 \pi\, r_0 \,\cosh \beta},\qquad  S_0 = V_0 \frac{\Omega_8}{4G_{10}} r_0^{8}  \cosh \beta,\qquad \mu_0 = \tanh \beta,
\end{eqnarray}
where $\Omega_{n}=\frac{2 \pi ^{\frac{n+1}{2}}}{\Gamma \left(\frac{n+1}{2}\right)}$ is the area of a unit radius S$^n$, $V_0$ is the D$0$-brane worldvolume. These quantities satisfy the thermodynamic first law $\mathrm dM_0 = T_0 \mathrm dS_0 +\mu_0 \mathrm dQ_0$ and the Smarr relation $7\,M_0=8\,T_0 S_0+7\,\mu_0 Q_0$.

One arrives at the conjectured duality \cite{Itzhaki:1998dd} between type IIA string theory and SYM$_{(1+0)}$ by taking the corresponding decoupling limit of the type IIA supergravity solution \eqref{nonExtDp}. To do so, we must complete the relationships between quantities in \eqref{nonExtDp} and stringy quantities $g_s$ and $\ell_s$, and then match them to SYM quantities $g_{\mathrm YM}$, $U$, and $K_0$. The coupling constants are already related via \eqref{gYM} and $16 \pi  G_{10}\equiv (2 \pi )^7 g_s^2 \ell_s^8$.  The relation between charges is given by 
\begin{equation}\label{defKp}
K_0\equiv \frac{(2 \pi )^7 g_s^2 Q_0}{7 V_0 \Omega_{8} \ell _s^{2}}= \frac{(2 \pi)^{7} g_s N}{7 \,\Omega_8\ell_s^{3}}\;,
\end{equation}
which can be obtained by matching the 0-brane charge, computed in the string frame via
\begin{equation}\label{defQp}
Q_0=\frac{V_0}{(2\pi)^7 \ell_s^8}\int_{S^{8}}\star \mathrm{d}A_{(1)}\,,
\end{equation}
with the charge of $N$ D$0$-branes $Q_0\equiv N \,\mathfrak{t}_0 V_0$, where $\mathfrak{t}_0=\ell_s^{-1}g_s^{-1}$ is the D$0$-brane tension.  As previously discussed, the energy is given by $U=\frac{r}{\ell_s^2}$ (and $U_0=\frac{r_0}{\ell_s^2}$). To summarise, the decoupling limit is given by \cite{Itzhaki:1998dd} 
\begin{equation}\label{NHlim}
 \ell_s\to 0, \qquad g_{\rm YM}^2={\rm fixed}, \qquad {U}\equiv \frac{r}{\ell_s^2}={\rm fixed}, \qquad {U}_0\equiv \frac{r_0}{\ell_s^2}={\rm fixed}, \qquad 
 K_0 ={\rm fixed}.
\end{equation}
Applying this decoupling limit to \eqref{nonExtDp} yields the near-horizon limit of $N$ non-extremal D$0$-branes\footnote{Note that we gauge away the constant term in $A_{(1)}$ and used the fact that $\beta\to\infty$ in the decoupling limit. Note that $\beta$ itself can be related to SYM quantities in this limit via \eqref{defKp}.}:
\begin{eqnarray}\label{NHnonExtDp}  
&& {\mathrm ds}^2= \ell_s^2\Bigg\{ - \frac{U^{\frac{7}{2}}}{\sqrt{d_0} g_{\rm YM} \sqrt{N}}\left( 1-\frac{U_0^{7}}{U^{7}} \right) {\mathrm d}t^2  + \frac{\sqrt{d_0} g_{\rm YM} \sqrt{N}}{U^{\frac{7}{2}}}\Bigg[
\frac{{\mathrm d}U^2}{1-\frac{U_0^{7}}{U^{7}}} 
+U^2 {\mathrm d}\Omega_8^2\Bigg] \Bigg\},  \nonumber \\
&& e^{\phi}=(2\pi)^{2}\,g_{\rm YM}^2\, \frac{\left(d_0 g_{\rm YM}^2 N \right)^{\frac{3}{4}}}{U^{\frac{21}{4}}} , \qquad \hbox{with} \:\: d_0\equiv 2^{7} \pi ^{\frac{9}{2}} \Gamma \left(\frac{7}{2}\right). \nonumber \\
&& A_{(1)}=g_s (2\pi)^{-2}\ell_s\, \frac{U^{7}}{d_0 g_{\rm YM}^4 N}  {\mathrm d}t\,,
\end{eqnarray}

The energy above extremality, $\mathcal{E}_0\equiv M_0-Q_0$, the entropy $\mathcal{S}_0$ and temperature $\mathcal{T}_0$ of these near-horizon solutions are \cite{Itzhaki:1998dd,Harmark:1999xt,Harmark:2004ws,Kanitscheider:2008kd,Assel:2015nca}\footnote{Note that the energy \eqref{NHnonExtDpThermo} of \eqref{NHnonExtDp} agrees with the energy computed using holographic renormalization if we further use the supersymmetric limit to fix a constant that is left undetermined by the holographic renormalization procedure \cite{Kanitscheider:2008kd,Assel:2015nca}.} 
\begin{align}
\label{NHnonExtDpThermo}
&\mathcal{E}_0 = V_0\frac{9}{56 \pi ^2} \frac{1}{d_0}   \frac{1}{ g_{\rm YM}^4}  U_0^{7},  \nonumber\\
&\mathcal{S}_0 =V_0 \frac{1}{7 \pi} \frac{\sqrt{N}}{\sqrt{d_0}} \frac{1}{g_{\rm YM}^3}U_0^{\frac{9}{2}},\qquad \qquad
\mathcal{T}_0 = \frac{7}{4 \pi } \frac{1}{\sqrt{d_0 N}} \frac{1}{g_{\rm YM}}U_0^{\frac{5}{2}},
\end{align}
which obey the thermodynamic first law $\mathrm d\mathcal{E}_0 = \mathcal{T}_0 \mathrm d\mathcal{S}_0 $ and the Smarr relation $\mathcal{E}_0=\frac{1}{2}\frac{9}{7} \,\mathcal{T}_0 \mathcal{S}_0$.

As we have mentioned, classical supergravity is only valid when curvature scales are much smaller than the string scale and for small dimensionless string coupling. The curvature is given by $\alpha^\prime R\sim1/{g_{\mathrm{eff}}}$, and the effective string coupling goes as $e^\phi\sim g_{\mathrm{eff}}^{7/2}/N$.  Together, these give \cite{Itzhaki:1998dd}
\begin{equation}\label{sugravalidity}
1\ll g_{\mathrm{eff}}^2\ll N^{4/7}\;.
\end{equation}
Thus, the validity of classical supergravity requires the dual SYM theory to be strongly coupled and $N$ to be large. Note that $g_{\mathrm{eff}}$ depends on the energy $U$ so, at fixed $g_{\mathrm{YM}}$, the validity of the supergravity theory depends upon the energy (or temperature). For future use, it is convenient to rewrite the validity window \eqref{sugravalidity} in terms of the temperature:
\begin{equation}\label{sugravalidityT:D0inIIA}
\lambda ^{\frac{1}{3}} N^{-\frac{10}{21}}\ll  \mathcal T_0 \ll \lambda ^{\frac{1}{3}} 
\qquad \Leftrightarrow \qquad N^{-\frac{10}{21}}\ll  \mathcal \tau \ll 1\,, \quad \hbox{with}\quad \tau\equiv \mathcal{T}_0/\lambda^{1/3}\,,
\end{equation}
where, as justified previously, we use the t'Hooft coupling $\lambda\equiv g_{YM}^2 N$ to work with the dimensionless temperature $\tau$.
Note that the lower bound in  \eqref{sugravalidityT:D0inIIA} ensures that the dilaton (at the horizon) is small and thus string coupling corrections are suppressed (i.e. it corresponds to the upper bound in \eqref{sugravalidity}).
On the other hand, the upper bound in \eqref{sugravalidityT:D0inIIA} is required to have small curvature in string units (so that $\alpha'$ corrections are negligible), i.e. it corresponds the lower bound in \eqref{sugravalidity}. 

\subsection{Oxidation of IIA phases to 11 dimensional supergravity\label{sec:IIAto11sugra}}
 
We are interested  on the duality between  $(0+1)$ dimensional $SU(N)$ SYM theory (a.k.a. super quantum mechanics or matrix quantum mechanics or BFSS theory) and supergravity \cite{Banks:1996vh,Itzhaki:1998dd}. 

Summarizing the previous subsection, there is a duality between  $(0+1)$ dimensional $SU(N)$ SYM theory (BFSS theory) and, in the low energy classical limit, IIA supergravity with 1-form gauge field sourcing D0-branes.  
The thermodynamics of this system is controlled by two dimensionless parameters: $N$ and $\tau=T/\lambda^{1/3}$, where $T$ is the temperature. 
Under the decoupling limit \eqref{NHlim}, the $(0+1)$ dimensional BFSS theory lives on the worlvolume  of a stack of $N$ non-extremal D0-branes described, in the string frame of type IIA supergravity, by  \eqref{NHnonExtDp}. 
This IIA supergravity description is only valid in the window of temperatures \eqref{sugravalidityT:D0inIIA}; in particular, for $\tau\gg N^{-\frac{10}{21}}$.
However, for reasons that will become clear soon, we would like to probe the physics of the system for lower temperatures, $\tau< N^{-\frac{10}{21}}$. 

Fortunately, at large $N$ this low temperature regime is accessible through 11-dimensional supergravity. Indeed, recall that type IIA supergravity is a dimensional reduction of 11-dimensional supergravity, i.e. the classical low energy limit of M-theory. Conversely, the near-horizon limit of D0-branes, namely \eqref{NHnonExtDp}, can be oxidated and has a 11-dimensional supergravity description. Crucially, the latter description is valid for temperatures lower than \eqref{sugravalidityT:D0inIIA}, as we discuss next. 

To provide this description, let the 11-dimensional compact direction, parametrising the M-theory circle,  be $z\sim z+L_{11}$ with length $L_{11}=2\pi g_s\ell_s$ such that the 11-dimensional Newton's gravitational constant is $G_{11}=L_{11}G_{10}=16 \pi^7 g_s^3 \ell_s^9\equiv 16 \pi^7 \ell_P^9$ where $\ell_P\equiv g_s^{1/3} \ell_s$ is the (11-dimensional) Planck length. A standard Kaluza-Klein oxidation procedure along this $z$-direction guarantees that if the graviton ${\mathrm{d}S}^{\,2}_{\hbox{\tiny(E)}}$ (in the Einstein frame), the gauge field $A_t $ and the dilaton $\phi$ describe altogether the NH D0-brane solution of type IIA supergravity then  
\begin{equation}\label{KKuplift}
{\mathrm ds}^2_{11}= e^{-\frac{1}{6}(\phi-\phi_\infty)}{\mathrm{d}s}^{\,2}_{\hbox{\tiny(E)}}+e^{\frac{4}{3}(\phi-\phi_\infty)} \left( {\mathrm d}z+A_t \, {\mathrm d}t \right)^2, 
\end{equation}
provides the 11-dimensional supergravity description of the solution. This graviton field \eqref{KKuplift} solves the field equations \eqref{eqs:Msugra} of 11-dimensional supergravity which reduce simply to $R_{\mu\nu}=0$. That is to say, since the 3-form gauge potential $A_{(3)}$ of 11-dimensional supergravity vanishes in our system, 11-dimensional supergravity reduces simply to 11-dimensional vacuum Einstein gravity; see \eqref{actionMsugra}-\eqref{eqs:Msugra} with $G_{(4)}=\mathrm{d}A_{(3)}=0$.   In more detail, we start with fields $\mathrm{d}s^2$, $A_t$ and $\phi$ $-$ given in \eqref{NHnonExtDp} $-$ that describe a stack of $N$ (non-)extremal D0-branes. Since in \eqref{NHnonExtDp} the graviton $\mathrm{d}s^2$ is in the string frame we then rewrite it in the Einstein frame, ${\mathrm{d}S}^{\,2}_{\hbox{\tiny(E)}}= e^{-\frac{1}{2}(\phi-\phi_\infty)} \mathrm{d}s^2$, before inserting our solution into \eqref{KKuplift} (recall that $e^{\phi_{\infty}}=g_s$ and the dilaton and gauge field are the same in both frames). In these conditions \eqref{KKuplift} yields a solution that can be rewritten in a more familiar form. For that  we take $U=r/\ell_s^2$ and $U_0=r_0/\ell_s^2$ and we apply a $U(1)$ gauge transformation that shifts $A_t$ in  \eqref{NHnonExtDp} into a gauge where it vanishes at the horizon, 
$A_t\to  A_t =\frac{g_s (2\pi)^{-2}\ell_s}{d_0 g_{\rm YM}^4 N} \left( U^7-U_0^7\right)$. This yields 
\begin{equation}\label{UniformD0uplifted}
{\mathrm ds}^2_{11}= -f(r)\,\mathrm{d}t\left( \frac{r_0^7}{\mathcal{R}^7}\,\mathrm{d}t -2 \mathrm{d}z \right) +\frac{\mathrm{d}r^2}{f(r)} + r^2\mathrm{d}\Omega^2_{8}+\frac{\mathcal{R}^7}{r^7}\,\mathrm{d}z^2
\end{equation}
where we have defined $f(r)$ and $\mathcal{R}$ as
\begin{equation}\label{UniformD0uplifted:aux}
f(r)=1-\frac{r_0^7}{r^7}\,,\qquad \frac{\mathcal{R}^7}{\ell_s^7}=60\pi^3g_s N\,, \qquad \hbox{and}\quad  \frac{r_0^5}{\ell_s^5}=\frac{120\pi^2}{49}\left(2\pi g_s N \right)^{5/3}\tau^2
\end{equation}
which follows from the definition \eqref{NHnonExtDpThermo} for the temperature of the system $\mathcal{T}_0=\lambda^{1/3}\tau$ in terms of $r_0$. This solution is the 11-dimensional supergravity description a stack of $N$ (non-)extremal D0-branes uniformly extended along the 11-dimensional circle.  Onwards, we will often refer to it as simply the `{\it uniform D0 phase}' (dual to the `{\it uniform BFSS phase}').

\begin{figure}[th]
\centering
\includegraphics[width=1.0\textwidth]{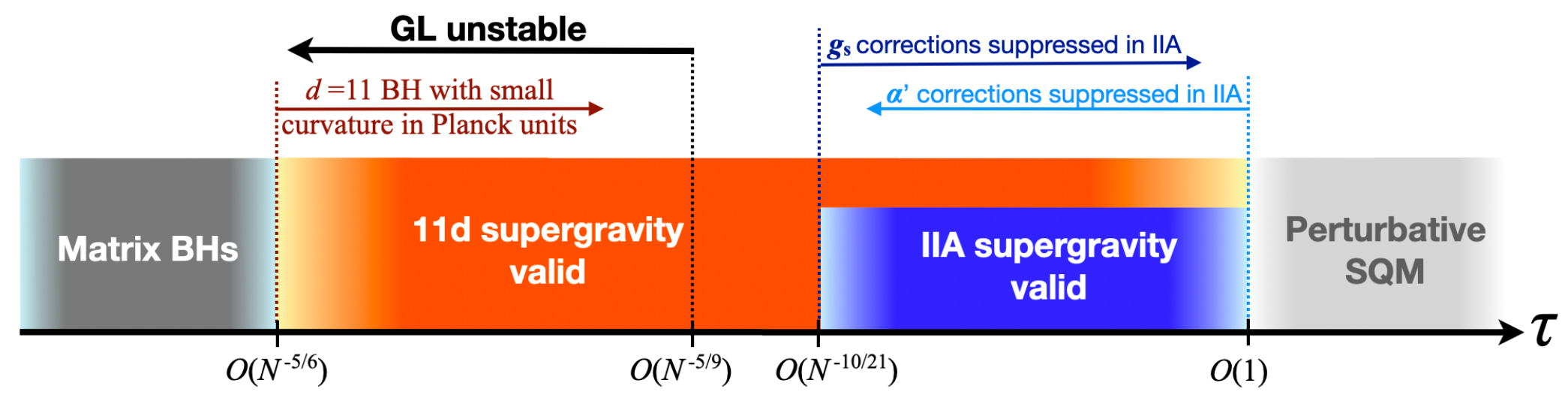}
\caption{Schematic regime of validity for matrix black holes (BHs), $d=11$ and IIA supergravities and perturbative super quantum mechanics (SQM). Bright colors describe the region where the given theory is valid and the increasingly faded colors represent the regions where it becomes an increasingly poor approximation. Note that when type IIA supergravity is valid, so is $d=11$ supergravity (they both have the same upper cut-off temperature where $\alpha'$ corrections start being non-negligible and both supergravities stop being a good approximation to type IIA  string theory and M-theory, respectively). However, the 11-dimensional supergravity description is valid for lower temperatures $\tau$ than type IIA supergravity. 
Here, we take $N\gg 1$, which is required for  the Gregory-Laflamme (GL) transition at $\mathcal{O}(N^{-5/9})$ to be visible within 11-dimensional  supergravity.}\label{Fig:validitySUGRA}
\end{figure}

The crucial advantage of uplifting the type IIA solution (that describes the near-horizon limit of a stack of D0-branes) to 11-dimensional supergravity is that the system is now described by \eqref{UniformD0uplifted} which has a regime of validity that is not restricted to \eqref{sugravalidityT:D0inIIA}. The key observation here is that the supergravity description of M-theory, and thus \eqref{UniformD0uplifted}, is valid for much smaller temperatures that the lower limit of  \eqref{sugravalidityT:D0inIIA} assuming, as we have been doing, that $N$ is large. Indeed, the 11-dimensional supergravity approximation is valid while the gravitational curvature of the solution at hand stays well below the 11-dimensional Planck scale. Consider then the uplifted uniform D0-brane solution in \eqref{UniformD0uplifted}. Its horizon curvature, characterized by the Kretschmann scalar, scales as $r_0^{-4}$. To ensure the validity of the supergravity description, the dimensionless combination $r_0^{-4}\ell_P^4$ must be sufficiently small. Using the last relation for $r_0(\tau) $ in \eqref{UniformD0uplifted:aux}, together with $\ell_P = g_s^{1/3}\ell_s$, leads to the lower bound $\tau\gg N^{-5/6}$. We have also evaluated the same requirement using the 11-dimensional `localized' phase  (which will be introduced only in the next section) but $-$ as we shall see  after presenting \eqref{ThermoSYMpertLoc2} $-$  this yields the lower bound  $\tau \gg N^{-4/3}$). So the bound above for the uniform phase is the most stringent (for large $N$) if we wish to keep both solutions under control. Summarizing, the 11-dimensional description \eqref{UniformD0uplifted} of the near-horizon limit of a stack of  uniform D0-branes (and actually of non-uniform and localized solutions that will be introduced later) is valid for the window of temperature 
\begin{equation}\label{sugravalidityT:D0inM}
 N^{-\frac{5}{6}}\ll  \mathcal \tau \ll 1\,, \qquad \hbox{with}\quad \tau\equiv \mathcal{T}/\lambda^{1/3}\,,
\end{equation}
where the upper bound ensures that curvatures remain small in string units, and the lower
bound ensures that curvatures are small compared to the eleven-dimensional Planck length.

For large $N$, the lower limit  of $N^{-5/6}$ in \eqref{sugravalidityT:D0inM} can be much smaller than the one of $N^{-10/21}$ in  \eqref{sugravalidityT:D0inIIA}. That is, the 11-dimensional supergravity description is valid for much smaller temperatures than the type IIA description and allows to explore physics of the system that might occur only at low temperatures: the window of temperatures where the type IIA and 11-dimensional supergravity descriptions are valid are schematically represented in Fig.~\ref{Fig:validitySUGRA}. 
This is particular relevant for us since \eqref{UniformD0uplifted} is expected to suffer from an instability for temperatures that are not captured by the IIA supergravity description but are probed by the 11-dimensional supergravity description, as we explain next.

To see this, the first important observation is that \eqref{UniformD0uplifted} effectively describes a uniform black string solution with momentum in 11-dimensions. To show this is the case, we start with the standard 11-dimensional (uniform) black string line element ${\rm d}s^2=-f(r){\rm d}T^2+\frac{{\rm d}r^2}{f(r)}+r^2 \mathrm{d}\Omega_8^2+{\rm d}Z^2$ and apply the Carrolian coordinate transformation (footnote \ref{foot:Carrol} justifies the nomenclature choice)  
\begin{equation}\label{Carrollian:transf0}
T=\frac{r_0^{7/2}}{\mathcal{R}^{7/2}}\,t -\frac{\mathcal{R}^{7/2}}{r_0^{7/2}}\,z\,, \qquad Z=\frac{\mathcal{R}^{7/2}}{r_0^{7/2}}\,z \,,
\end{equation}
which indeed yields \eqref{UniformD0uplifted}. The second relevant observation is that uniform black strings are known to be afflicted by the Gregory-Laflamme instability when there is a hierarch of scales between its length and transverse lengthscales \cite{Gregory:1993vy,Gregory:1994bj,Kleihaus:2007dg,Dias:2009iu,Dias:2010eu,Dias:2010maa,Dias:2011jg,Dias:2022mde}.
We expect this behavior to occur well within the regime where the 11-dimensional supergravity description is valid. Deep in this regime, the entropy should scale linearly with $N$.
  This in turn allows us to estimate the scaling at which we expect the Gregory-Laflamme instability to set in as \cite{Horowitz:1997fr,Itzhaki:1998dd}
\begin{equation}\label{GLcondU0}
\frac{S}{N} \sim \frac{1}{N}\frac{r_0^9}{4 G_{11}} \sim \frac{1}{N}\frac{(\ell_s^2 U_0)^9}{L_{11} G_{10}} \sim \frac{1}{N} \frac{U_0^9 \ell_s^9}{g_s^3} \lesssim 1 \quad \Leftrightarrow \quad U_0 \lesssim g_{YM}^{2/3} N^{1/9},
\end{equation}
where we have used $r_0 = \ell_s^2 U_0$, $L_{11} = 2 \pi g_s \ell_s$, and $G_{11} = L_{11} G_{10} = 16 \pi^7 g_s^3 \ell_s^9$.


As before, we can translate this condition \eqref{GLcondU0} into a bound for the temperature using the expression $\mathcal{T}_0(U_0)$ in  \eqref{NHnonExtDpThermo} and $\mathcal{T}_0=\lambda^{1/3}\tau$. One finds that the uniform black string should be Gregory-Laflamme unstable for a temperature of order 
\begin{equation}\label{GLcondT}
\tau \lesssim \tau_{\hbox{\tiny GL}} \qquad \hbox{with} \quad \tau_{\hbox{\tiny GL}}\sim N^{-\frac{5}{9}}\,. \:\: 
\end{equation}
This critical temperature $ \tau_{\hbox{\tiny GL}}$ is also schematically pinpointed in Fig.~\ref{Fig:validitySUGRA}. We see that, for large $N$, this instability is present for temperatures where the system is described by 11-dimensional supergravity but where the type IIA supergravity is not a good approximation, as previously anticipated. In section~\ref{sec:GL} we will study perturbations of the uniform D0 phase \eqref{UniformD0uplifted}-\eqref{UniformD0uplifted:aux} and conclude that the onset of the Gregory-Laflamme instability occurs at $\varepsilon \simeq 1.71 \, N^{4/9}$, i.e. $\tau\simeq 0.59  \, N^{-\frac{5}{9}}$, which indeed is in line with the expectation \eqref{GLcondT} and its interpretation: see \eqref{GL:energy} and \eqref{GL:temp}. 

This has far-reaching consequences. Since the solution describing a stack of D0 branes uniformly extended along the 11-dimensional direction is unstable, there should be another solution that describes the stable phase (and with higher entropy for given mass and charge). Namely, as typically standard in Gregory-Laflamme systems, we expect that the system has non-uniform strings and localized black holes on the M-circle  besides the uniform phase. And in $d=11$ we can expect that the localized black hole solution is the phase that should dominate the microcanonical ensemble for sufficiently small energies, and the canonical ensemble for sufficiently small temperatures. Our aim in section~\ref{sec:NumConstruction} will be to construct numerically the non-uniform and localized phases to confirm some of the expectations above (and find surprising results). But before doing so we will confirm that the uniform solution is indeed unstable to the Gregory-Laflamme instability in section~\ref{sec:GL} and we will identify the map between 11-dimensional supergravity solutions and BFSS thermal states in section~\ref{sec:MapEinsteinBFSS}.

\section{The Gregory-Laflamme instability on the uniform BFSS dual phase \label{sec:GL}}

One may ask how our results align with the dynamics of the original Gregory–Laflamme instability \cite{Gregory:1993vy,Lehner:2010pn,Figueras:2022zkg}, which shows that the uniform black string becomes unstable at sufficiently low energies. To clarify this, it is crucial to note that the Carrollian transformation~\eqref{Carrollian:transf0} does not preserve the dynamics, even though it maps stationary solutions. In particular, the linear analysis of \cite{Gregory:1993vy} expands the metric functions into Fourier modes of the form $e^{-{\rm i}\,\Omega\,T+2\pi n \, {\rm i}\,Z}$. However, under~\eqref{Carrollian:transf0}, these Fourier modes are mapped to modes in $t$ and $z$ that fail to respect the required periodicity in $z$. Recall also that for the unstable Gregory–Laflamme modes, one has $\Omega = {\rm i}\,\hat{\Omega}$ with $\hat{\Omega}>0$.

To investigate the linear stability of the line element given in \eqref{UniformD0uplifted}, we employ the following metric ansatz:
\begin{multline}
{\rm d}s^2=-\frac{r^7}{\mathcal{R}^7}f(r)Q_1(t,r,z)\left[{\rm d}t+Q_6(t,r,z){\rm dr}\right]^2+\frac{Q_2(t,r,z)\left[{\rm d}r+Q_7(t,r,z){\rm d}z\right]^2}{f(r)}
\\
+\frac{\mathcal{R}^7}{r^7}Q_3(t,r,z)\left[{\rm d}z+\frac{r^7}{\mathcal{R}^7}f(r)Q_4(t,r,z){\rm d}t\right]^2+r^2Q_5(t,r,z){\rm d}\Omega^2_8
\end{multline}
where we take
\begin{equation}
\begin{aligned}
&Q_I(t,r,z)=1+\epsilon \,e^{- {\rm i}(\omega t-k z)}q_I(r)\,\quad \text{for}\quad I=1,\ldots,5\,,
\\
&Q_I(t,r,z)=\epsilon \, e^{- {\rm i}(\omega t-k z)}q_I(r)\,\quad \text{for}\quad I=6,7\,,
\end{aligned}
\end{equation}
and expand the equations of motion to first order in $\epsilon$. At this stage no gauge has yet been imposed, so all seven functions $q_I$ remain unfixed.

A generic infinitesimal diffeomorphism $\xi$, compatible with the imposed $SO(9)$ symmetry, is given by
\begin{equation}
\xi=\left[{\xi_t(r){\rm d}t}+{\xi_r(r){\rm d}r}+{\xi_z(r){\rm d}z}\right]e^{-{\rm i}(\omega t-k z)}\,,
\end{equation}
under which, our several metric perturbations transform as $q_I \to q_I+ \delta q_I$ with
\begin{subequations}
\begin{align}
& \delta q_1(r)=\xi _r(r) \left[f'(r)+\frac{7 f(r)}{r}\right]+2 {\rm i} \xi _t(r) \left[\frac{\omega  \mathcal{R}^7}{r^7 f(r)}+k\right]-\frac{2 {\rm i} \xi _z(r) \left[k r^7 f(r)+\omega  \mathcal{R}^7\right]}{\mathcal{R}^7}\,,
\\
&
 \delta q_2(r)=\xi _r(r) f'(r)+2 f(r) \xi _r'(r)\,,
\\
&
 \delta q_3(r)=-\frac{7 f(r) \xi _r(r)}{r}+\frac{2 {\rm i} k r^7 \xi _z(r)}{\mathcal{R}^7}\,,
\\
&
 \delta q_4(r)=\xi _r(r) \left[f'(r)+\frac{7 f(r)}{r}\right]-{\rm i} \xi _z(r) \left[\frac{\omega }{f(r)}+\frac{2 k r^7}{\mathcal{R}^7}\right]+\frac{{\rm i} k \xi _t(r)}{f(r)}\,,
\\
&
 \delta q_5(r)=\frac{2 f(r) \xi _r(r)}{r}\,,
\\
&
 \delta q_6(r)=\frac{{\rm i} \omega  \mathcal{R}^7
   \xi _r(r)}{r^7 f(r)}-\frac{\mathcal{R}^7 (f(r)-1) \left[r f'(r)+7 f(r)\right] \xi _t(r)}{r^8 f(r)^2}
\nonumber \\
 & \hspace{1.5cm}  +\frac{\left[r f'(r)+7 f(r)-7\right] \xi _z(r)}{r}-\frac{\mathcal{R}^7 \xi _t'(r)}{r^7 f(r)}\,,
\\
&
 \delta q_7(r)=\frac{\mathcal{R}^7 \left[r f'(r)+7 f(r)\right] \xi _t(r)}{r^8}-\frac{f(r) \left[r f'(r)+7 f(r)-7\right] \xi _z(r)}{r}
\nonumber \\
  & \hspace{1.5cm}  +{\rm i} k f(r) \xi _r(r)+f(r) \xi _z'(r)\,.
 \end{align}
\end{subequations}
Note, in particular, that $\xi_r$ appears algebraically in the transformation of $q_5$, which allows us to set $q_5=0$. The two remaining gauge degrees of freedom are then used to impose two linear relations among the other variables. Specifically, we impose
\begin{equation}
q_1(r) - q_3(r) - \left(2 - \frac{r_0^7}{r^7}\right) q_4(r) = 0 \, .
\end{equation}
This condition is admissible because the combination
\begin{equation}
\hat{\xi}(r) \equiv \xi_t(r)\,(k r_0^7 - 2 \omega \mathcal{R}^7) - r_0^7 \omega \, \xi_z(r)
\end{equation}
enters only algebraically in the variation
\begin{equation}
\delta q_1(r) - \delta q_3(r) - \left(2 - \frac{r_0^7}{r^7}\right) \delta q_4(r) 
= \mathrm{i} \, \frac{\hat{\xi}(r)}{\,r_0^7 - r^7\,} \, .
\end{equation}
For the final gauge condition, which fixes $\xi_z$ (and thereby removes all residual gauge freedom), we impose
\begin{equation}
q_6(r) + \frac{r^7 r_0^7 \omega \mathcal{R}^7}{\bigl(r^7 - r_0^7\bigr)^2 \bigl(k r_0^7 - 2 \omega \mathcal{R}^7\bigr)} \, q_7(r) = 0 \, .
\end{equation}
This choice is consistent because
\begin{multline}
\delta q_6(r) + \frac{r^7 r_0^7 \omega \mathcal{R}^7}{\bigl(r^7 - r_0^7\bigr)^2 \bigl(k r_0^7 - 2 \omega \mathcal{R}^7\bigr)} \, \delta q_7(r)=\frac{7 r_0^{14} \omega  \mathcal{R}^7\left(k r_0^7-\omega  \mathcal{R}^7\right)}{r^{15} f(r)^2 \left(k r_0^7-2 \omega  \mathcal{R}^7\right){}^2} \xi_z(r) 
\\
+\frac{7 r_0^7 \mathcal{R}^7 \left(k
   r_0^7-\omega  \mathcal{R}^7\right)}{r^{15} f(r)^2 \left(k r_0^7-2 \omega  \mathcal{R}^7\right)^2} \hat{\xi }(r)-\frac{\mathcal{R}^7}{r^7 f(r) \left(k r_0^7-2 \omega  \mathcal{R}^7\right)} \hat{\xi }'(r)+\frac{2 i
   \omega  \mathcal{R}^7 \left(k r_0^7-\omega  \mathcal{R}^7\right)}{r^7 f(r) \left(k r_0^7-2 \omega  \mathcal{R}^7\right)}\xi_r(r)  \, ,
\end{multline}
where $\xi_z(r)$ again appears only algebraically, and $\hat{\xi}(r)$ and $\xi_r(r)$ has already been fixed by the previous conditions.

We now change to new variables, $h_r$, $h_v$, $h_z$ and $h_t$, which automatically respect the relations detailed above, and allow for an easier presentation of our results (recall that $q_5=0$):
\begin{equation}
\begin{aligned}
 &q_1(r)=-\frac{r^7 h_t(r)}{r_0^7 f(r)}-\frac{r^7 f(r) h_z(r)}{r_0^7}\,,
 \\
 &q_2(r)=f(r) h_r(r)\,,
 \\
 &q_3(r)=\frac{r^7 h_t(r)}{r_0^7}+\frac{r^7 h_z(r)}{r_0^7}\,,
 \\
 &q_4(r)=-\frac{r^7 h_t(r)}{r_0^7 f(r)}-\frac{r^7 h_z(r)}{r_0^7}\,,
 \\
 &q_6(r)=\frac{{\rm i} \omega  \mathcal{R}^7 h_v(r)}{r^7-r_0^7}\,,
 \\
 &q_7(r)=-\frac{{\rm i} f(r) h_v(r) \left(k r_0^7-2 \omega  \mathcal{R}^7\right)}{r_0^7}\,.
\end{aligned}
\end{equation}
The perturbed Einstein equation yields
\begin{subequations}
\begin{align}
& h_v(r)=\frac{r^8 f(r) h_r(r)}{7 r_0^7}\,,
\\
& h_r(r)=-\frac{r_0^{14} \left(k^2 r^9+28 \mathcal{R}^7\right)-2 k r^9 r_0^7 \omega  \mathcal{R}^7+r^9 \omega ^2 \mathcal{R}^{14}}{72 r^7 r_0^7 \mathcal{R}^7 f(r)^3}h_t(r)
\nonumber \\
& \hspace{1.5cm }-\frac{4 r^{16}\omega ^2 \mathcal{R}^7+112 r^7 r_0^{14}-63 r_0^{21}}{288 r^{14} r_0^7 f(r)^3}h_z(r) \,,
\end{align}
\end{subequations}%
together with two first-order equations for $h_t(r)$ and $h_z(r)$. These can be reduced to a single second-order equation for $h_z(r)$, expressed as
\begin{subequations}
\begin{multline}
h_z''(r)+\frac{28 \left(8 r^7-r_0^7\right) \mathcal{R}^7+\tilde{k}^2 \left(17 r^9 r_0^7-10 r^{16}\right)}{r^8 f(r) \left(\tilde{k}^2 r^9+28\mathcal{R}^7\right)}h_z'(r)
\\
-\frac{\tilde{k}^2 r_0^7 \left(\tilde{k}^2 r^9-35 \mathcal{R}^7\right)}{f(r) \mathcal{R}^7 \left(\tilde{k}^2 r^9+28 \mathcal{R}^7\right)} h_z(r)+\frac{\omega ^2\mathcal{R}^7}{r_0^7 f(r)^2} h_z(r)=0
\end{multline}
with
\begin{equation}
\tilde{k}\equiv k-\frac{\mathcal{R}^7}{r_0^7}\omega\,.
\end{equation}
\end{subequations}%

To proceed, we impose the appropriate boundary conditions at the horizon $r=r_0$ and at asymptotically large $r$.
The equation for $h_z(r)$ possesses a regular singular point at $r=r_0$, with two possible behaviours near the horizon,
\begin{equation}
h_{z}(r) \sim (r-r_0)^{\pm \frac{{\rm i}\, \mathcal{R}^{7/2} \omega}{7 r_0^{5/2}}} \left[A^{\pm}+\mathcal{O}(r-r_0)\right]\, ,
\label{eq:lower}
\end{equation}
with $A^{\pm}$ a constant. To determine which branch ensures regularity at the future event horizon, we introduce ingoing Eddington--Finkelstein coordinates,
\begin{equation}
{\rm d}v = {\rm d}t + \frac{\mathcal{R}^{7/2}}{r^{7/2}} \frac{{\rm d}r}{f(r)},
\qquad
{\rm d}z = {\rm d}\hat{z} + \frac{r^{7/2}}{\mathcal{R}^{7/2}} {\rm d}r \, ,
\end{equation}
in which the background metric takes the form
\begin{equation}
{\rm d}s^2 = -\frac{r^7}{\mathcal{R}^7} f(r)\, {\rm d}v^2
+ 2 \frac{r^{7/2}}{\mathcal{R}^{7/2}} \, {\rm d}v\, {\rm d}r
+ \frac{\mathcal{R}^7}{r^7} \left[ {\rm d}z + \frac{r^7}{\mathcal{R}^7} f(r)\, {\rm d}v \right]^2
+ r^2 {\rm d}\Omega_8^2 \, ,
\end{equation}
which is manifestly smooth at $r=r_0$. Regularity in these coordinates, which extend smoothly across the future event horizon (i.e. enforce ingoing boundary conditions), selects the lower sign in \eqref{eq:lower}, that is to say, we take $A^+=0$.

At asymptotically large $r$, we require the solution to represent a purely outgoing wave.  
The equation for $h_z(r)$ has an irregular singular point at $r=+\infty$, where the asymptotic behaviour is
\begin{equation}
h_z(r) = 
\exp\!\left[\pm\,{\rm i}\sqrt{\frac{k r_0^2 \bigl(2 \omega \mathcal{R}^7 - k r_0^7\bigr)}{\mathcal{R}^7}} \, \frac{r}{r_0}\right]
\left(\frac{r}{r_0}\right)^{5}
\left[B^{\pm} + \mathcal{O}\!\left(r^{-1}\right)\right] \, ,
\end{equation}
with $B^{\pm}$ a constant.  

Allowing both $\omega$ and $k$ to take complex values, one finds that only the choice of the upper sign yields a convergent solution in the Laplace sense (see \cite{Leaver:1986gd}). This sign is moreover the one conventionally associated with an outgoing wave at asymptotically large $r$. This choice implies taking $B^-=0$.

To proceed, we introduce the compact radial coordinate
\begin{equation}
r = \frac{r_0}{1-y} \, ,
\end{equation}
together with a redefinition of variables,
\begin{equation}
h_z(r) =
\left(1 - \frac{r_0}{r}\right)^{-\frac{{\rm i}\, \mathcal{R}^{7/2} \omega}{7 r_0^{5/2}}}
\exp\!\left[{\rm i}\sqrt{\frac{k r_0^2 \bigl(2 \omega \mathcal{R}^7 - k r_0^7\bigr)}{\mathcal{R}^7}} \, \frac{r}{r_0}\right]
\left(\frac{r}{r_0}\right)^{5}
\, q\!\left(1 - \frac{r_0}{r}\right) \, ,
\end{equation}
where $q$ is now a function of $y$ only.  
The boundary conditions for $q$ at $y=0$ and $y=1$ reduce to simple Robin conditions that follow directly from the equation of motion and which we do not display explicitly here.

Lastly, we recall that $k$ is quantised in units of $L_{11}$,  
\begin{equation}
k = \frac{2\pi n}{L_{11}} = \frac{n}{g_s \ell_s} \, , \qquad n \in \mathbb{Z} \, .
\end{equation}
In addition, $\mathcal{R}$ and $r_0$ can be related to $\tau$, $N$, $\ell_s$, and $g_s$ through \eqref{UniformD0uplifted:aux}. Upon substituting these relations, the equation for $q(y)$ depends only on the combinations  
$\tau N^{5/9}$, $\varpi \equiv \omega N^{5/9} / \lambda^{1/3}$, and $n$.  

\begin{figure}
    \centering
     \includegraphics[width=0.6\textwidth]{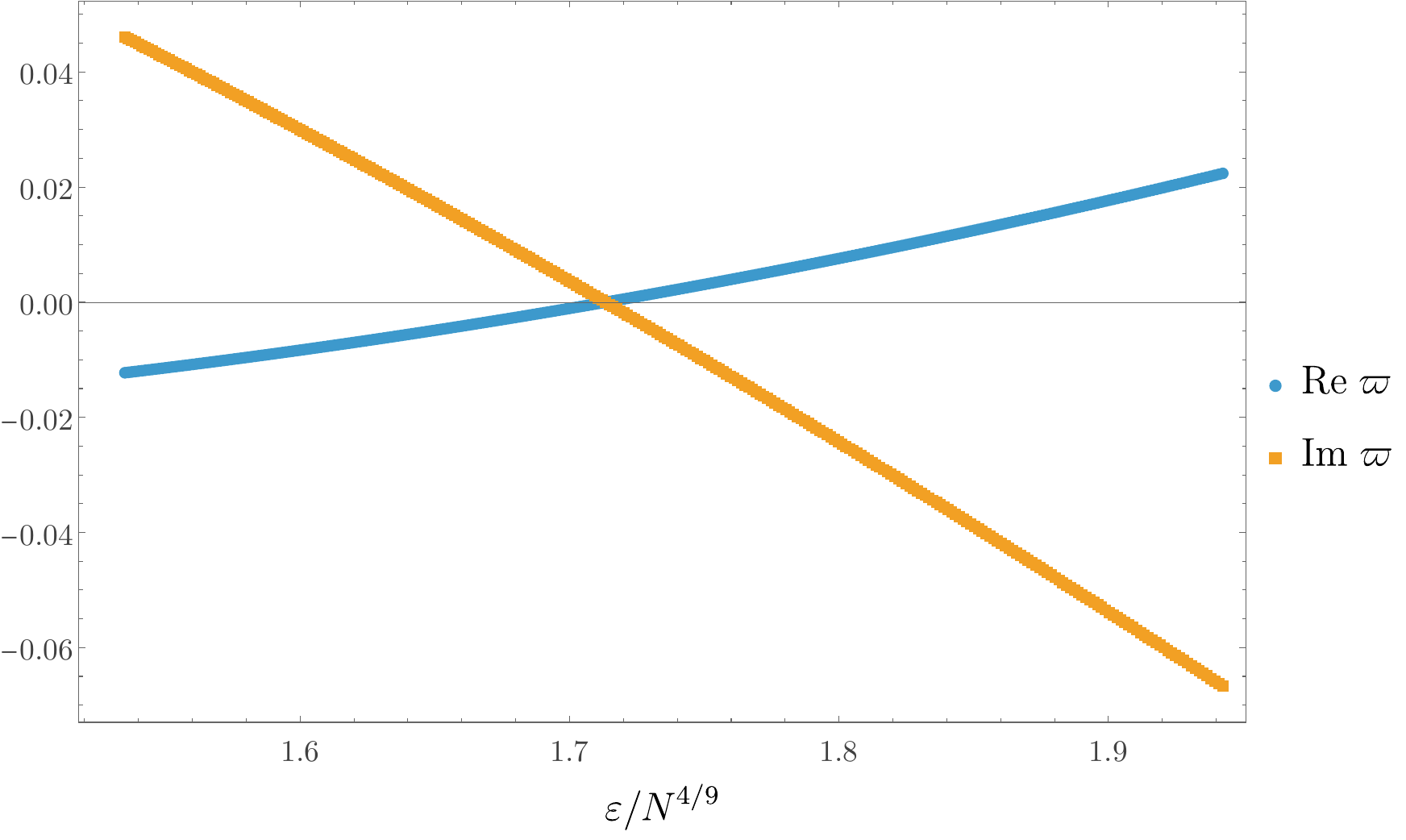}
\caption{Quasinormal mode frequency $\varpi$ as a function of $\varepsilon / N^{4/9}$. The real part is shown by blue disks and the imaginary part by orange squares. The instability sets in at $\varepsilon / N^{4/9}=1.7133(8)$, coinciding with the onset of the non-uniform phase. The behaviour reflects a second-order transition characterised by an exchange of stability: the uniform phase becomes unstable to the Gregory-Laflamme mode, while the non-uniform phase acquires stability.}
\label{fig:QNMs-BFSS}
\end{figure}

It is convenient to replace $\tau N^{5/9}$ by $\varepsilon / N^{4/9}$, a convention we adopt when presenting the plots.  For fixed $n$ and $\varepsilon / N^{4/9}$, the determination of $\varpi$ reduces to a generalised Sturm--Liouville problem,   which we solve numerically using a spectral collocation method on a Gauss-Lobatto grid. In Fig.~\ref{fig:QNMs-BFSS} we display the real (blue disks) and imaginary (orange squares) parts of $\varpi$ as functions of $\varepsilon / N^{4/9}$.  An instability is manifest for $\varepsilon < \varepsilon_{\hbox{\tiny GL}}$ (corresponding to ${\rm Im}(\varpi) > 0$), with equality signaling the onset of the novel non-uniform phase reported here namely at
\begin{equation}\label{GL:energy}
\varepsilon_{\hbox{\tiny GL}} N^{-4/9} = 1.7133(8).
\end{equation}
The above bound on the energy can be translated into a bound on the temperature, with the instability setting in when $\tau <\tau_{\hbox{\tiny GL}} $ with
\begin{equation}\label{GL:temp}
\tau_{\hbox{\tiny GL}} N^{5/9} = 0.5928(3).
\end{equation}
As promised, note that this value is indeed in line with the expectation \eqref{GLcondT} and its interpretation. 

In the microcanonical ensemble, the instability lies within the coexistence region, where the uniform and non-uniform phases overlap. As the transition is of second order, the results indicate an exchange of stability: the uniform phase becomes unstable to the Gregory-Laflamme mode, while the non-uniform phase emerges precisely at the onset of the instability and is stable.

\section{Map between gravitational vacuum solutions and BFSS thermal states \label{sec:MapEinsteinBFSS}}

Before embarking on the endeavour of finding possible non-uniform and localized D0 phases that emerge from the uniform D0 phase \eqref{UniformD0uplifted} due to the Gregory-Laflamme instability of the latter, there are a few observations in order that should prove to be very valuable to guide our search.

First note that asymptotically, i.e. at large $r$, the solution \eqref{UniformD0uplifted}-\eqref{UniformD0uplifted:aux} describing a stack of $N$ (non-)extremal D0-branes uniformly extended along the 11-dimensional circle approaches the metric:
\begin{equation} \label{asympMetricBFSSdual}
{\mathrm d}s^2\big|_{\infty}= -\frac{r^7}{\mathcal{R}^7}{\rm d}t^2+{\rm d}r^2+\frac{\mathcal{R}^7}{r^7}\left(\mathrm{d}z+\frac{r^7}{\mathcal{R}^7}{\rm d}t\right)^2+r^2\mathrm{d}\Omega_8^2\,,
\end{equation}
where, recall, $\mathrm{d}\Omega_8^2$ is the line element of a unit radius sphere $S^8$, $z\sim z+L_{11}$, with $L_{11}=2\pi g_s\ell_s$, parametrises the M-theory circle and $\mathcal{R}$ is given in \eqref{UniformD0uplifted:aux}.

The above geometry is an exact $pp-$wave solution in $d=11$. The easiest way to see this is to perform the following change of coordinates
\begin{equation}
{\rm d}t=-{\rm d}u\quad\text{and}\quad {\rm d}z={\rm d}v+\frac{2r^7}{\mathcal{R}^7}{\rm d}u\,,
\end{equation}
which brings \eqref{asympMetricBFSSdual} to
\begin{equation} \label{asympMetricBFSSdualpp}
{\mathrm d}s^2\big|_{\infty}= 2{\rm d}u{\rm d}v+\frac{\mathcal{R}^7}{r^7}{\rm d}v^2+{\rm d}r^2+r^2{\rm d}\Omega^2_8 \equiv 2{\rm d}u{\rm d}v+H(\mathbf{x}){\rm d}v^2+{\rm d}\mathbf{x}^2\,,
\end{equation}
with $H(\mathbf{x})=\mathcal{R}^7/|\mathbf{x}|^7$. This is precisely the metric of a $pp-$wave solution in Brinkmann coordinates \cite{Brinkmann1925}. What is somewhat unusual compared to the standard instances of $pp$-waves in general relativity is that $v$ is a null coordinate that is also periodic, inheriting the periodicity of the Kaluza-Klein circle $z$. A more detailed discussion of these and related issues can be found in \cite{Berenstein:2002jq,Blau:2002dy}.

A key observation is that if the theory has some other solution besides the uniform D0 phase (e.g. the expected non-uniform and localized D0 solutions), we want them to preserve the same asymptotics as the uniform D0 phase, i.e. to approach asymptotically the $pp-$wave background  \eqref{asympMetricBFSSdual}. This is because in the dual BFSS theory, all thermal phases must have: 1) the same number $N$ of D0-branes ($N$ is proportional to the momentum in \eqref{asympMetricBFSSdual}), 2) the same M-theory circle (parametrized by $z$) with length $L_{11}=2\pi g_s\ell_s$, and 3) the same generator $\partial_t$ of time translations as \eqref{asympMetricBFSSdual}. This fact will be a guide for our search and will ultimately allow us to find the map between solutions of 11-dimensional supergravity and BFSS thermals states in this section. 

More concretely, in subsection~\ref{sec:vacuumBHs} we describe the thermodynamics  of  asymptotically ${\mathbb R}^{(1,9)}\times S^1_{\hbox{\tiny $L$}}$ black strings/holes (a.k.a. Kaluza-Klein black holes) that are solutions of 11-dimensional supergravity with no gauge field, i.e. of  vacuum Einstein gravity. Then, in subsection~\ref{sec:ThermoUnifPertBFSS} we start by describing how we can apply a judiciously chosen Carrolian transformation to asymptotically  ${\mathbb R}^{(1,9)}\times S^1_{\hbox{\tiny $L$}}$ solutions of 11-dimensional supergravity to bring them into a Carrollian frame where the  asymptotics is described by the required $pp-$wave background \eqref{asympMetricBFSSdual}. This is the dual BFSS frame in the sense that it is in this frame that 11-dimensional supergravity is dual to the BFSS theory. In section~\ref{sec:BFSSthermo}, we use this fact to find the map that generates the thermodynamics of BFSS thermal states from the thermodynamics of 11-dimensional supergravity black objects with the $pp-$wave asymptotics \eqref{asympMetricBFSSdual}. This means that after this map is found, we will simply need to find the non-uniform and localized solutions of vacuum Einstein gravity (i.e. of 11-dimensional supergravity with vanishing 3-form gauge field) $-$ see Fig.~\ref{fig:sketch} $-$ and then apply the Carrollian map to immediately find the thermodynamics  of the (non-)uniform and localized BFSS thermal states dual to the 11-dimensional SUGRA (non-)uniform and localized D0 phases with $pp-$wave asymptotics.    
As an application example to a solution that is known in closed analytical form, in section~\ref{sec:ThermoUnifPertBFSS}, we start by determining the BFSS thermodynamics of the uniform thermal state that is dual to the uniform D0 phase \eqref{UniformD0uplifted}. As another example, we will recall that, in previous literature \cite{Harmark:2003yz}, there is an analytical matching asymptotic expansion description for localized black holes of $d=11$ Einstein gravity with ${\mathbb R}^{(1,9)}\times S^1_{\hbox{\tiny $L$}}$ asymptotics that gives the thermodynamics of these solutions in terms of the perturbative expansion parameter $r_0/L$ (where $r_0$ is the horizon radius and $L$ the $S^1$ length). Naturally, still in section~\ref{sec:ThermoUnifPertBFSS}, we then apply the Carrollian map  of section~\ref{sec:BFSSthermo} to generate the associated perturbative expressions for the dual BFSS localized state (remarkably, in section~\ref{sec:Results} we will find this perturbative result provides an excellent match to the exact numerical localized solution that we will find in section~\ref{sec:localized}).


\subsection{Static vacuum asymptotically ${\mathbb R}^{(1,9)}\times S^1_{\hbox{\tiny $L$}}$ solutions of Einstein gravity \label{sec:vacuumBHs}}

We are interested on all possible static, axially symmetric solutions of vacuum Einstein gravity, $R_{\mu\nu}=0$, that are asymptotically ${\mathbb R}^{(1,9)}\times S^1_{\hbox{\tiny $L$}}$ (a.k.a. Kaluza-Klein asymptotics). In the Schwarzschild gauge (whereby the radius of the $S^8$ is $r$, the aerial radius), the metric of any such solution takes the form 
\begin{equation} \label{solnSG:vacAllStatic}
{\mathrm d}s^{\,2} = - A \,{\mathrm d}T^2 + \frac{{\mathrm d}r^2}{B}  + r^2
{\mathrm d}\Omega^2_{8}  + C \, L^2\,{\mathrm d}Z^{\,2},
\end{equation}
where $Z\sim Z+1$ parametrises the compact circle $S^1$ with length $L$, ${\mathrm d}\Omega^2_{8}$ is the metric of a unit radius $S^{8}$ (parametrized by the angles $x^i$), and $A, B, C$ are functions of $(r, Z)$ which approach unity at large radial direction $r$ and are periodic in $Z$ with period $1$. We further assume that the solution is a (non-extremal) black hole and therefore the metric functions $A$ and $B$ vanish linearly at the horizon location $r=r_0$, i.e. $A(r,Z)=(r-r_0)\mathcal{A}(r,Z)$ and $B(r,Z)=(r-r_0)\mathcal{B}(r,Z)$ where $\mathcal{A}(r,Z)$ and $\mathcal{B}(r,Z)$  are non-vanishing functions at the horizon.

The Kaluza-Klein ${\mathbb R}^{(1,9)}\times S^1_{\hbox{\tiny $L$}}$ asymptotics implies that the functions $A(r,Z)$, $B(r,Z)$ and $C(r,Z)$  admit an asymptotic Taylor expansion around $r=\infty$ of the form $Q_j=1-  q_j(Z)\,\frac{L^{d-4}}{r^{d-4}}  + \cdots$ where $Q_j\in\{A,B,C\}$ (at higher orders we also have radially exponentially decaying terms). Since $q_j(Z)$ is a periodic function of $Z$ with period $1$ we can expand it as a sum of Fourier modes 
$e^{i\,2\pi\, n \, Z}$ for integer $n$. A Taylor expansion of the Einstein equation about the boundary $r=\infty$ then dictates that the only Fourier mode that is excited in the first subleading term is $n=0$ and that the functions have the specific off-boundary expansion 
\begin{equation}\label{SG:KKdecay}
A=  1 -  a(\xi_0)\,\frac{L^{7}}{r^{7}}  + \cdots\,, \quad
B= 1 -   \big[a(\xi_0)-c(\xi_0) \big] \,\frac{L^{7}}{r^{7}}  + \cdots\,, \quad C= 1 + c(\xi_0)\,\frac{L^{7}}{r^{7}}  + \cdots  \;,
\end{equation}
where, for a given dimensionless horizon radius $\xi_0\equiv r_0/L$, $a(\xi_0)$ and $c(\xi_0)$ are free constant parameters that depend on the solution at hand, \emph{i.e.} they cannot be determined by an expansion at infinity alone, and are fixed after solving the equations of motion subject to some interior boundary condition (in the present case, regularity at the horizon).  The dots in~\eqref{SG:KKdecay} represent sub-leading terms that decay as $r^{-14} e^{-\frac{2\pi}{L}\, r}$ or faster and that also depend on $Z$. Only the sub-leading terms displayed in \eqref{SG:KKdecay} contribute to the energy and tension that we now discuss.

We can now compute the energy and tension along $Z$ of static solutions described by  \eqref{solnSG:vacAllStatic}.
For that we can use the covariant Noether charge formalism (also known as the covariant phase method)  \cite{wald1993black,iyer1994some,wald2000general}, which is a manifestly covariant Hamiltonian formalism that allows one to define and compute the Noether charges of generic solutions of gravitational theories.  Formulae that give the  charges of 11-dimensional supergravity and of vacuum Einstein gravity can be found in section 3 of  \cite{Dias:2019wof}. Borrowing these, one finds that the dimensionless energy and tension of  of static solutions described by  \eqref{solnSG:vacAllStatic} are given by\footnote{For reference, in the uniform black string solution one has $a(\xi_0)=\xi_0^{7}$ and $c(\xi_0)=0$.} 
\begin{align}\label{SGETz:static}
& \mathcal{E}^{(0)}\equiv \frac{G_{11}}{L^8}\,E^{(0)}=\frac{\Omega_{8}}{16\pi}  \big[ 8a(\xi_0)-c(\xi_0)\big], \nonumber \\
&  \mathcal{T}_{\hbox{\tiny $Z$}}^{(0)}\equiv \frac{G_{11}}{L^7}\,  T_{\hbox{\tiny $Z$}}^{(0)}= \frac{\Omega_{8}}{16\pi}  \big[a(\xi_0)-8c(\xi_0)\big]. 
\end{align}
where $ \Omega_{n}=\int_{S^{n}} d\Omega_{n}=\frac{2\pi^{(n+1)/2}}{\Gamma[(n+1)/2]}$ is the volume of a  $n$-sphere $S^{n}$, and $a(\xi_0)$ and $c(\xi_0)$ are the asymptotic decays defined in \eqref{SG:KKdecay}.
 For later use, note that in \eqref{SGETz:static} we have added the superscript $^{(0)}$ to emphasize that these expressions refer to the thermodynamic quantities of the vacuum solutions described by \eqref{solnSG:vacAllStatic} in the $\{ T,r,Z,x^i \}$ coordinate frame. 

We can also compute the temperature and entropy of solutions described by \eqref{solnSG:vacAllStatic}. The vector field $\partial_{\hbox{\tiny $T$}}$ is the horizon generator (i.e. $|\partial_{\hbox{\tiny $T$}}|_{r=r_0}=0$) and its norm at the asymptotic boundary is 1, $|\partial_{\hbox{\tiny $T$}}|_{r\to\infty}=1$. Thus, the temperature of \eqref{solnSG:vacAllStatic} is simply the surface gravity divided by $2\pi$. On the other hand, the horizon entropy is the area of the horizon divided by $4 G_{11}$. Altogether the dimensionless temperature and entropy of static solutions described by \eqref{solnSG:vacAllStatic} are 
\begin{align}\label{SG:TS:static}
\mathcal{T}_{\hbox{\tiny $H$}}^{(0)}&\equiv L \,T_{\hbox{\tiny $H$}}^{(0)} = \frac{1}{4\pi}\,\sqrt{A'(\xi_0)}\sqrt{B'(\xi_0)}\,, \nonumber\\
\mathcal{S}_{\hbox{\tiny $H$}}^{(0)} &\equiv   \frac{G_{11}}{L^9}\, S_{\hbox{\tiny $H$}}^{(0)}
=\frac{\Omega_{8}}{4}\,  \xi_0^8    \sqrt{C(\xi_0)} \,.
\end{align}

The thermodynamic quantities \eqref{SGETz:static} and \eqref{SG:TS:static} satisfy the following first law of thermodynamics and Smarr relation: 
\begin{equation}\label{SGStaticFirstlawSmarr}
\begin{array}{cc}
 \bigg\{ & 
\begin{array}{cc}
\!\! \mathrm dE^{(0)}= T_{\hbox{\tiny $H$}}^{(0)}\,\mathrm{d} S_{\hbox{\tiny $H$}}^{(0)}+T_{\hbox{\tiny $Z$}}^{(0)}\,\mathrm{d}L    \\
\!\!  8\,E^{(0)} =9\, T_{\hbox{\tiny $H$}}^{(0)}\,S_{\hbox{\tiny $H$}}^{(0)}+T_{\hbox{\tiny $Z$}}^{(0)}\,L     \\
\end{array}
  \\
\end{array}
, \qquad \hbox{and} \qquad 
\begin{array}{cc}
 \bigg\{ & 
\begin{array}{cc}
\!\!  \mathrm d \mathcal{E}^{(0)}= \mathcal{T}_{\hbox{\tiny $H$}}^{(0)} \,\mathrm{d} \mathcal{S}_{\hbox{\tiny $H$}}^{(0)}     \\
\!\!  8\, \mathcal{E}^{(0)} =9\, \mathcal{T}_{\hbox{\tiny $H$}}^{(0)}  \,\mathcal{S}_{\hbox{\tiny $H$}}^{(0)}     \\
\end{array}
.  \\
\end{array}
\end{equation}
Using \eqref{SGETz:static} and \eqref{SG:TS:static} we can also compute the Helmoltz free energy $\mathcal{F}^{(0)}=\mathcal{E}^{(0)}-  \mathcal{T}_{\hbox{\tiny $H$}}^{(0)}  \mathcal{S}_{\hbox{\tiny $H$}}^{(0)}$, which satisfies the first law of thermodynamics, $\mathrm d\mathcal{F}^{(0)}=-\mathcal{S}_{\hbox{\tiny $H$}}^{(0)} \,\mathrm d\mathcal{T}_{\hbox{\tiny $H$}}^{(0)}$.

It should be noted that we have checked that our thermodynamic quantities \eqref{SGETz:static}-\eqref{SGStaticFirstlawSmarr}, computed using the covariant Noether charge formalism, agree with the ones first obtained using the ADM approach \cite{Harmark:2003dg}.

\subsection{Carrollian map from Kaluza-Klein into $pp-$wave asymptotics \label{sec:CarrolMap}}

Recall that the Carrollian transformation  \eqref{Carrollian:transf0} allows to rewrite the 11-dimensional black string solution in a frame where its asymptotics is that of a $pp-$wave solution \eqref{asympMetricBFSSdual}. We can now find the analogue of the transformation \eqref{Carrollian:transf0} that allows to rewrite {\it any}  asymptotically ${\mathbb R}^{(1,9)}\times S^1_{\hbox{\tiny $L$}}$ (a.k.a. Kaluza-Klein) vacuum solution  \eqref{solnSG:vacAllStatic} of Einstein gravity in a form where the solution asymptotes to the  $pp-$wave geometry \eqref{asympMetricBFSSdual}; of course it should agree with  \eqref{Carrollian:transf0}  for the uniform solution. Recall the discussion below \eqref{asympMetricBFSSdual} which justifies the need to mandatorily have this asymptotics \eqref{asympMetricBFSSdual} for any solution dual to the BFSS theory: all must have same number $N$ of D0 branes, same M-circle length, $L_{11}=2\pi g_s\ell_s$, and same time generator $\partial_t$.

We find that this is achieved applying to \eqref{solnSG:vacAllStatic} the Carrollian transformation (footnote \ref{foot:Carrol} justifies the nomenclature choice) 
\begin{align}\label{SG:CarrollianTransf}
& T=\frac{L^{7/2}}{\mathcal{R}^{7/2}}\,\gamma\,t -  \frac{1}{\gamma}\,\frac{\mathcal{R}^{7/2}}{L^{7/2}}\,z\,, \qquad Z= \frac{1}{\gamma}\, \frac{\mathcal{R}^{7/2}}{L^{7/2}}\,\frac{z}{L}\,,  \nonumber \\
& \hbox{with} \quad \gamma=\sqrt{a(\xi_0)+c(\xi_0)} =\frac{1}{\pi ^{3/2}}\sqrt{\frac{15}{2}} \sqrt{\mathcal{E}^{(0)}-\mathcal{T}_{\hbox{\tiny $Z$}}^{(0)}}
\end{align}
where $a(\xi_0),c(\xi_0)$ were introduced in \eqref{SG:KKdecay} and $\mathcal{E}^{(0)}, \mathcal{T}_{\hbox{\tiny $Z$}}^{(0)}$ are defined in \eqref{SGETz:static}. Note that the factor $\gamma$, and thus the Carrollian transformation, depend on the energy and tension of the solution at hand. In particular, for the uniform black string solution one has $\gamma=r_0^{7/2}/L^{7/2}$ since $a=\xi_0^7$ and $c=0$ and thus \eqref{SG:CarrollianTransf}  reduces to \eqref{Carrollian:transf0}. But $\gamma$ is different for the non-uniform and localized solutions and thus  \eqref{Carrollian:transf0} is not valid; one must use \eqref{SG:CarrollianTransf} to get the $pp-$wave asymptotics  \eqref{asympMetricBFSSdual}.  
To be  clear, with the given judicious choice of $\gamma$, the application of the Carrollian transformation \eqref{SG:CarrollianTransf} to {\it any} static solution described by \eqref{solnSG:vacAllStatic} (independently of whether $Z$ is a symmetry direction or not)   yields a solution of vacuum Einstein gravity, namely
\begin{subequations}\label{SGsoln:vacAllBoosted}
\begin{align}
& {\mathrm d}s^{2} =  -\frac{r^7}{\mathcal{R}^7} \mathcal{V}(r,z)\, \mathrm{d}t^2
 +\frac{\mathrm{d}r^2}{  \mathcal{B}(r,z)}   + r^2\mathrm{d}\Omega^2_{8}
 +  \frac{\mathcal{R}^7}{r^7}\,\mathcal{Z}(r,z)\left(  \mathrm{d}z + \frac{r^7}{\mathcal{R}^7}\mathcal{W}(r,z) \,\mathrm{d}t \right)^2, \\
& \hbox{with} \nonumber\\
& \mathcal{V}=\gamma^2\, \frac{L^7}{r^7}\, \frac{A \,C}{C-A}\,, \qquad
  \mathcal{B}=B\,,\qquad
 \mathcal{Z}= \frac{1}{\gamma ^2} \, \frac{ r^7}{L^7} (C-A)\,,  \qquad
  \mathcal{W}=\gamma^2\, \frac{L^7}{r^7}\, \frac{A}{C-A}\,,
\end{align}
\end{subequations}
 which has the asymptotic form \eqref{asympMetricBFSSdual}  of a $pp-$wave solution because the functions $\mathcal{V},\mathcal{B},\mathcal{Z}$ and $\mathcal{W}$ all approach 1 as $r\to\infty$ for our choice \eqref{SG:CarrollianTransf} of $\gamma$; recall that functions $A,B,C$ were introduced in  \eqref{solnSG:vacAllStatic}.
Note that, for the uniform black string solution, one has  $A(r,z)=B(r,z)=f(r), C(r,z)=1$ and \eqref{SGsoln:vacAllBoosted} reduces to \eqref{UniformD0uplifted} that describes a stack of $N$ (non-)extremal D0-branes uniformly extended along the 11-dimensional circle. 

It is important to observe that the proper length of the circle $S^1$ changes under the Carrollian transformation~\eqref{SG:CarrollianTransf}: in the Carrollian frame $\{t,r,z,x^i\}$ it  is no longer $L$  but  $L \,\gamma\, \frac{L^{7/2}}{\mathcal{R}^{7/2}}$ since $z\sim z+L \,\gamma \, \frac{L^{7/2}}{\mathcal{R}^{7/2}}$. One way to see this must be the case is to recall that any periodic function along $Z$ $-$ and thus periodic along $z$ after the Carrollian transformation $-$  admits a Fourier series representation where the relevant part of the argument of the Fourier modes transforms under the Carrollian boost as $2\pi \,Z\to  2\pi\,z \,  \frac{1}{L\,\gamma}  \frac{\mathcal{R}^{7/2}}{L^{7/2}}$. But the Fourier mode is periodic in $z$, i.e.  
$\exp\big[i\, 2\pi \,  \frac{1}{L\,\gamma}  \frac{\mathcal{R}^{7/2}}{L^{7/2}} \left(z+ \Delta z \right)\big]=\exp\big[ i\, 2\pi \,  \frac{1}{L\,\gamma}  \frac{\mathcal{R}^{7/2}}{L^{7/2}}  \,z \big] $,  if and only if $z$ has period 
\begin{equation}\label{DeltaZ}
 \Delta z =\gamma \, \frac{L^{9/2}}{\mathcal{R}^{7/2}} =  \frac{\sqrt{15}}{\sqrt{2}}  \frac{1}{\pi^{3/2}}  \frac{L^{9/2}}{\mathcal{R}^{7/2}} \sqrt{\mathcal{E}^{(0)}-\, \mathcal{T}_{\hbox{\tiny $Z$}}^{(0)}}
 \,.
 \end{equation}
 That is to say, under the Carrollian transformation, the length of the circle $S^1$, which in the original frame is $L$, becomes $\Delta z$.
We want to fix $\Delta z \equiv L_{11}=2\pi g_s\ell_s$ so that it matches the M-circle length of 11-dimensional supergravity and we can  use \eqref{DeltaZ} to find the original length $L$ pior to the Carrollian transformation as a function of $L_{11}$.

Next, we want to find how the thermodynamic quantities of the static solution changes under the Carrollian transformation \eqref{SG:CarrollianTransf}. Using again the covariant phase (Noether) formalism \cite{wald1993black,iyer1994some,wald2000general}, this time applied to \eqref{SGsoln:vacAllBoosted}, one finds that the energy and tension of asymptotically $pp-$wave solutions described by  \eqref{SGsoln:vacAllBoosted} are 
 \begin{align}\label{SG:ETzPCarrollian}
 E&=\Delta z\,\frac{9 L^{7}\,\Omega_{8}}{32\pi G_{11}}  \big[ a(\xi_0)-c(\xi_0)\big] \nonumber \\
     & 
         = \frac{L^8}{G_{11}} \frac{\Delta z}{L} \frac{1}{2}\left( \mathcal{E}^{(0)} + \mathcal{T}_{\hbox{\tiny $Z$}}^{(0)}  \right)  , \nonumber \\
T_z & = - \frac{ L^{7}\,\Omega_{8}}{32\pi G_{11}}  \big[5 \,a(\xi_0)+ 23 \, c(\xi_0)\big] 
          \nonumber \\
     & 
        =    \frac{L^7}{G_{11}}\left(  \frac{3}{2}\,\mathcal{T}_{\hbox{\tiny $Z$}}^{(0)} - \frac{1}{2} \mathcal{E}^{(0)} \right),
\end{align}
where the energy  $ \mathcal{E}^{(0)}$  and tension $ \mathcal{T}_{\hbox{\tiny $Z$}}^{(0)} $ in the static frame are given by \eqref{SGETz:static} and $\Delta z$ is given by \eqref{DeltaZ}.
On the other hand, the temperature and entropy of asymptotically $pp-$wave solutions described by  \eqref{SGsoln:vacAllBoosted} are 
\begin{align}\label{SG:TSCarrollian}
T_H&= \frac{\Delta z}{L}\,T_{\hbox{\tiny $H$}}^{(0)}\,,
\nonumber\\
S_H&=   S_{\hbox{\tiny $H$}}^{(0)},
\end{align}
where the thermodynamic quantities $T_{\hbox{\tiny $H$}}^{(0)}$ and $S_{\hbox{\tiny $H$}}^{(0)}$ are defined in \eqref{SG:TS:static}.\footnote{It should be noted that for the uniform string one has $a(\xi_0)=\xi_0^7$ and $c(\xi_0)=0$ and \eqref{SG:ETzPCarrollian}-\eqref{SG:TSCarrollian} with $\Delta z \equiv L_{11}=2\pi g_s\ell_s$  boil down to \eqref{NHnonExtDpThermo}, as it should be.}

The Carrollian thermodynamic quantities \eqref{SG:ETzPCarrollian}  and \eqref{SG:TSCarrollian} obey the following first law and Smarr relations:
\begin{equation}\label{SG:CarrollianFirstlawSmarr}
\begin{array}{cc}
 \bigg\{ & 
\begin{array}{cc}
\!\! \mathrm dE = T_H \, \mathrm dS_H+T_z\,\mathrm{d} \Delta z\,,    \\
\!\!  \frac{23}{9}\,E=2T_H\,S_H+T_z\,\Delta z \,.   \\
\end{array}
\end{array}
\end{equation}
It is straightforward to check that if the static thermodynamic quantities obey the first law and Smarr relations \eqref{SGStaticFirstlawSmarr} then the Carrollian thermodynamic quantities  \eqref{SG:ETzPCarrollian}  and \eqref{SG:TSCarrollian} automatically obey \eqref{SG:CarrollianFirstlawSmarr}, as it must be the case.
\subsection{BFSS thermodynamics from 11-dimensional SUGRA thermodynamics \label{sec:BFSSthermo}}

Ultimately, the main lesson from the above discussions is the following.  Now that we have the map  \eqref{SG:ETzPCarrollian}  and \eqref{SG:TSCarrollian} between the thermodynamic quantities of the asymptotically $pp-$wave (Carrollian)  solutions and their static Kaluza-Klein partners, $E=E(\mathcal{E}^{(0)}, \mathcal{T}_{\hbox{\tiny $Z$}}^{(0)})$ and $T_z=T_z(\mathcal{E}^{(0)}, \mathcal{T}_{\hbox{\tiny $Z$}}^{(0)})$, we do not need to find explicitly the Carrollian solutions. We just need to find the static Kaluza-Klein solutions and compute their thermodynamics \eqref{SGETz:static} and \eqref{SG:TS:static}, i.e. $\mathcal{E}^{(0)},\mathcal{T}_{\hbox{\tiny $Z$}}^{(0)},\mathcal{T}_{\hbox{\tiny $H$}}^{(0)},\mathcal{S}_{\hbox{\tiny $H$}}^{(0)}$. The thermodynamic quantities $E,T_z,T_H,S_H$  of the Carrollian 11-dimensional supergravity solutions follow simply from the application of the map \eqref{SG:ETzPCarrollian}  and \eqref{SG:TSCarrollian}.  
 
The final task is to use the proposed holographic duality of \cite{Itzhaki:1998dd} to translate the thermodynamics of the 11-dimensional supergravity solutions \eqref{SGsoln:vacAllBoosted} with $pp-$wave asymptotics  \eqref{asympMetricBFSSdual} into the thermodynamics of the dual BFSS field theory.
This step is straightforward. Indeed, the thermodynamics of a BFSS thermal state dual to a Carrollian supergravity black hole phase  \eqref{SGsoln:vacAllBoosted}  is simply obtained by taking the dimensionfull supergravity quantities and find build dimensionless quantities using the t'Hooft coupling $\lambda\equiv g_{\rm YM}^2N$.  
Namely, the dimensionless energy, temperature and entropy of the dual BFSS phases are then given by
\begin{equation}\label{QFTthermoMap0}
\begin{split}
\varepsilon &=  \frac{E} {\lambda^{1/3}} = N^{\frac{4}{9}} \,\frac{3}{7}\left( \frac{6^7 \pi^{14}}{25} \right)^{1/9}  \frac{a(\xi_0)-c(\xi_0)}{\left[a(\xi_0)+c(\xi_0)\right]^{7/9}}\,, \\
\tau&=  \frac{T_H} {\lambda^{1/3}} =  N^{-\frac{5}{9}} 
\,\left( \frac{2^7 \pi^{5}}{15^2} \right)^{1/9}  
\,\left[a(\xi_0)+c(\xi_0)\right]^{2/9} \, \mathcal{T}_{\hbox{\tiny $H$}}^{(0)}(\xi_0),  \\
\sigma &=S_H=N\, \frac{15}{\pi^2}\, \frac{15\, \mathcal{S}_{\hbox{\tiny $H$}}^{(0)}(\xi_0)}{\pi^2 \left[a(\xi_0)+c(\xi_0)\right]}\,, \\
\mathfrak{f} &= \varepsilon-\tau\,\sigma =N^{\frac{4}{9}} \, \frac{1}{7}\left( \frac{6^7}{25 \pi^{13}}\right)^{1/9}
\frac{3\pi^3 \left[ a(\xi_0)-c(\xi_0)\right] -35\,\mathcal{T}_{\hbox{\tiny $H$}}^{(0)}(\xi_0) \, \mathcal{S}_{\hbox{\tiny $H$}}^{(0)}(\xi_0)}{\left[a(\xi_0)+c(\xi_0)\right]^{7/9}}\,,
\end{split}
\end{equation}
where $a(\xi_0),c(\xi_0)$ and $ \mathcal{T}_{\hbox{\tiny $H$}}^{(0)}(\xi_0), \mathcal{S}_{\hbox{\tiny $H$}}^{(0)}(\xi_0)  $  are either asymptotic or horizon quantities,  introduced in  \eqref{SG:KKdecay} and \eqref{SG:TS:static}, that uniquely characterize the static asymptotically ${\mathbb R}^{(1,9)}\times S^1_{\hbox{\tiny $L$}}$ supergravity solution \eqref{solnSG:vacAllStatic} at hand.
To get  \eqref{QFTthermoMap} we used the holographic dictionary relations  $\lambda\equiv g_{YM}^2 N$, $g_{\rm YM}^2\equiv (2 \pi )^{-2} g_s \ell_s^{-3}$ and $G_{11}=G_{10}L_{11}$ (with $G_{10}=8 \pi^6 g_s^2 \ell_s^8$ and $L_{11}=2\pi g_s \ell_s$) which were introduced in section~\ref{sec:SQM}. We can also use \eqref{SGETz:static}-\eqref{SG:TS:static}
to express  $a(\xi_0),c(\xi_0)$  as a function of $ \mathcal{E}^{(0)}, \mathcal{T}_{\hbox{\tiny $Z$}}^{(0)}$ which allows to rewrite 
 \eqref{QFTthermoMap0} as (of course, this also follows directly from  \eqref{SG:ETzPCarrollian}  and \eqref{SG:TSCarrollian}): 
\begin{equation}\label{QFTthermoMap}
\begin{split}
\varepsilon &=  \frac{E} {\lambda^{1/3}} =  N^{\frac{4}{9}} \left(32 \pi ^8\right)^{1/9}\,\frac{ \mathcal{E}^{(0)}+ \mathcal{T}_{\hbox{\tiny $Z$}}^{(0)} }{\left(\mathcal{E}^{(0)} -\mathcal{T}_{\hbox{\tiny $Z$}}^{(0)}\right)^{7/9}} \,,\\
\tau&=  \frac{T_H} {\lambda^{1/3}} =  N^{-\frac{5}{9}} \, \left(\frac{32}{\pi}\right)^{1/9}\, \mathcal{T}_{\hbox{\tiny $H$}}^{(0)}\left(\mathcal{E}^{(0)}-\mathcal{T}_{\hbox{\tiny $Z$}}^{(0)}\right)^{2/9} , \\
\sigma &=S_H=  N  \, \frac{2\pi\, \mathcal{S}_{\hbox{\tiny $H$}}^{(0)}}{\mathcal{E}^{(0)}-\mathcal{T}_{\hbox{\tiny $Z$}}^{(0)}} \,,\\
\mathfrak{f} &= \varepsilon-\tau\,\sigma = N^{\frac{4}{9}} \left(32 \pi ^8\right)^{1/9}
\,\frac{ \mathcal{E}^{(0)}+ \mathcal{T}_{\hbox{\tiny $Z$}}^{(0)}- 2 \,\mathcal{T}_{\hbox{\tiny $H$}}^{(0)}\,\mathcal{S}_{\hbox{\tiny $H$}}^{(0)}  }{\left(\mathcal{E}^{(0)} -\mathcal{T}_{\hbox{\tiny $Z$}}^{(0)}\right)^{7/9}}  \,,
\end{split}
\end{equation}
These relations are key entries of the holographic dictionary of the D0-brane/BFSS duality conjectured in  \cite{Itzhaki:1998dd} and thus they constitute one of the main findings of the present study.

There is an alternative way to find the thermodynamic quantities \eqref{QFTthermoMap}. That is to use Kaluza-Klein holography for non-conformal D$p$-branes (including the case $p=0$ of interest here) developed in \cite{Kanitscheider:2008kd}. For that we would need to start with a 11-dimensional supergravity solution with the $pp-$wave asymptotics \eqref{asympMetricBFSSdual}; e.g. the (non-)uniform or localized solutions. Then we would do a Kaluza-Klein dimensional reduction down to IIA using the Kaluza-Klein ansatz \eqref{KKuplift}. This yields a solution of type IIA supergravity which is not asymptotically AdS$_2 \times S^8$. However, one can apply a conformal transformation that move the analysis into the so-called `dual frame' \cite{Kanitscheider:2008kd} where the solution is now asymptotically AdS$_2 \times S^8$. From here, one can apply Kaluza-Klein holographic renormalization to compute the expectation values of the dual BFSS operators. This should reproduce the thermodynamics \eqref{QFTthermoMap} and it would be interesting to complete this exercise. This would have the added value of also providing expectation values of relevant operators.\footnote{A caveat is that localized and non-uniform BFSS phases typically exist in regimes of temperatures that are captured by the 11-dimensional supergravity description but not by IIA supergravity so one should analyse if the holographic renormalization procedure, that starts from the IA description, would be valid for these temperatures; see Fig.~\ref{Fig:validitySUGRA}.} 

Summarising, given the thermodynamics $\{ \mathcal{E}^{(0)}, \mathcal{T}_{\hbox{\tiny $Z$}}^{(0)}, \mathcal{T}_{\hbox{\tiny $H$}}^{(0)}, \mathcal{S}_{\hbox{\tiny $H$}}^{(0)} \}$  of a vacuum asymptotically  ${\mathbb R}^{(1,9)}\times S^1_{\hbox{\tiny $L$}}$  gravitational solution, the thermodynamics $\{\varepsilon,\tau,\sigma,\mathfrak{f}\}$ of the dual super-quantum mechanics BFSS theory follows directly from the map \eqref{QFTthermoMap}. 
So, onwards, our strategy is to find $\{ \mathcal{E}^{(0)}, \mathcal{T}_{\hbox{\tiny $Z$}}^{(0)}, \mathcal{T}_{\hbox{\tiny $H$}}^{(0)}, \mathcal{S}_{\hbox{\tiny $H$}}^{(0)} \}$ for the Einstein gravity solutions that describe 
the  uniform, non-uniform, and localized phases. Then, we simply use \eqref{QFTthermoMap} to find the phase diagram of BFSS thermal states and discuss which of them dominates the microcanonical or canonical ensembles.
\subsection{Thermodynamics for the uniform and perturbative localized BFSS phases}
\label{sec:ThermoUnifPertBFSS}

Let us apply the duality map \eqref{QFTthermoMap} discussed in the previous two subsections to phases for which one has an analytical or perturbative solution. 
The only asymptotically ${\mathbb R}^{(1,9)}\times S^1_{\hbox{\tiny $L$}}$ vacuum Einstein solution known entirely in closed form is the uniform black string, which has horizon topology $S^8\times S^1$.  Its line element is  \eqref{solnSG:vacAllStatic} with $A=B= 1-\frac{r_0^7}{r^7}$ and $C=1$. 
 It follows that the series expansion at infinity \eqref{SG:KKdecay} simply yields $a(\xi_0)=\xi_0^7$ and $c(\xi_0)=0$. 
 Plugging this directly into  \eqref{solnSG:vacAllStatic}-\eqref{SG:TS:static} one finds that the thermodynamics of the 1-parameter family of uniform black strings is
\begin{equation}
\begin{split} \label{ThermoGRunif}
& \qquad \mathcal{E}^{(0)} \equiv  \frac{G_{11}}{L^8}\, E^{(0)}=\frac{16 \,\pi^3}{105}\,\xi_0^7 , \qquad \qquad
\mathcal{S}_{\hbox{\tiny $H$}}^{(0)}  \equiv \frac{G_{11}}{L^{9}}\, S_{\hbox{\tiny $H$}}^{(0)} =\frac{8 \pi^4}{105} \,\xi_0^8\,,\qquad \\
& \qquad  \mathcal{T}_{\hbox{\tiny $H$}}^{(0)} \equiv T_{\hbox{\tiny $H$}}^{(0)}\, L= \frac{7}{4 \pi }\,\frac{1}{\xi_0}\,,\qquad  \qquad \qquad
 \mathcal{F}^{(0)} \equiv \frac{G_{11}}{L^8}\, F^{(0)} = \frac{2 \pi^3}{105}\,\xi_0^7\,.
\end{split}
\end{equation}
It will be useful for us to express the entropy of the uniform string as a function of its energy and the free energy in terms of the temperature:
\begin{equation} \label{ThermoGRunif2}
\begin{split}
\mathcal{S}_{\hbox{\tiny $H$}}^{(0)}( \mathcal{E}^{(0)} ) & =  \frac{105^{1/7} \pi^{4/7} }{2^{11/7}}\big( \mathcal{E}^{(0)}\big)^{8/7},\quad\qquad \\
\mathcal{F}^{(0)}(\mathcal{T}_{\hbox{\tiny $H$}}^{(0)}) & = \frac{117649}{122880 \, \pi ^4} \,
 \frac{1}{(\mathcal{T}_{\hbox{\tiny $H$}}^{(0)})^7}\,.
\end{split}
\end{equation}

However, we are mainly interested in the BFSS thermodynamics dual to \eqref{ThermoGRunif}. Using the duality map \eqref{QFTthermoMap0} this yields straightforwardly:
\begin{equation}\label{QFTthermoMapUniform}
\begin{split}
\varepsilon &=  \frac{E} {\lambda^{1/3}} = N^{\frac{4}{9}} \,\frac{3}{7}\left( \frac{6^7 \pi^{14}}{25} \right)^{1/9}  \xi_0^{14/9}\,, \qquad \sigma =S_H=N\, \frac{8}{7}\,\pi^2\,\xi_0\,, \\
\tau&=  \frac{T_H} {\lambda^{1/3}} =  N^{-\frac{5}{9}} 
\,\frac{7}{2\left( 30\,\pi^2 \right)^{2/9}} \,\xi_0^{5/9},  \qquad
\mathfrak{f}  =-N^{\frac{4}{9}} \frac{1}{7}\left(  \frac{10^7 \pi^{14}}{9}  \right)^{1/9} \xi_0^{14/9}\,, 
\end{split}
\end{equation}
We can rewrite the entropy as a function of the energy and the free energy $\mathfrak{f}$ as a function of the temperature $\tau$ for the uniform BFSS phase:
\begin{equation}
\label{ThermoSYMunif}
 \sigma(\varepsilon)=\frac{4  \sqrt{2} \, 5^{1/7}\pi }{3^{8/7} \, 7^{5/14}}
 \,\bigg(\frac{\varepsilon}{N^{4/9}}\bigg)^{9/14} \! N\,, \quad\qquad 
\mathfrak{f}(\tau)= -\frac{1}{21} \left(\frac{120 \pi ^2}{49}\right)^{7/5}
\, \left( N^{\frac{5}{9}} \tau \right)^{14/5} \! N^{\frac{4}{9}}   \,.
\end{equation}

Another asymptotically ${\mathbb R}^{(1,9)}\times S^1_{\hbox{\tiny $L$}}$ solution of vacuum Einstein gravity is a black hole localized on the circle $S^1$ with horizon topology $S^9$. When energies are low compared to the circle size $L$, the geometry near the horizon resembles that of an asymptotically flat 11-dimensional Schwarzschild-Tangherlini black hole.  At larger energies, the presence of the circle deforms the horizon.  When these deformations are small, they can be captured perturbatively through an expansion in $ \xi_0=r_0/L \ll 1$, where $r_0$ is the horizon radius and $L$ is the size of $S^1$ \cite{Harmark:2003yz,Gorbonos:2004uc,Harmark:2004ws,Dias:2007hg}. In particular, thermodynamic quantities (within vacuum Einstein) can be found in section 6 of \cite{Harmark:2003yz}:\footnote{Note that \cite{Harmark:2003yz} sets ${L}=2\pi$.  Here, we express their results in terms of the dimensionless parameter $\xi_0=r_0/{L}$. The volume of a unit radius $S^{9}$ is $\Omega_9=\frac{1}{12}\,\pi ^5$ and $\zeta(8)=\pi^8/9450$ where $\zeta (s)=\sum _{k=1}^{\infty } k^{-s}$ is the Riemann zeta function.}
\begin{equation}
\begin{split}
\label{thermoLocGRpert}
&   \mathcal{E}^{(0)}  \equiv \frac{G_{11}}{L^8}\,  E^{(0)} = \frac{9\,\Omega_{9}}{16\pi} \,\xi_0^{8} 
\left( 1 + \frac{\zeta(8)}{2} \,\xi_0^{8} 
+ \mathcal{O} ( \xi_0^{16} ) \right) \,,
\qquad  \\
&  \mathcal{S}_{\hbox{\tiny $H$}}^{(0)}  \equiv   \frac{G_{11}}{L^9}\, S_{\hbox{\tiny $H$}}^{(0)}=  \frac{\Omega_{9}}{4}\, \xi_0^{9} 
\left( 1 + \frac{9\,\zeta(8)}{8}\, \xi_0^{8} 
+ \mathcal{O}  ( \xi_0^{16} ) \right),   \\
&     \mathcal{T}_{\hbox{\tiny $H$}}^{(0)}  \equiv   T_{\hbox{\tiny $H$}}^{(0)} L = \frac{2}{\pi\, \xi_0} 
\left( 1 - \frac{9\,\zeta(8)}{8} \xi_0^{8} 
+ \mathcal{O} ( \xi_0^{16} ) \right)\,, 
\qquad  \\
& \mathcal{F}_{\hbox{\tiny $H$}}^{(0)} \equiv \frac{G_{11}}{L^8}  F^{(0)} = \frac{\,\Omega_{9}}{16\pi} \,\xi_0^{8} 
{\biggl (} 1 + \frac{9 \zeta(8)}{8} \,\xi_0^{8} 
+ \mathcal{O} ( \xi_0^{16} ) {\biggr )}  \,,
\end{split}
\end{equation}
which we can use to write   
\begin{equation} \label{thermoLocGRpert2}
\begin{split} 
 &\qquad \mathcal{S}_{\hbox{\tiny $H$}}^{(0)}( \mathcal{E}^{(0)} ) = \frac{2^{11/4} \sqrt{\pi }}{3^{17/8}} \big(  \mathcal{E}^{(0)} \big)^{9/8}\left( 1+\frac{2 \pi ^4}{1575} \, \mathcal{E}^{(0)} +\mathcal{O}\big(( \mathcal{E}^{(0)})^2\big) \right) ,
  \\
& \qquad \mathcal{F}^{(0)}(\mathcal{T}_{\hbox{\tiny $H$}}^{(0)})  = \frac{4}{3\, \pi^4}\frac{1}{\big(\mathcal{T}_{\hbox{\tiny $H$}}^{(0)}\big)^8}\left( 1-\frac{64}{525} \frac{1}{\big(\mathcal{T}_{\hbox{\tiny $H$}}^{(0)}\big)^8} +\mathcal{O}\big((\mathcal{T}_{\hbox{\tiny $H$}}^{(0)})^{-16}\big) \right).
\end{split}
\end{equation}
For larger energies, the localized solutions can only be obtained numerically, as we will do in section~\ref{sec:localized}. In (later) Fig.~\ref{Fig:thermoGR}  we  compare our numerical results (black dots) with the (dashed yellow) perturbative results \eqref{ThermoSYMpertLoc} and find that the expressions \eqref{ThermoSYMpertLoc} give an excellent approximation, even well beyond the regime where they may be expected to be valid.

The thermodynamics of the localized BFSS state dual to \eqref{thermoLocGRpert} is obtained using the duality map \eqref{QFTthermoMap0} yielding:
\begin{equation}
\begin{split} \label{ThermoSYMpertLoc}
& \varepsilon(\xi_0)=N^{\frac{4}{9}} \left(\frac{3^2 \pi ^{16}}{2^7}\right)^{\frac{1}{9}} \,\xi_0^{16/9}\left(1+ \frac{13 \pi ^8}{17010}\, \xi_0^8 + \mathcal{O} ( \xi_0^{16} ) \right) \,,
\qquad  \\
& \sigma(\xi_0)=  N \,\frac{8 \pi^2}{9}\,\xi_0 \left(1+
\frac{37 \pi ^8}{75600}\,\xi_0^8+ \mathcal{O}  ( \xi_0^{16} ) \right),  \\
& \tau(\xi_0) =N^{-\frac{5}{9}} \left(\frac{6}{\pi }\right)^{2/9} \,\xi_0^{7/9}\left(1-\frac{137 \pi ^8}{680400}\,
\xi_0^8+ \mathcal{O} ( \xi_0^{16} )\right) \,,
\qquad  \\
&
\mathfrak{f}(\xi_0)= -N^{\frac{4}{9}}  \frac{7 \pi ^{16/9}}{2^{7/9} 3^{16/9}} \,\xi_0^{16/9} \left(1-\frac{193 \pi ^8}{595350} \xi_0^8+ \mathcal{O} ( \xi_0^{14} )\right)  \,.
\end{split}
\end{equation}
To discuss the relevance of the localized phase to the microcanonical and canonical ensembles, it is fundamental to use \eqref{ThermoSYMpertLoc} to write the entropy $\sigma$ as a function of the energy $\varepsilon$ and the free energy $\mathfrak{f}=\varepsilon-\tau\sigma$ as a function of the temperature $\tau$. This yields for the localized BFSS state:
\begin{equation}
\begin{split} \label{ThermoSYMpertLoc2}
& \sigma(\varepsilon) = N\, \frac{2^{55/16} \pi }{3^{17/8}}
 \,\left(\frac{\varepsilon}{N^{4/9}}  \right)^{9/16}\left( 1+ \frac{1}{3150 \sqrt{2}}\,\left(\frac{\varepsilon}{N^{4/9}}\right)^{9/2}\  +\mathcal{O}(\varepsilon^{9}) \right),  \\
& \mathfrak{f}(\tau) = - N^{\frac{4}{9}}  \frac{7 \pi ^{16/7}}{2^{9/7} 3^{16/7}}
\,\left( N^{\frac{5}{9}} \, \tau\right)^{16/7}\left[ 1+ \frac{\pi ^{72/7}}{264600\ 6^{2/7}}
\left( N^{\frac{5}{9}}  \, \tau\right)^{72/7} +\mathcal{O}(\tau^{144/7})\right] \,.
\end{split}
\end{equation}
For larger energies, the localized solutions can only be obtained numerically, as we will in the next section. In (later) Figs.~\ref{Fig:thermoGR}$-$\ref{fig:BFSSphaseDiagrams} we will compare our numerical results (black disks) with the (dashed yellow) perturbative results \eqref{ThermoSYMpertLoc}. Again, we will conclude that the perturbative expressions \eqref{ThermoSYMpertLoc} give an excellent approximation, even (well) beyond the regime where they may be expected to be valid.

\section{Non-uniform and localized phases} \label{sec:NumConstruction}
In this section we numerically construct the non-uniform and localized phases (recall that the analytical results for the localized phase in the previous section are, \`a priori, only valid for small energies).  In this section, we will adopt the language of 11-dimensional vacuum Einstein gravity with ${\mathbb R}^{(1,9)}\times S^1_{\hbox{\tiny $L$}}$ asymptotics where the $S^1_{\hbox{\tiny $L$}}$ has circumference of size $L$. Afterwards, in  Section \ref{sec:Results}, we use the map \eqref{QFTthermoMap} to read off the BFSS thermodynamics of the holographic dual $(1+0)$-dimensional SYM theory. 

To generate our numerical solutions we choose to use the DeTurck method \cite{Headrick:2009pv,Figueras:2011va,Wiseman:2011by,Dias:2015nua}. This method requires that we first choose a reference metric $\overline g$.  This metric need not be a solution to the Einstein equation, but must contain the same symmetries and causal structure as the desired solution.  With the reference metric chosen, the DeTurck method then modifies the Einstein equation $R_{\mu\nu}=0$ to
\begin{equation}\label{EdeT}
R_{\mu\nu}-\nabla_{(\mu}\xi_{\nu)}=0\;,\qquad \xi^\mu \equiv g^{\alpha\beta}[\Gamma^\mu_{\alpha\beta}-\overline{\Gamma}^\mu_{\alpha\beta}]\;,
\end{equation}
where $\Gamma$ and $\overline{\Gamma}$ define the Levi-Civita connections for $g$ and $\bar g$, respectively. Unlike $R_{\mu\nu}=0$, this equation yields PDEs that are elliptic in character. But after solving these PDEs, we must verify that $\xi^\mu=0$ to confirm that $R_{\mu\nu}=0$ is indeed solved.\footnote{The condition $\xi^\mu=0$ also fixes all gauge freedom in the metric.}  Fortunately, the results of \cite{Figueras:2011va} have proven that static solutions to \eqref{EdeT} must satisfy $\xi^\mu=0$. Nevertheless, we will still monitor $\xi^\mu$ as a measure of numerical accuracy.

\subsection{Non-uniform black strings} \label{sec:SetupNonUnif}

\begin{figure}[ht]
\centering
\includegraphics[width=.6\textwidth]{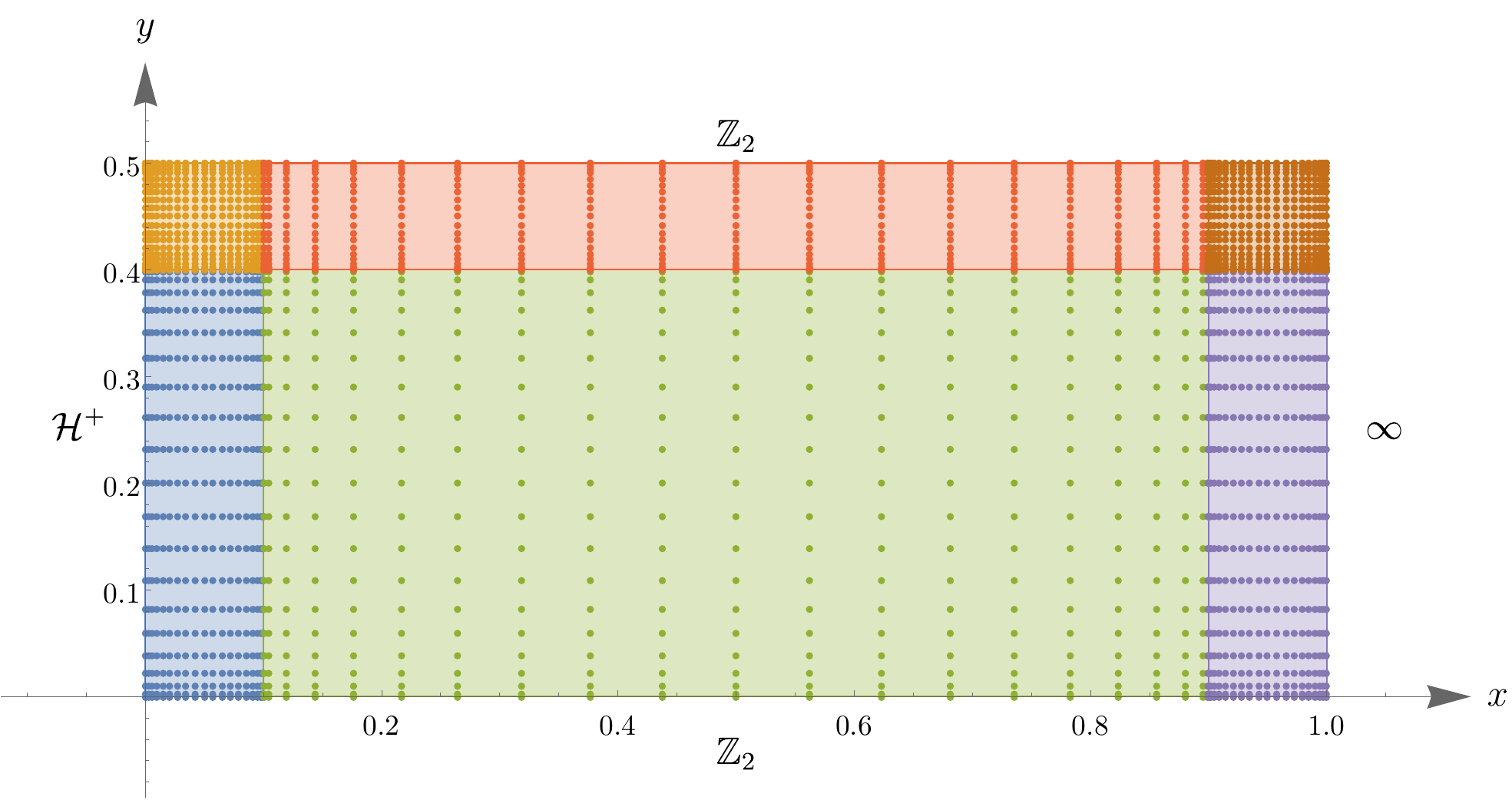}
\caption{Integration domain for the non-uniform strings with six patches and displaying regions where higher resolution is required.}\label{fig:appen}
\end{figure}

The non-uniform black strings we seek are asymptotically ${\mathbb R}^{(1,9)}\times S^1_{\hbox{\tiny $L$}}$ and have horizon topology $S^8\times{S}^1$ (see Fig.~\ref{fig:sketch}).  They are static and axisymmetric, and so only depend upon a periodic coordinate $y$ (say) and a radial coordinate $x$ (say).  
Constructing the non-uniform black string solutions is relatively ``simple" in the sense that the domain of integration naturally fits within a square, which allows to use a single coordinate system to cover the full integration domain. As an ansatz for solving the Einstein-DeTurck equation one chooses
\begin{subequations}
\begin{equation} \label{nonU:ansatz}
{\rm d}s^2=L^2\left[-G(x)x^2 Q_1{\rm d}\tilde{T}^2+\frac{4\,x_+^2\,Q_2 {\rm d}x^2}{G(x)(1-x^2)^4}+\frac{x_+^2\,Q_5}{(1-x^2)^2}{\rm d}\Omega^2_8+Q_4\left({\rm d}y+Q_3 {\rm d}x\right)^2\right]
\end{equation}
where
\begin{equation}
G(x)=\sum_{i=0}^6(1-x^2)^i\,,
\end{equation}
\end{subequations}%
and the functions $Q_i$, which depend only on the coordinates  $\{x,y\}$, are to be found solving the Einstein-DeTurck equation. In \eqref{nonU:ansatz}, the horizon is at $x=0$, spatial infinity is at $x=1$ and we have  $\mathbb{Z}_2$ axes of symmetry at $y=0$ and $y=1/2$ which reflect the fact that we have a periodic identification along the circle along which the string is extended (we use the $\mathbb{Z}_2$  symmetry to halve the integration domain); see Fig.~\ref{fig:appen} with this integration domain.  

For the DeTurck reference metric, we must choose one that has the same symmetries and causal structure as the non-uniform strings we search for. It is then natural to choose uniform black string solution, namely
\begin{equation}\label{nonU:dTreference}
Q_1=Q_2=Q_4=Q_5=1\quad\text{and}\quad Q_3=0\,.
\end{equation}
which indeed describes the uniform black string $-$ namely, \eqref{solnSG:vacAllStatic} with $A=B=1 -  \frac{r_0^{7}}{r^{7}}$ and $C=1$ $-$ under the coordinate transformation $\tilde{T}=T/L$, $x=\left( 1- \frac{r_0}{r}\right)^{1/2}$ and $y=Z$ with $x_+ \equiv r_0/L$.

To have a well-posed boundary value problem we now have to impose the appropriate physical boundary conditions.
At spatial infinity ($x=1$) we require that all the functions $Q_i$ approach the reference metric functions \eqref{nonU:dTreference}. At the $\mathbb{Z}_2$ axes of symmetry, $y=0$ and $y=1/2$, we demand that
\begin{equation}
\left.\frac{\partial Q_i}{\partial y}\right|_{y=0,1/2}=0\,,\quad\text{for}\quad i=1,2,4,5\,, \qquad\qquad Q_3\big|_{y=0,1/2}=0\,.
\end{equation}
Finally, regularity at the horizon bolt ($x=0$) requires that
\begin{equation}
\left.\frac{\partial Q_i}{\partial x}\right|_{x=0}=0\,,\quad\text{for}\quad i=1,2,4,5\,, \qquad\qquad Q_3\big|_{x=0}=0\,,
\end{equation}
which also imply, due to regularity, $Q_1(0,y)=Q_2(0,y)$.

So, the problem was relatively simple to setup as a well defined boundary-value problem. However, it is rather difficult to solve it numerically when we reach the regime where the black strings are very deformed, namely near the conical topology-changing point \cite{Kol:2002xz,Kol:2003ja,Emparan:2019obu,Emparan:2024mbp} where the non-uniform strings merge with the localized black holes (see phase diagrams of section~\ref{sec:Results-Grav}). Indeed, near the merger, the non-uniform string wants to pinch the $S^8$, transitioning to a localized black hole with $S^9$ topology. Given that our domain of integration naturally lives on a square $(x,y)\in(0,1)\times(0,1/2)$, the pinch of the black string will occur either at $(x,y)=(0,0)$ or $(x,y)=(0,1/2)$, with both solutions being equivalent. For dominant non-uniform strings (note that excited strings can have several nodes fitting inside the circle) these are the only two possibilities. Naturally, near the pinch, the solutions develop very large gradients that are challenging and have to be handled with great care.

Our strategy to deal with the large gradients is to break the integration domain into six sub-domains (patches) and resort to Coon maps, as detailed in \cite{Dias:2015nua}, to link the several patches together. Fig.~\ref{fig:appen} assumes that the pinching is occurring near $(x,y)=(0,1/2)$ and displays the six patches we use. It is at the top corners that one needs particularly high resolution to resolve the large gradients and this justifies the dense numerical grid we use there.

A second challenge that we face is that  reading off the energy near spatial infinity is known to be difficult in higher dimensions. Our strategy to circumvent this is to we use a different set of variables in the two patches boundary  at $x=1$. In order to motivate this change of coordinates, we solve the Einstein DeTuck equation asymptotically to find the asymptotic expansion of the $Q_i$'s near spatial infinity:
\begin{align}\label{nonU:asympQi}
Q_1(x,y)&=1+(1-x)^7 \delta _1-\frac{7}{2} (1-x)^8 \delta _1+\frac{21}{4} (1-x)^9 \delta _1+\ldots\nonumber
\\
Q_2(x,y)&=1+(1-x)^7 \delta _2-\frac{7}{2} (1-x)^8 \delta _2-8 (1-x)^9 \delta _5+\ldots\nonumber
\\
Q_3(x,y)&=(1-x)^6 \delta _3-3 (1-x)^7 \delta _3+\frac{15}{4} (1-x)^8 \delta _3-\frac{5}{2} (1-x)^9 \delta _3+\ldots
\\
Q_4(x,y)&=1+(1-x)^7 \delta _4-\frac{7}{2} (1-x)^8 \delta _4+\frac{21}{4} (1-x)^9 \delta _4+\ldots\nonumber
\\
Q_5(x,y)&=1+(1-x)^7 \delta _2-\frac{7}{2} (1-x)^8 \delta _2+\left(\delta _5+\frac{189 \delta _2}{32}\right) (1-x)^9+\ldots\nonumber
\end{align}
where the $\ldots$ include two types of contributions. The first comes from higher order polynomial contributions of the form $(1-x)^{n}$ with $n\geq10$. The second contribution is due to non-perturbative terms in $1-x$ of the form $e^{-\frac{2\pi\,n}{1-x}}\cos(2\pi n y)$ (for $i=1,2,4,5$) and $e^{-\frac{2\pi\,n}{1-x}}\sin(2\pi n y)$ (for $i=3$). The coefficients $\delta_i$ are independent of $y$ and will later parametrise the energy and tension of the non-uniform black string.

The requirement that the DeTurck vector must vanish, $\xi=0$, to ultimately have solutions of Einstein-DeTurck that are also solutions of Einstein gravity demands that the following condition is necessarily satisfied:
\begin{equation}
\delta _1+7 \delta _2+\delta _4=0,
\label{eq:appde}
\end{equation}
which we will use to check the accuracy of our numerics (and also below to compute thermodynamic quantities).

The asymptotic expansion \eqref{nonU:asympQi} straightforwardly motivates the following field redefinitions:
\begin{equation}\label{nonU:widehatQ}
Q_i=1+(1-x^2)^7\widehat{Q}_i\quad\text{for}\quad i=1,2,4,5\,, \qquad\qquad Q_3=(1-x^2)^6\widehat{Q}_3\,,
\end{equation}
so that the asymptotic boundary conditions, translated into the $\widehat{Q}_i$'s, read: 
\begin{equation}
\left.\frac{\partial \widehat{Q}_i}{\partial x}\right|_{x=1}=0
\end{equation}
An advantage of these field redefinitions is that we can extract all the constants $\delta_i$ with at most two derivatives with respect to $x$. Actually, to compute $\delta_i$ with $1,2,3,4$ we do not even need to take any derivative at all. This considerably reduces the numerical error in the final energy and tension.

 To compute the energy $\mathcal{E}^{(0)}$ and tension $\mathcal{T}_{\hbox{\tiny $Z$}}^{(0)}$ densities we follow \cite{Harmark:2003yz,Dias:2007hg,Kraus:1999di}. This yields:
 \begin{equation}\label{nonU:EnergyTension}
 \mathcal{E}^{(0)}=\frac{\pi ^3 x_+^7}{6720}\left(1024-8 \delta _1-\delta _4\right)\qquad\text{and}\qquad\mathcal{T}_{\hbox{\tiny $Z$}}^{(0)}=\frac{\pi ^3 x_+^7 }{6720}\left(128-\delta _1-8 \delta _4\right)\,,
 \end{equation}
 where we used \eqref{eq:appde} to eliminate $\delta_2$ from the final expressions. Note that in the remaining four patches, we still use the original functions $Q_i$; it is only in the two patches nearer to  $x=1$ that we replace $Q_i$ with the $\widehat{Q}_i$ as defined in \eqref{nonU:widehatQ}.

 Finally, the entropy density $\mathcal{S}_{\hbox{\tiny $H$}}^{(0)}$ and temperature $\mathcal{T}_{\hbox{\tiny $H$}}^{(0)}$ are respectively given by
 \begin{equation}\label{nonU:EntropyTemp}
\mathcal{S}_{\hbox{\tiny $H$}}^{(0)}=\frac{16\pi^4}{105}\,x_+^8 \int_0^{1/2}\sqrt{Q_4(0,y)}\,Q_5(0,y)^4{\rm d}y\qquad\text{and}\qquad \mathcal{T}_{\hbox{\tiny $H$}}^{(0)}=\frac{7}{4 \pi} =\frac{1}{x_+}\,.
 \end{equation} 
 We can test the accuracy of our thermodynamic quantities \eqref{nonU:EnergyTension}-\eqref{nonU:EntropyTemp} by checking that they obey the first law and Smarr relations \eqref{SGStaticFirstlawSmarr} with an error smaller than $0.01\%$ (in the worst case scenario).
 
The thermodynamic quantities  $\{ \mathcal{E}^{(0)},\mathcal{T}_{\hbox{\tiny $Z$}}^{(0)},\mathcal{T}_{\hbox{\tiny $H$}}^{(0)},\mathcal{S}_{\hbox{\tiny $H$}}^{(0)}\}$ of non-uniform black strings will be presented in section~\ref{sec:Results-Grav}.

The BFSS thermodynamics $\{ \varepsilon, \sigma, \tau, \mathfrak{f} \}$ for the non-uniform BFSS phase, that will be presented in section~\ref{sec:Results-BFSS},  is obtained from $\{ \mathcal{E}^{(0)},\mathcal{T}_{\hbox{\tiny $Z$}}^{(0)},\mathcal{T}_{\hbox{\tiny $H$}}^{(0)},\mathcal{S}_{\hbox{\tiny $H$}}^{(0)}\}$ in \eqref{nonU:EnergyTension}-\eqref{nonU:EntropyTemp} via the supergravity/BFSS map \eqref{QFTthermoMap}. 

\begin{figure}[ht]
\centering
\includegraphics[width=.47\textwidth]{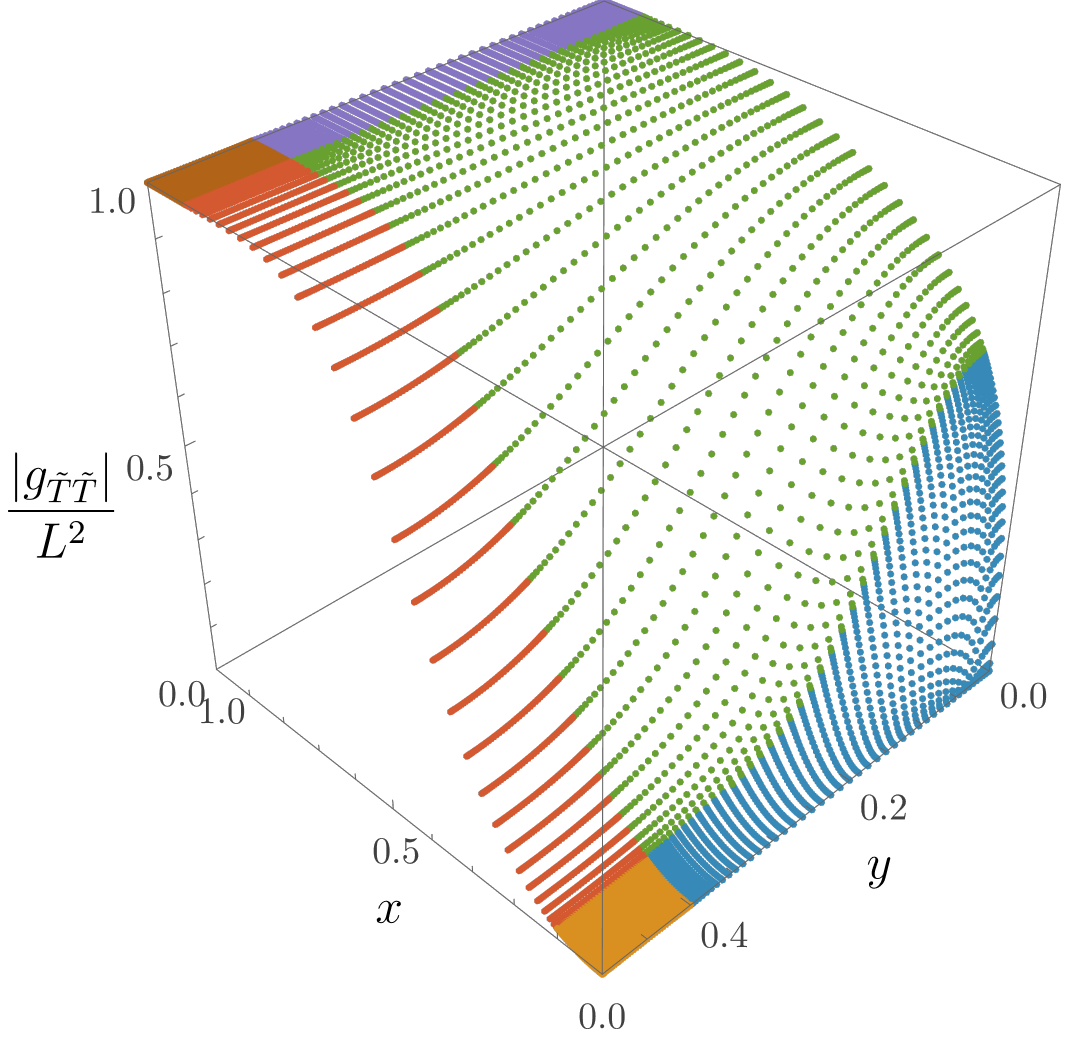}
\caption{The gauge-invariant metric component $\lvert g_{\tilde{T}\tilde{T}} \rvert / L^2$ is shown in the $\{x,y\}$ coordinates for $x_+ \approx 0.411377$. The steep gradients near $x=0$ illustrate why six coordinate patches are necessary. In the near-horizon region, where $\lvert g_{\tilde{T}\tilde{T}} \rvert / L^2$ is nearly constant, a different set of variables is employed, allowing for an accurate extraction of the energy and tension of the non-uniform string (see Eq.~\ref{nonU:widehatQ}).}\label{Fig:nonuni}
\end{figure}

Figure~\ref{Fig:nonuni} shows the gauge-invariant metric component $\lvert g_{\tilde{T}\tilde{T}} \rvert / L^2$ in the $\{x,y\}$ coordinates for a non-uniform string with $x_+ \approx 0.411377$. The steep gradients near $x=0$ highlight the necessity of using six coordinate patches to resolve the geometry accurately. In the far region, where $\lvert g_{\tilde{T}\tilde{T}} \rvert / L^2$ is approximately constant, a different set of variables is employed. This allows for a precise extraction of the energy and tension of the non-uniform string, as described in \eqref{nonU:widehatQ}). For reference, in the uniform string, the system varies with $x$ but shows no variation along $y$.

\subsection{Localized black holes} \label{sec:localized}
Within 11-dimensional vacuum Einstein gravity, localized black holes are asymptotically  ${\mathbb R}^{(1,9)}\times S^1_{\hbox{\tiny $L$}}$ black holes with horizon topology $S^9$  (see Fig.~\ref{fig:sketch}). For small energies, a perturbative construction of these solutions is available \cite{Harmark:2003yz,Gorbonos:2004uc,Harmark:2004ws,Dias:2007hg}, as reviewed in subsection \ref{sec:ThermoUnifPertBFSS}. For higher energies, one must necessarily resort to numerical methods which we present here. The perturbative results will provide a valuable check on our numerics, while the numerical results will assess the regime of validity of the perturbative expansions \eqref{thermoLocGRpert}-\eqref{ThermoSYMpertLoc}.

Like non-uniform strings, localized black holes, are also static and axisymmetric.  However, the non-uniform strings contain a horizon that covers the entire axis, while the axis is partially exposed in localized black holes. This introduces a fifth boundary in the integration domain which complicates the construction of localized black holes. A suitable reference metric for localized black holes therefore must contain an axis, a topologically $S^9$ horizon, and asymptote to ${\mathbb R}^{(1,9)}\times S^1_{\hbox{\tiny $L$}}$.  Furthermore, there is a periodic coordinate containing a $\mathbb Z_2$ symmetry, which we use to halve the integration domain.   To accommodate these five boundaries, we work with two different coordinate systems. One of these is adapted to to the the structure of the asymptotic region, while the other is adapted to the topology of the horizon.

Let us now design our reference metric, beginning with the coordinates adapted to the asymptotic region. Our starting point is the ${\mathbb R}^{(1,9)}\times S^1_{\hbox{\tiny $L$}}$ solution \eqref{solnSG:vacAllStatic} with $A=B=C=1$:
\begin{equation} \label{MinkXcircle1}
\mathrm ds^2_{{\mathbb R}^{(1,9)}\times S^1_{\hbox{\tiny $L$}}}=-\mathrm dT^2+\mathrm dr^2+r^2\mathrm d\Omega_8+L^2dZ^2\;,
\end{equation}
where $Z\in(-\frac{1}{2},\frac{1}{2})$ is periodic, $Z\sim Z +1$. Scale out $L$ by using the redefinitions $T=\pi\, t/L$, $r=\pi \,R/L$, and $Z=\theta/\pi$  to get
\begin{equation}\label{MinkXcircle2}
\mathrm ds^2_{{\mathbb R}^{(1,9)}\times S^1_{\hbox{\tiny $L$}}}=\frac{L^2}{\pi^2}\bigg(-\mathrm dt^2+\mathrm dR^2+\mathrm d\theta^{\,2}+R^2\mathrm d\Omega_8\bigg)\;,
\end{equation}
where $\theta\in(-\pi/2,\pi/2)$ is periodic, $\theta\sim \theta+\pi$. Then perform a further change of coordinates $R=\rho\sqrt{2-\rho^2}/(1-\rho^2)$ and $\theta=2\arcsin(\xi/\sqrt 2)$ to
\begin{equation}\label{MinkXcircle3}
\mathrm ds^2_{{\mathbb R}^{(1,9)}\times S^1_{\hbox{\tiny $L$}}}=\frac{L^2}{\pi^2}\bigg[-\mathrm dt^2+\frac{4\mathrm d\rho^2}{(2-\rho^2)(1-\rho^2)^4}+\frac{4\mathrm d\xi^2}{2-\xi^2}+\frac{\rho^2(2-\rho^2)}{(1-\rho^2)^2}\mathrm d\Omega_8\bigg]\;.
\end{equation}
Now the coordinate ranges are the more convenient $\rho\in[0,1]$ and $\xi\in[-1,1]$.  Since we will be exploiting the $\mathbb Z_2$ symmetry in $\xi$, we will instead take $\xi\in[0,1]$ and demand reflection symmetry at $\xi=0$ and $\xi=1$; the circle $S^1_{\hbox{\tiny $L$}}$ starts at $\xi=0$ and `reaches its halve' at $\xi=1$. There is an axis at $\rho=0$ and asymptotic infinity is at $\rho=1$.  

This analysis invites us to choose the following DeTurck reference metric in the far region:
\begin{equation}\label{reffar}
\overline{\mathrm ds}^2=\frac{L^2}{\pi^2}\bigg\{-m\,\mathrm dt^2+g\bigg[\frac{4\mathrm d\rho^2}{(2-\rho^2)(1-\rho^2)^4}+\frac{4\mathrm d\xi^2}{2-\xi^2}+\frac{\rho^2(2-\rho^2)}{(1-\rho^2)^2}\mathrm d\Omega_8\bigg]\bigg\}\;,
\end{equation}
where $m$ and $g$ are functions of $\rho$ and $\xi$ that we need to specify (as we shall do later).  We have chosen this particular form \eqref{reffar} for a number of reasons.  
First, the reference metric must have the same causal structure and symmetries as the localized black hole solution we look for. This is the case if $m(\rho,\xi)$ is chosen to have a horizon (more later). Second,  note that the $\mathrm d\rho^2$ and $\mathrm d\xi^2$ components in the metric originate directly from $\mathrm dr^2+L^2 \mathrm dZ^2$, which is manifestly flat. It is therefore straightforward to transform \eqref{reffar} to other orthogonal coordinates. Our aim is to perform such a coordinate transformation where we can easily choose $m$ and $g$ so that the reference metric describes a black hole, as stated above.  Finally, when the black hole has small energy (high temperature), we would like the reference geometry near the horizon to resemble asymptotically flat 11-dimensional Schwarzschild-Tangherlini, as expected of the final solution. With the reference metric in the form \eqref{reffar}, it is easier to accommodate this by using Schwarzschild-Tangherlini in isotropic coordinates (as we do below).

\begin{figure}[ht]
\centering
\includegraphics[width=.5\textwidth]{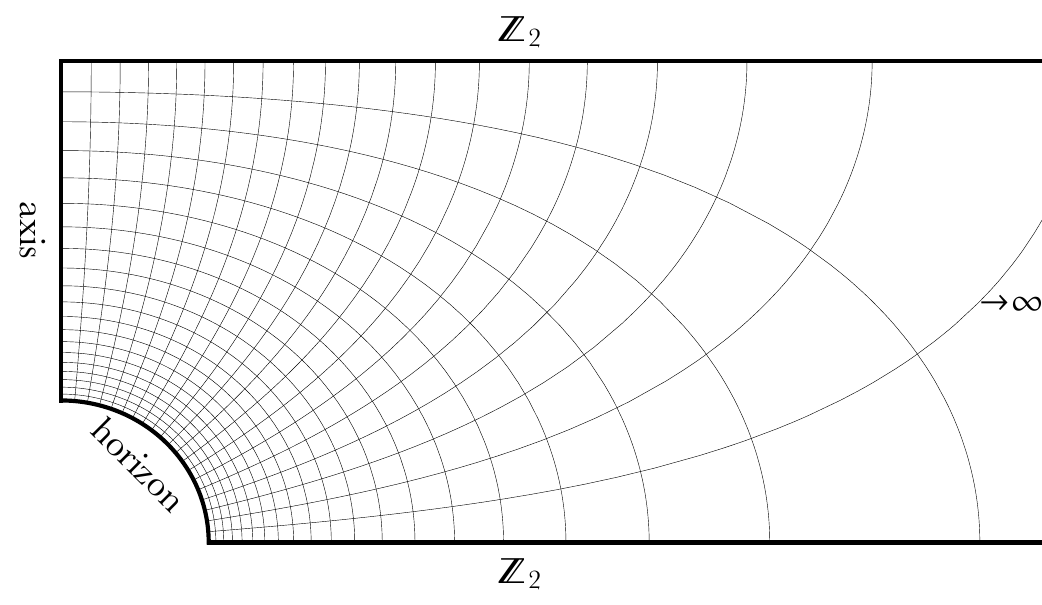}
\caption{Sketch of integration domain in $\{R,\theta\}$ coordinates. The $\{\rho,\xi\}$ coordinates we use are directly related to these coordinates via $R=\rho\sqrt{2-\rho^2}/(1-\rho^2)$ and $\theta=2\arcsin(\xi/\sqrt 2)$.  The grid lines are lines of constant $x$ and constant $y$.}\label{Fig:coord}
\end{figure}

The far-region coordinates $(\rho,\xi)$ are effectively elliptic coordinates that are often employed to describe 2-dimensional flat space. The latter can also be written in bipolar coordinates  $(x,y)$. 
These two coordinate charts can be obtained from a conformal mapping of 2-dimensional Cartesian coordinates. These conformal mappings also allow to obtain the coordinate transformation, and its inverse,  between the elliptic and bipolar coordinates, namely:
\begin{align}\label{coordmap}
   & \hspace{-0.5cm} x=\sqrt{1-\frac{\sinh {\Bigl( }\frac{\rho  \sqrt{2-\rho ^2}}{1-\rho ^2}{\Bigr) }}{\sqrt{\xi^2\left(2-\xi ^2\right)+\sinh^2{\Bigl( }\frac{\rho  \sqrt{2-\rho^2}}{1-\rho ^2}{\Bigr) }}}},  \qquad y=\frac{y_0\left(1-\xi ^2\right)}{\sqrt{\xi^2\left(2-\xi ^2\right)+\sinh^2{\Bigl( }\frac{\rho  \sqrt{2-\rho^2}}{1-\rho ^2}{\Bigr) }}} \,;\nonumber \\
   & \hspace{-0.5cm}  \rho=\sqrt{1-\frac{1}{\sqrt{1+{\rm arcsinh}^2{\Bigl( }\frac{y_0 \left(1-x^2\right)}{\sqrt{y^2+y_0^2  \, x^2 \left(2-x^2\right)}}{\Bigr) }}}}, \qquad \xi=\sqrt{1-\frac{y}{\sqrt{y^2+y_0^2 \, x^2 \left(2-x^2\right)}}}. 
\end{align}
In these new coordinates, the DeTurck reference metric \eqref{reffar} becomes
\begin{align}\label{refnear}
      \overline{ds}^2= \frac{L^2}{\pi^2} {\bigg \{} -m\,{\mathrm d}t^2+g {\biggl [} \frac{y_0^2}{h} \left( \frac{{\mathrm d}y^2}{y^2+y_0^2} +\frac{4\mathrm dx^2}{2-x^2}\right) +\, s \left(1-x^2\right)^2  {\mathrm d}\Omega_8^2 {\biggr ]}  {\bigg \}}\;,
\end{align}
where we assume that $y_0>0$, $m$ and $g$ transform as scalars, and we have defined
\begin{equation}\label{hsdef}
 h=y^2 + y_0^2 x^2 \left(2-x^2\right) \,,\qquad s=\frac{1}{\left(1-x^2\right)^2}{\rm arcsinh}^2\left[\frac{y_0 \left(1-x^2\right)}{\sqrt{y^2+y_0^2 \,x^2 \left(2-x^2\right)}}\right].
\end{equation}
Note that $s$ is positive definite and regular, even at $x=1$ while $h$ is positive except at $x=y=0$, where it vanishes. In these new coordinates, the axis is at $x=1$ and asymptotic infinity is at the coordinate `point' $x=y=0$.  The locations $\xi=0$ and $\xi=1$ where we require reflection symmetry have been mapped to $x=0$ and $y=0$, respectively. Finally, the location $\rho=0$, $\xi=0$ has been mapped to $y\to\infty$. We have introduced the constant $y_0$  in \eqref{coordmap}, and thus in \eqref{refnear}-\eqref{hsdef}, in anticipation of placing a horizon at $y=1$.  That is to say, $y_0$ controls the location of the horizon in the $\{\rho, \xi\}$ coordinate system. A sketch of the integration domain and grid lines of constant $x$ and $y$ are displayed in Fig. \ref{Fig:coord}.

As promised, we now motivate and choose the functions $m$ and $g$ that appear in \eqref{reffar} and \eqref{refnear}.  These functions must satisfy a number of requirements. Our asymptotics requires that they must approach $1$ at $x=y=0$.  The reflection symmetries requires that  both $m$ and $g$ are even functions of $x$ and $y$. Moreover, the choice of $m$ and $g$ must permit a regular horizon at $y=1$. Finally, for small $y_0$,  the geometry near the horizon  should resemble asymptotically flat Schwarzschild-Tangherlini in isotropic coordinates. The latter, in $d$-dimensions, can be written as
\begin{equation}\label{schwiso}
\mathrm ds^2_{\mathrm{Schw}}=-\left(\frac{1-y^{d-3}}{1+y^{d-3}}\right)^2\mathrm dt^2+y_0^2(1+y^{d-3})^{\frac{4}{d-3}}\left(\frac{\mathrm dy^2}{y^4}+\frac{1}{y^2}\mathrm d\Omega_{d-2}\right)\;.
\end{equation}
Note that these coordinates look similar to the $y_0\to0$ limit of our reference metric \eqref{refnear}. This suggests that we take
\begin{equation}\label{mgdef}
m=\frac{1}{1+y_0^2x^2(2-x^2)}\left(\frac{1-y^6}{1+y^6}\right)^2\;,\qquad g=(1+y^6)^{2/3}\;,
\end{equation}
where we choose to use the $d=9$ metric components from \eqref{schwiso} rather than $d=11$ because, after trial and error, this turns out to improve the numerical convergence of our solutions. With respect to  \eqref{schwiso}, the extra $x$ dependence in the numerator of $m$ in \eqref{mgdef} is introduced to fix regularity of the horizon. Finally, note that these functions  \eqref{mgdef} can be easily mapped back to $\{\rho,\xi\}$ coordinates through \eqref{coordmap}.  

Summarizing the analysis so far, we have a DeTurck reference metric in two coordinate systems \eqref{reffar}, and \eqref{refnear}, where the auxiliary functions are given in \eqref{hsdef} and \eqref{mgdef}, and the far/near coordinates are related by \eqref{coordmap}. This DeTurck reference metric can now be `dressed up' to produce the final metric ansatz to search for localized black holes:
\begin{align} \label{localizedansatz}
      {\mathrm ds}^2 &= \frac{L^2}{\pi^2} {\biggl \{} -m \,\tilde{f}_1{\mathrm d}t^2+g {\biggl [} \frac{4 \tilde{f}_2\,{\mathrm d}\rho^2 }{(2-\rho^2)(1-\rho^2)^4}
+\frac{4\tilde{f}_3}{2-\xi^2}\left( d\xi -\tilde{f}_5 \frac{\xi(2-\xi^2)(1-\xi^2)\rho}{(1-\rho^2)^2}d\rho \right)^2  \nonumber \\
      & \quad\quad\quad\quad\quad\quad\quad\quad\quad\quad 
+\tilde{f}_4\frac{\rho^2(2-\rho^2)}{(1-\rho^2)^2} \, {\mathrm d}\Omega_8^2  {\biggr ]}    {\biggr \}},\nonumber \\
&
= \frac{L^2}{\pi^2} {\Biggl (}- m\,f_1\,{\mathrm d}t^2+g {\biggl \{} \frac{y_0^2}{h} \bigg[ \frac{f_2\,{\mathrm d}y^2}{y^2+y_0^2} 
\nonumber \\
      & \hspace{4.7cm}
+\frac{4 f_3}{2-x^2} \bigg( {\mathrm d}x -f_5\,\frac{x \left(2-x^2\right) \left(1-x^2\right) y \left(1-y^2\right)}{h}\,{\mathrm d}y  \bigg)^2\bigg]  \nonumber \\
      &\hspace{4cm} +f_4\, s \left(1-x^2\right)^2  {\mathrm d}\Omega_8^2 {\biggr \}}  {\Biggr )}\;,  
\end{align}
where the known functions $h,s, m, g$,  given by \eqref{hsdef} and \eqref{mgdef},  are treated as scalars, transforming between the far/near coordinate systems as \eqref{coordmap}. Moreover, $\tilde f_i$ are unknown functions of $\{\rho,\xi\}$, and $f_i$ are unknown functions of $\{x,y\}$ that we need to solve for. By construction, we recover the DeTurck reference metric \eqref{reffar} and \eqref{refnear} when we set ${\tilde f}_{1,2,3,4}=1, {\tilde f}_5=0$ and  $f_{1,2,3,4}=1, f_5=0$, respectively. 

We need to solve the Einstein-de Turck equations motions \eqref{EdeT} for the localized black hole ansatz \eqref{localizedansatz} subject to a set of boundary conditions. At the asymptotic boundary $\rho=1$, the solution must approach the DeTurck reference metric \eqref{reffar}, so we impose the Dirichlet conditions ${\tilde f}_{1,2,3,4}=1, {\tilde f}_5=0$.  At the horizon $y=1$, regularity of the solution implies that $f_i$ must obey certain Robin boundary conditions (whose expressions are long and unilluminating). At the remaining boundaries, $f_i$ and $\tilde{f}_i$ must obey Neumann conditions either due to regularity at the axis $\rho=0$ (or $x=1$), or reflection symmetry at $\xi=0$ (or $x=0$) and at $\xi=1$ (or $y=0$).

\begin{figure}[ht]
\centering
\includegraphics[width=.47\textwidth]{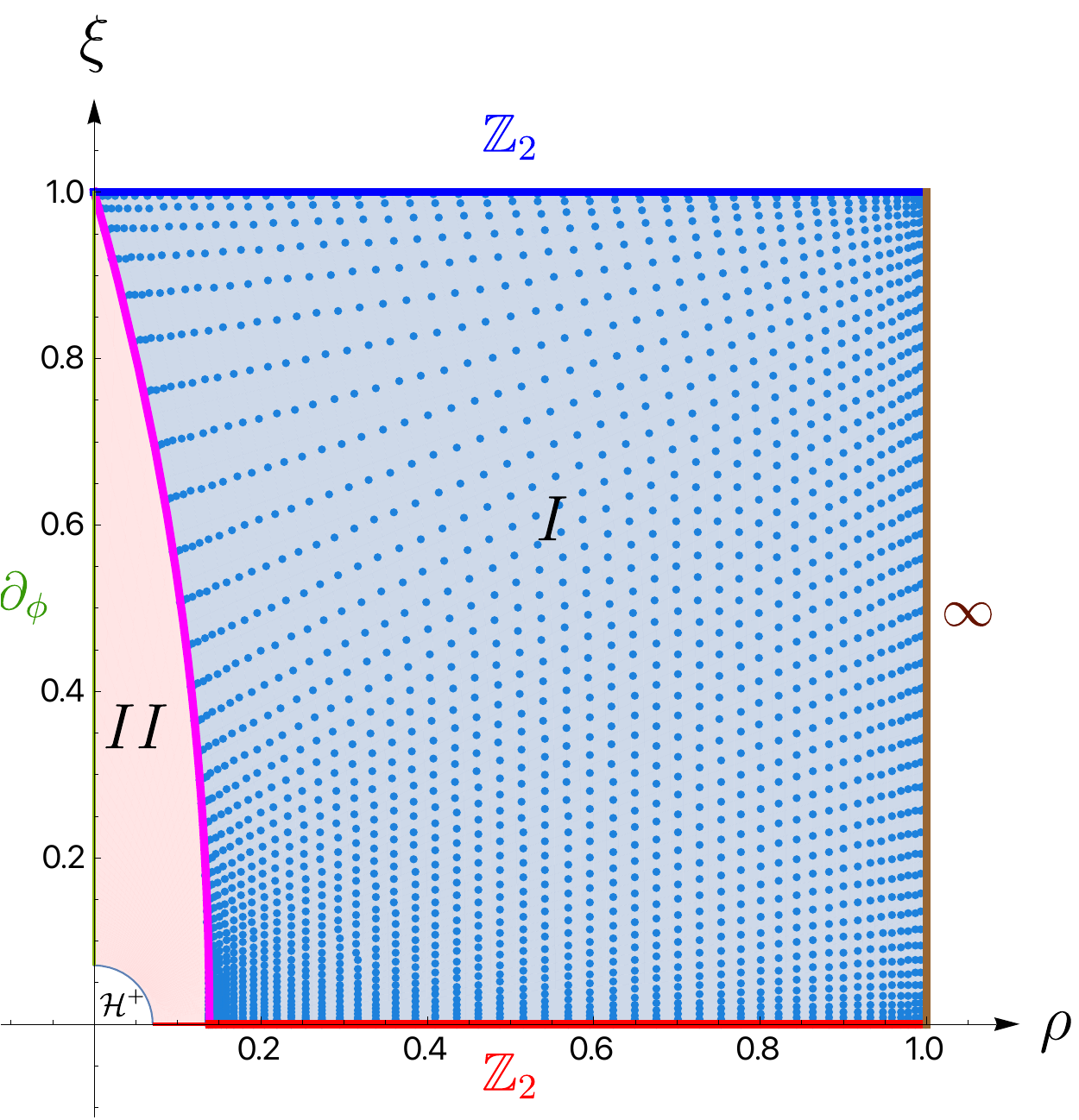}\hspace{0.5cm}
\includegraphics[width=.47\textwidth]{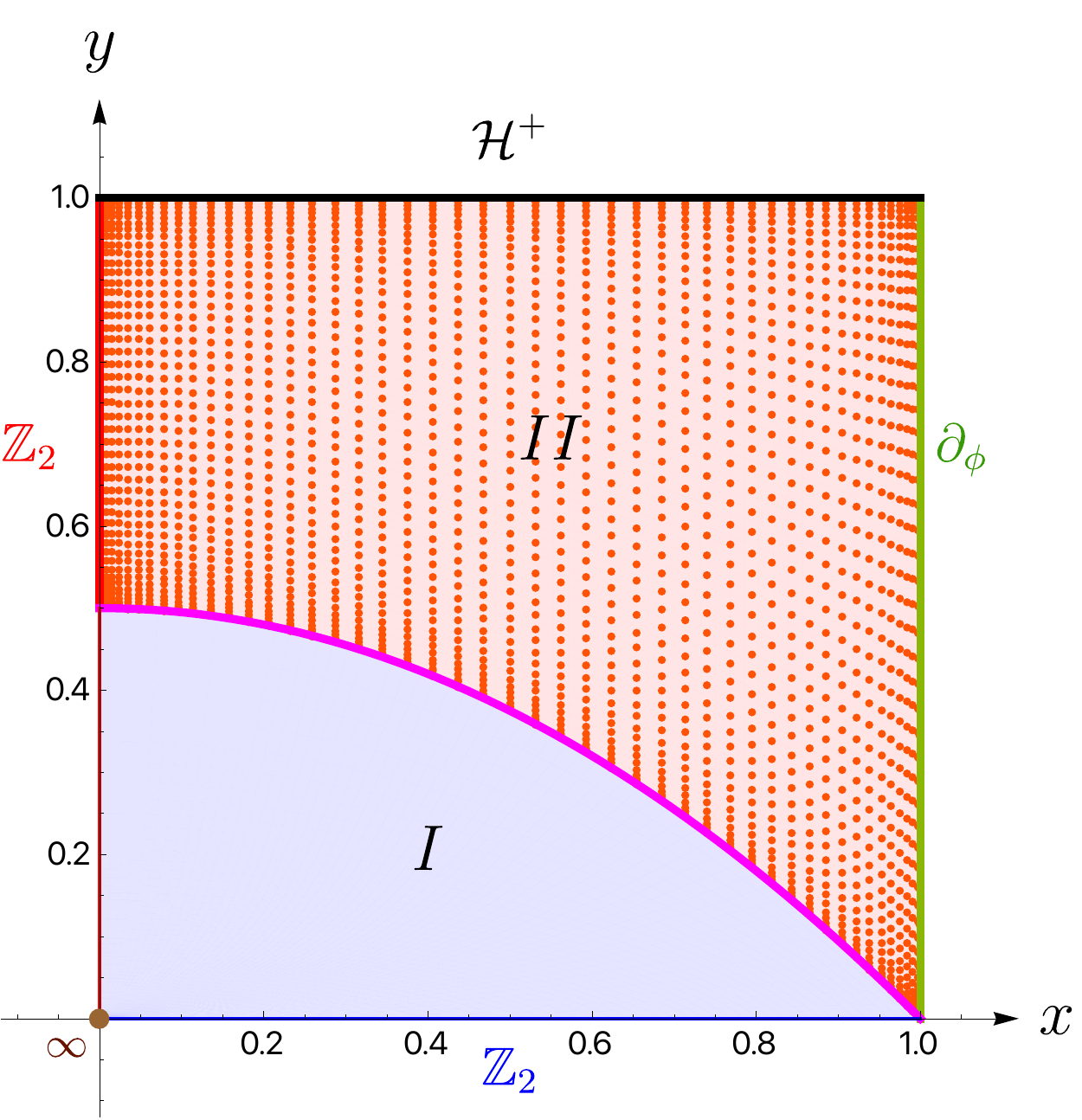}
\caption{Integration domain with two patches $I$ and $II$.  Chebyshev-Gauss-Lobatto grids with $50\times 50$ points are placed using transfinite interpolation. {\bf Left panel:} Patch $I$ (in the far region) uses  $\{\rho,\xi\}$ coordinates and patch $II$ (near the horizon in the quarter circle in the lower left) is mapped from $\{x,y\}$ coordinates using \eqref{coordmap}.  {\bf Right panel:} Patch II (near the horizon) uses  $\{x,y\}$ coordinates.}\label{Fig:patches}
\end{figure}

\begin{figure}[ht]
\centering
\includegraphics[width=.47\textwidth]{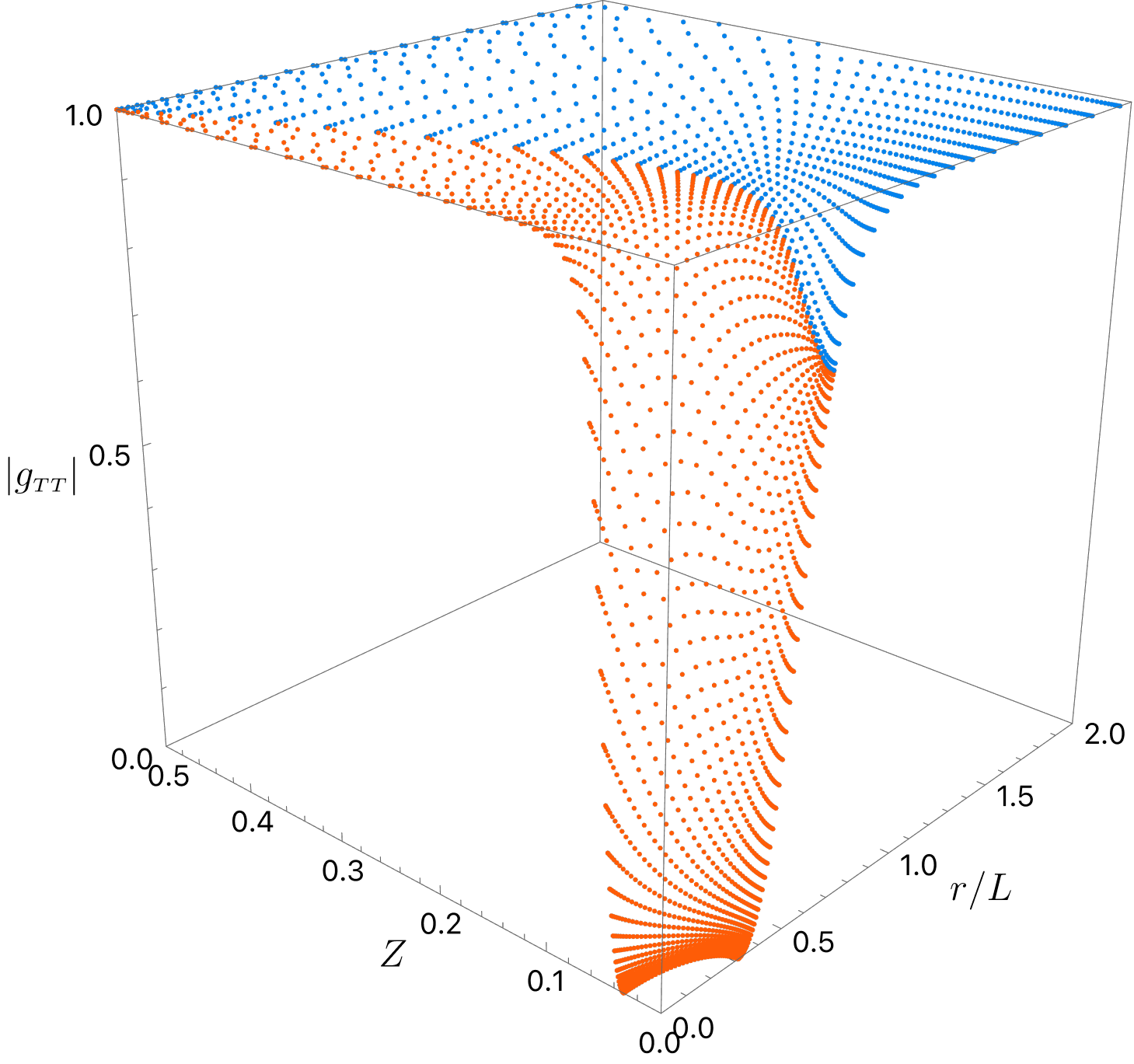}\hspace{0.5cm}
\includegraphics[width=.47\textwidth]{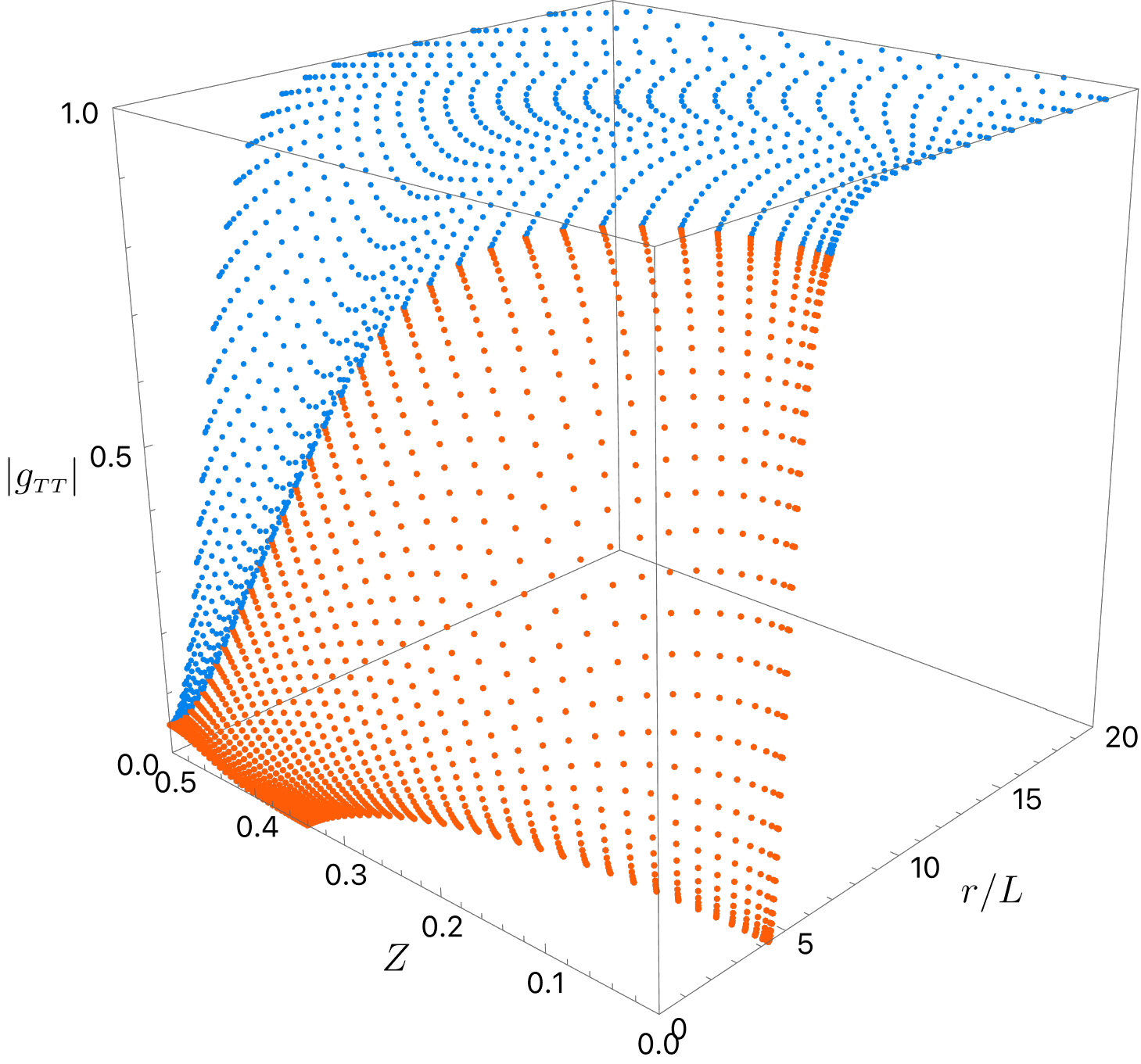}
\caption{The gauge invariant metric component $|g_{\hbox{\tiny $TT$}}|$, in the $\{r,Z \}$ coordinate chart,   for the localized black hole with  $y=0.1$ (i.e. $\mathcal{T}_{\hbox{\tiny $H$}}^{(0)}\simeq 11.9649$ and $\mathcal{E}^{(0)}\simeq  2.607\times 10^{-10}$)  ({\bf left panel}) and  with $y_0=1.8275$ (i.e. $\mathcal{T}_{\hbox{\tiny $H$}}^{(0)}\simeq 1.35713550$ and $\mathcal{E}^{(0)}\simeq 0.01047193$)  ({\bf right panel}), using the same color code as in Fig.~\ref{Fig:patches}. Recall that  the circle $S^1_{\hbox{\tiny $L$}}$ starts at $Z=0$ and `reaches its halve' at $Z=1/2$ ($\xi=1$). Further note that $r$ extends all the way to infinity but `quickly' approaches $g_{\hbox{\tiny $TT$}}\sim 1$ at small values of $r/L$}\label{Fig:LOCgtt}
\end{figure}

As in the non-uniform case, we solve the boundary value problem numerically using a Newton-Raphson algorithm.  To discretise the numerical grid, we split the integration domain into two patches as shown in Fig.~\ref{Fig:patches}: in the (far region) patch $I$ we use  $\{\rho,\xi\}$ coordinates while in the (near horizon) patch $II$ we use  $\{x,y\}$ coordinates. In each patch we use Chebyshev-Gauss-Lobatto $N \times N$ grids using transfinite interpolation (these methods are reviewed in \cite{Dias:2015nua}; the results we present have up to $N=50$). The patching boundary  (magenta curve) between patches $I$ and $II$ is given by $y=k_0 (1-x^2)$, and we have freedom to choose $k_0=1/2$ (the value we used to get our numerical results). 
At this patch boundary, we require that: 1) the line elements given by \eqref{localizedansatz} do match, and 2) the normal derivative of all metric functions across the patch boundary do match (see this is indeed the case in the examples of Fig.~\ref{Fig:LOCgtt} to be discussed soon).  

Given $f_i(x,y)$, the entropy and temperature of the localized black holes are: 
\begin{align}\label{thermoLocGRnum}
&\mathcal{S}_{\hbox{\tiny $H$}}^{(0)}= \int_0^1 \mathrm dx\,\frac{256\,y_0 \,{\rm arcsinh}^8{\bigl( }\frac{y_0 \left(1-x^2\right)}{\sqrt{1+y_0^2\,x^2 \left(2-x^2\right)}} {\bigr) }}{105 \pi^5 \sqrt{2-x^2}\sqrt{1+y_0^2\,x^2 \left(2-x^2\right)}} \,\sqrt{f_3(x,1)}\, f_4(x,1)^4\,,\nonumber \\
&\mathcal{T}_{\hbox{\tiny $H$}}^{(0)}= \frac{3}{2^{4/3}}\frac{\sqrt{1+y_0^2}}{y_0}.
\end{align}
Localized black holes are a one-parameter family of solutions. We can take this parameter to be $y_0$ for our numerical solutions which $-$ see \eqref{thermoLocGRnum} $-$ is essentially the temperature $\mathcal{T}_{\hbox{\tiny $H$}}^{(0)}$. Note that the $S^1_{\hbox{\tiny $L$}}$ length $L$ just sets a scale and drops out of the equations of motion. 
 We choose the reference metric with $y_0=1/10$ as a first seed for the Newton-Raphson algorithm, and then march our code for increasing (decreasing) values of $y_0$ up (down) to  $y_0=1.8275$  ($y_0=0.02$). 
Our choice of DeTurck reference metric has restricted our temperature range to $\mathcal{T}_{\hbox{\tiny $H$}}^{(0)} >3/2^{4/3}\approx 1.19$; this is \eqref{thermoLocGRnum} with $y_0=0$. Should we need to cross this value to lower temperatures, we would need to change our reference metric.
In practice we will not need to do so since we will find that localized black holes merge (and thus terminate) with non-uniform strings at a temperature that is in the vicinity of   $\mathcal{T}_{\hbox{\tiny $H$}}^{(0)}\sim 1.35 > 1.19$: see later discussion of \eqref{SG:ConicalTransition}. 
To get the Helmoltz free energy $\mathcal{F}^{(0)}$ we integrate the first law of thermodynamics, $\mathrm d\mathcal{F}^{(0)}=-\mathcal{S}_{\hbox{\tiny $H$}}^{(0)} \,\mathrm d\mathcal{T}_{\hbox{\tiny $H$}}^{(0)}$, and the energy is then $\mathcal{E}^{(0)}=\mathcal{F}^{(0)}+\mathcal{T}_{\hbox{\tiny $H$}}^{(0)}\mathcal{S}_{\hbox{\tiny $H$}}^{(0)}$.  The tension $\mathcal{T}_{\hbox{\tiny $Z$}}^{(0)}$ along the circle $S^1_{\hbox{\tiny $L$}}$ can be obtained from the Smarr relation \eqref{SGStaticFirstlawSmarr}, $ 8\,\mathcal{E}^{(0)} =9\,\mathcal{T}_{\hbox{\tiny $H$}}^{(0)}\,\mathcal{S}_{\hbox{\tiny $H$}}^{(0)}+\mathcal{T}_{\hbox{\tiny $Z$}}^{(0)}$.
The values of these thermodynamics quantities are in excellent agreement with the results obtained using the standadard Arnowitt-Deser-Misner formalism \cite{Harmark:2003yz,Dias:2007hg} or the covariant Noether charge formalism (a.k.a. covariant phase space method) \cite{Wald:1999wa,Dias:2019wof}. 

The thermodynamic quantities  $\{ \mathcal{E}^{(0)},\mathcal{T}_{\hbox{\tiny $Z$}}^{(0)},\mathcal{T}_{\hbox{\tiny $H$}}^{(0)},\mathcal{S}_{\hbox{\tiny $H$}}^{(0)}\}$ of 11-dimensional localized black holes will be presented in section~\ref{sec:Results-Grav}.
The BFSS thermodynamics $\{ \varepsilon, \sigma, \tau, \mathfrak{f} \}$ for the localized BFSS phase, that will be presented in section~\ref{sec:Results-BFSS}, is obtained from $\{ \mathcal{E}^{(0)},\mathcal{T}_{\hbox{\tiny $Z$}}^{(0)},\mathcal{T}_{\hbox{\tiny $H$}}^{(0)},\mathcal{S}_{\hbox{\tiny $H$}}^{(0)}\}$ via the supergravity/BFSS map \eqref{QFTthermoMap}. 

As a first flavour of how the localized black holes look like and how their properties change as we move along this 1-parameter family of solutions, we plot the norm of the gauge invariant metric component 
 $|g_{\hbox{\tiny $TT$}}|\equiv A(r,Z)$ $-$ as defined in \eqref{solnSG:vacAllStatic} $-$ as a function of the cylindrical coordinates $\{ r,Z\}$ used in  \eqref{solnSG:vacAllStatic}. For that, we take  \eqref{localizedansatz} and apply the coordinate transformation $r=r(\rho,\chi)$ and $Z=Z(\rho,\chi)$ as defined in the discussion of \eqref{MinkXcircle1}-\eqref{MinkXcircle3} or, equivalently,  $r=r(x,y)$ and $Z=Z(x,y)$ after using \eqref{coordmap}.
 This is done in the left panel of Fig.~\ref{Fig:LOCgtt} for a black hole with `large' temperature $\mathcal{T}_{\hbox{\tiny $H$}}^{(0)}\simeq 11.9649$ (i.e. $y=0.1$), that is for a localized black hole that is (effectively) `close' to the limit $\mathcal{T}_{\hbox{\tiny $H$}}^{(0)}\to \infty$ where the localized family reduces to the $d=11$ Schwarzschild-Tangherlini black hole with a perfectly round $S^9$ horizon topology\footnote{\label{foot:NoLargerT}We have generated solutions up to $\mathcal{T}_{\hbox{\tiny $H$}}^{(0)}\simeq 59.5394$ (i.e. down to $\mathcal{E}^{(0)}\simeq  7.060 \times 10^{-16}$ or $y_0=0.02$) which are included in the top panel of  Fig.~\ref{Fig:thermoGR}. They are not shown in the bottom panel of Fig.~\ref{Fig:thermoGR} because the plot would not show clearly the interesting structure at small temperatures. On the other hand, for this  $y_0=0.02$, $|g_{tt}|/L^2$ is qualitatively very similar to the left panel of Fig.~\ref{Fig:LOCgtt}.}. On the other hand, in the right panel of Fig.~\ref{Fig:LOCgtt} we plot  $|g_{\hbox{\tiny $TT$}}|\equiv A(r,Z)$ in the opposite small temperature limit, namely for $\mathcal{T}_{\hbox{\tiny $H$}}^{(0)}\simeq 1.35714$ (i.e. $y=1.8275$). This is the localized black hole (that we present) that is the closest to the merger with the non-uniform string branch of solutions. Thus, the horizon of this localized black hole is expected to be highly deformed, consistent with the fact that it is approaching the conical point (often called Kol's double-cone merger) where its $S^9$ horizon topology is at the verge of changing into the horizon topology $S^{8}\times S^1_{\hbox{\tiny $L$}}$ of a non-uniform string \cite{Kol:2002xz,Kol:2003ja,Emparan:2019obu,Emparan:2024mbp} (more in section~\ref{sec:Results-Grav}).  
 In these plots of Fig.~\ref{Fig:LOCgtt}, we use the same color code displayed in Fig.~\ref{Fig:patches} which identifies which coordinate chart $-$ $\{ \rho, \xi \}$ (blue points) or $\{ x, y \}$ (red points) $-$ we used to construct the solution in the particular region we look at. 

Comparing the transition from the left to the right panels of Fig.~\ref{Fig:LOCgtt}, we see that  $\big| g_{\hbox{\tiny $TT$}}\big|_{r=0,Z=1/2} \to 0$ as we increase $y_0$ (i.e., as we decrease the temperature or increase the energy)\footnote{This corresponds to $\big| g_{tt}\big|_{\rho=0,\xi=1} \to 0$ and  $\big| g_{tt}\big|_{x=1,y=0} \to 0$ as we increase $y_0$ (i.e., as we decrease the temperature or increase the energy).}. That is to say, starting from very large temperatures where localized black holes resemble the $d=11$ Schwarzchild-Tangherlini black hole, as we increase the black hole deformation and it approaches the merger with the non-uniform string, we see that the value of $g_{\hbox{\tiny $TT$}}$ pinches towards zero at the point where the $\partial_{\phi}$ and $\mathbb{Z}_2$ axes meet.
Moreover, we also see that the horizon is increasingly approaching $Z=1/2$ (note the evolution along $Z$-axis from left to right panel)\footnote{For our purposes (see Figs.~\ref{Fig:thermoGR}$-$\ref{fig:BFSSphaseDiagrams} and their discussion) it is not necessary to approach further the merger point, given that this would be computationally costly for little gain.}.
 These two features reflect the fact that we are approaching a topology-changing transition where the horizon topology $S^{9}$ will change into $S^{8}\times S^1_{\hbox{\tiny $L$}}$; the merger with the non-uniform branch occurs when the horizon exactly reaches $Z=1/2$ and thus the horizon extends along all the circle  $S^1_{\hbox{\tiny $L$}}$  \cite{Kol:2002xz,Kol:2003ja,Emparan:2019obu,Emparan:2024mbp} (see Fig.~\ref{fig:sketch}; more in section~\ref{sec:Results-Grav}).

\section{Phase diagrams and discussion of results \label{sec:Results}}
 
 We finally have all we need to present the BFSS phase diagram with its three BFSS phases, namely the (non-)uniform and localized 1-parameter families of thermal states.
For that, as a previous minimalistic summary of our work stated in the Introduction, we first need to find and present the thermodynamics the (non-)uniform and localized solutions of asymptotically ${\mathbb R}^{(1,9)}\times S^1_{\hbox{\tiny $L$}}$ vacuum Einstein gravity. We do this in section~\ref{sec:Results-Grav}. Then, in section~\ref{sec:Results-BFSS}, we apply the Carrollian thermodynamic map \eqref{QFTthermoMap}  to find (and present) the thermodynamics of the (non-)uniform and localized BFSS thermal states.  

\subsection{Supergravity (Einstein gravity) thermodynamics \& phase diagrams \label{sec:Results-Grav}}
\begin{figure} [ht]
    \centering
     \includegraphics[width=0.66\textwidth]{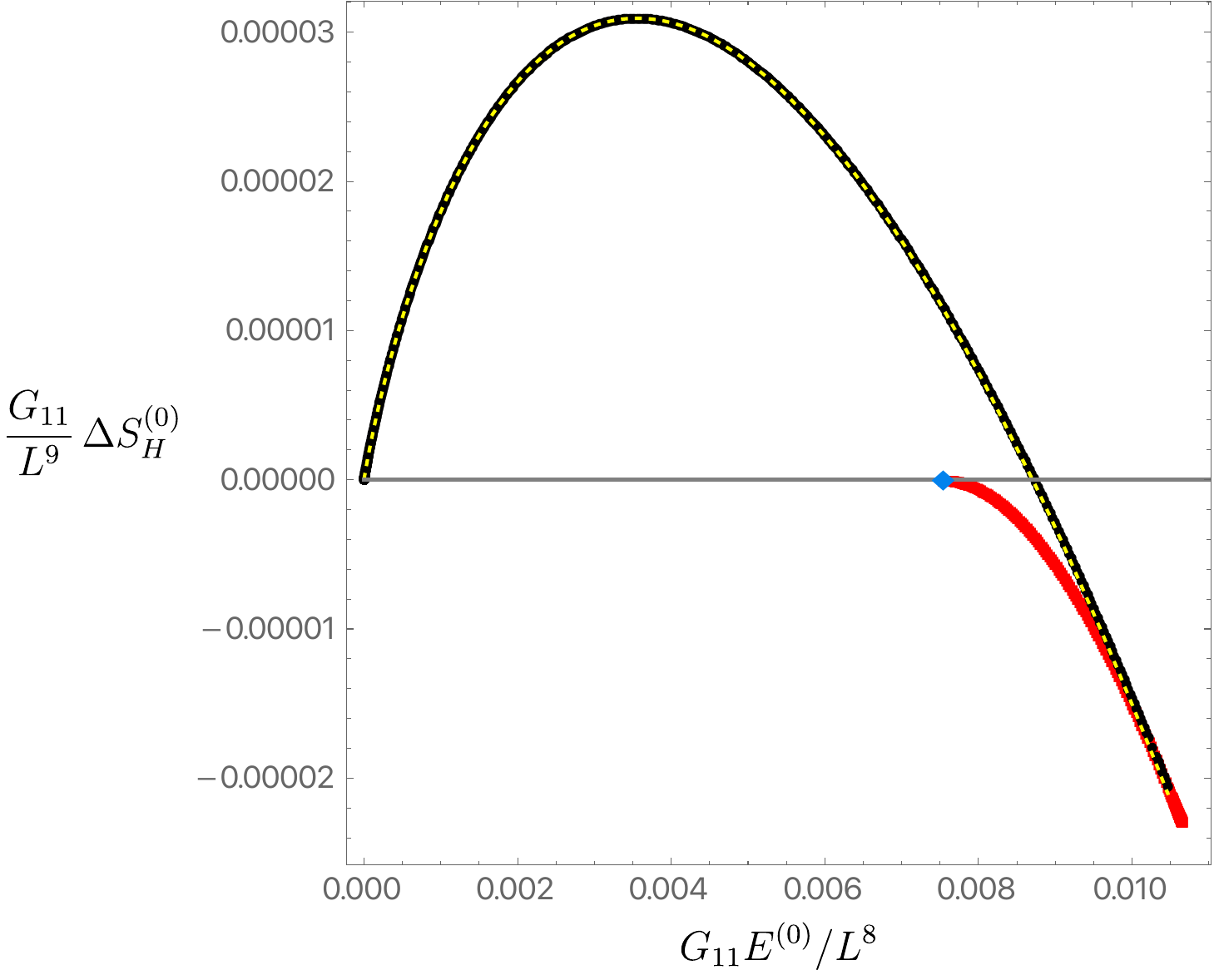}
     \\
    \includegraphics[width=0.66\textwidth]{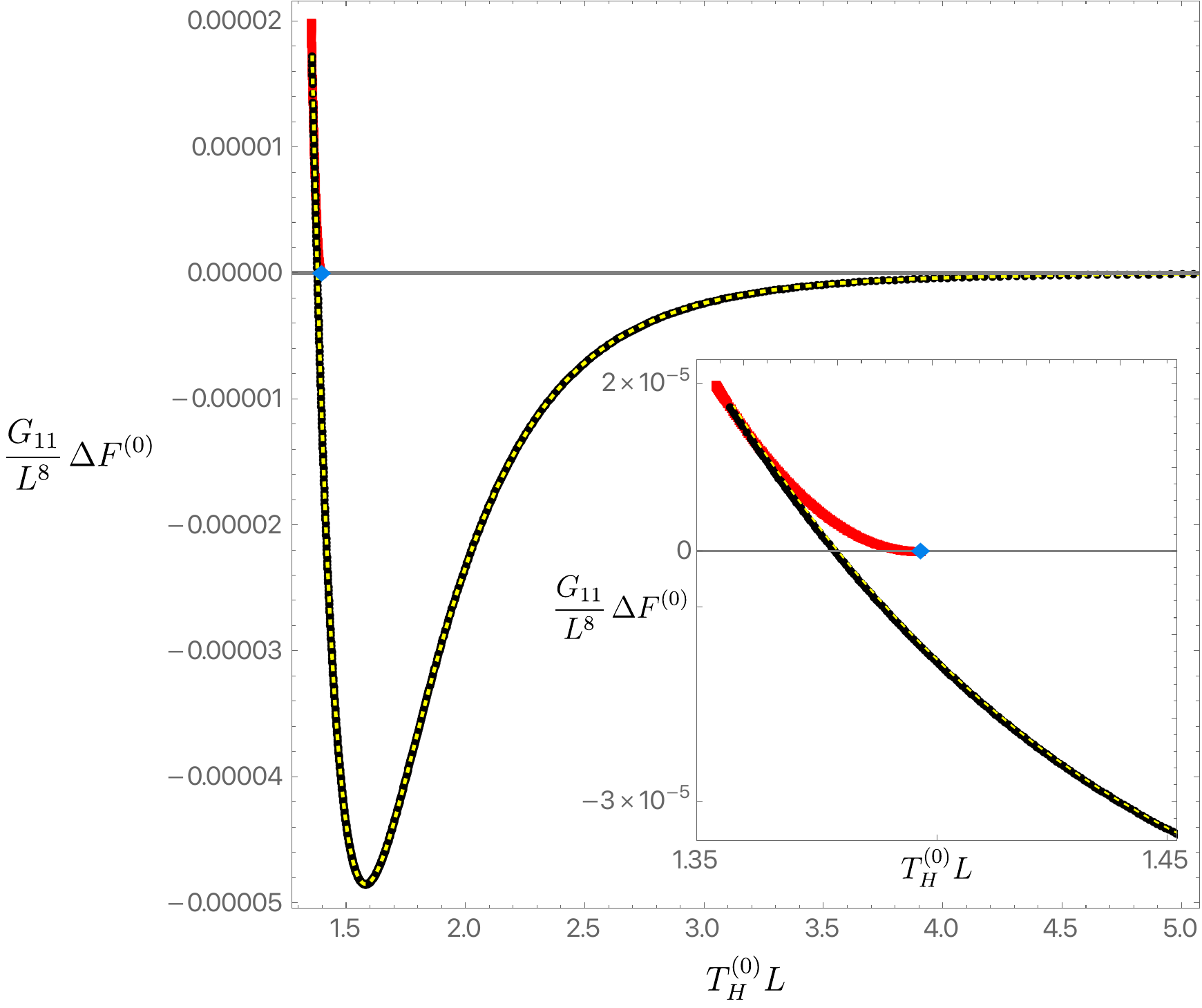}
\caption{{\bf Top panel:} Phase diagram of asymptotically ${\mathbb R}^{(1,9)}\times S^1_{\hbox{\tiny $L$}}$ vacuum Einstein solutions in the microcanonical ensemble. $\Delta \mathcal{S}^{(0)}_{\hbox{\tiny $H$}}(\mathcal{E}^{(0)})$ gives the entropy difference between a given solution and the uniform black string with the same energy $\mathcal{E}^{(0)}$.  
{\bf Bottom panel:} Phase diagram of asymptotically ${\mathbb R}^{(1,9)}\times S^1_{\hbox{\tiny $L$}}$ vacuum Einstein solutions in the canonical ensemble. $\Delta \mathcal{F}^{(0)}(\mathcal{T}^{(0)}_{\hbox{\tiny $H$}})$  gives the free energy difference between a given solution and the uniform black string with the same temperature $\mathcal{T}^{(0)}_{\hbox{\tiny $H$}}$.  
In these diagrams,  the black disks show exact numerical data for  the localized black hole family and the red squares describe the exact numerical data for the non-uniform black string family. Finally, the horizontal (gray) line with $\Delta \mathcal{S}^{(0)}_{\hbox{\tiny $H$}}=0$ (top panel) or $\Delta \mathcal{F}^{(0)}=0$ (bottom panel)  describes the uniform black string while the yellow dash curve (essentially on top of the black curve)  describes the perturbative approximation \eqref{thermoLocGRpert2} for the localized black hole \cite{Harmark:2003yz}.}\label{Fig:thermoGR}
\end{figure}

Start by recalling, from section~\ref{sec:vacuumBHs}, that all asymptotically ${\mathbb R}^{(1,9)}\times S^1_{\hbox{\tiny $L$}}$ static solutions of Einstein gravity in $d=11$ dimensions (that are also solutions of 11-dimensional supergravity with vanishing gauge field $C(3)$)  can be written as \eqref{solnSG:vacAllStatic}. This applies, in particular, to the uniform and non-uniform black strings as well as to the  Kaluza-Klein black holes localized on the $S^1_{\hbox{\tiny $L$}}$ (recall Fig.~\ref{fig:sketch}). Reading the asymptotic decays \eqref{SG:KKdecay} as well as the horizon quantities of the solutions, one obtains the thermodynmics quantities \eqref{SGETz:static}-\eqref{SG:TS:static} for any of the  three vacuum Einstein solutions. We can then plot the three 1-parameter families of solutions in a phase diagram and analyse their competition.

We can start with the microcanonical ensemble where, for a given dimensionless energy $\mathcal{E}^{(0)}\equiv \frac{G_{11}}{L^8}\,E^{(0)}$, the solution that is the most dominant is the one with the highest entropy $\mathcal{S}_{\hbox{\tiny $H$}}^{(0)} \equiv   \frac{G_{11}}{L^9}\, S_{\hbox{\tiny $H$}}^{(0)}$. Because the entropies of the three families of solutions are very close to each other (in certain regions where they co-exist), instead of $\mathcal{S}_{\hbox{\tiny $H$}}^{(0)}(\mathcal{E}^{(0)})$, for clarity of the diagrams we plot $\Delta \mathcal{S}_{\hbox{\tiny $H$}}^{(0)}(\mathcal{E}^{(0)})$ which measures the entropy difference between a given solution and the uniform black string with the same energy $\mathcal{E}^{(0)}$. The micro-canonical phase diagram of vacuum Einstein solutions is displayed in the top panel of Fig.~\ref{Fig:thermoGR}. The uniform phase $-$ as given by \eqref{ThermoGRunif2} $-$ is described by the horizontal gray line with  $\Delta \mathcal{S}_{\hbox{\tiny $H$}}^{(0)}=0$, the non-uniform phase by the red squares, and the localized phase by the black disks. Analysing the top panel of  Fig.~\ref{Fig:thermoGR} we first conclude that the non-uniform strings (red squares) never dominate the canonical ensemble. They bifurcate from (or merge with) the non-uniform strings at the onset point of the Gregory-Laflamme  instability  (uniform strings with $\mathcal{E}^{(0)}<\mathcal{E}^{(0)}_{\hbox{\tiny GL}}$ are unstable) already identified in section~\ref{sec:GL}, namely:
\begin{equation} \label{SG:microcanonicalGL}
\big\{ \mathcal{E}^{(0)}_{\hbox{\tiny GL}}, \mathcal{S}^{(0)}_{\hbox{\tiny GL}} \big\}\simeq
\big\{ 0.00754675,\, 0.00472457 \big\},
\end{equation}
which is represented by the blue diamond in Fig.~\ref{Fig:thermoGR}.
The next important conclusion is that for small energies, $\mathcal{E}^{(0)}< \mathcal{E}^{(0)}_c$, the localized black hole dominates the micro-canonical ensemble. But, at a critical energy $\mathcal{E}^{(0)} = \mathcal{E}^{(0)}_c$  there is a first order phase transition and, for  $\mathcal{E}^{(0)}> \mathcal{E}^{(0)}_c$, it is the uniform black string that has the highest entropy. More concretely this first order phase transition occurs when the localized solution has $ \Delta \mathcal{S}_{\hbox{\tiny $H$}}^{(0)}=0 $, i.e. at the point
\begin{equation} \label{SG:microcanonicalFirstOrTransition}
\big\{ \mathcal{E}^{(0)}_c, \mathcal{S}^{(0)}_c \big\}\simeq\big\{ 0.00873161,\, 0.00558142 \big\}. 
\end{equation}
In  Fig.~\ref{Fig:thermoGR} we also display the dashed yellow curve which describes the perturbative prediction \eqref{thermoLocGRpert2} for the localized black hole  \cite{Harmark:2003yz}. We see that this curve is essentially on top of the exact numerical black curve. In particular, the perturbative result  \eqref{thermoLocGRpert2} predicts that the first order phase transition should occur at $\big\{ \mathcal{E}^{(0)}_c, \mathcal{S}^{(0)}_c \big\} \big|_{\rm pert}\simeq\big\{ 0.00870640,\, 0.00556301\big\}$. 
 This corresponds to a relative error w.r.t. \eqref{SG:microcanonicalFirstOrTransition} of just $\sim0.3\%$. So the perturbative result \eqref{thermoLocGRpert2} turns out to be an excellent approximation for the whole range of energies where the localized black hole exists, which is quite remarkable (so, well beyond the region of very small energies $-$ for which  the localized black hole is a very small deformation of $d=11$ Schwarzschild black hole  $-$  where it should hold).

We can now discuss the canonical ensemble of static vacuum solutions of Einstein gravity with ${\mathbb R}^{(1,9)}\times S^1_{\hbox{\tiny $L$}}$ asymptotics, whose phase diagram is displayed in the bottom panel of Fig.~\ref{Fig:thermoGR}. This time we want to fix the dimensionless temperature $\mathcal{T}_{\hbox{\tiny $H$}}^{(0)} \equiv L \,T_{\hbox{\tiny $H$}}^{(0)}$
and the canonical potential is the Helmoltz free energy $\mathcal{F}^{(0)}\equiv \frac{G_{11}}{L^8}\,F^{(0)}$. For a given temperature, the solution with smallest free energy dominates the ensemble and we choose to plot the free energy difference $\Delta \mathcal{F}^{(0)}(\mathcal{T}_{\hbox{\tiny $H$}}^{(0)} )$ of a given solution to the one of the uniform phase with the same temperature $\mathcal{T}_{\hbox{\tiny $H$}}^{(0)}$. The colour code in the bottom panel is the same as for the top panel. We immediately conclude that non-uniform strings (red squares) also never dominate the canonical ensemble. They bifurcate from the uniform black string through a second order phase transition at the Gregory-Laflamme onset (blue diamond in Fig.~\ref{Fig:thermoGR}), namely:
\begin{equation} \label{SG:canonicalGL}
\big\{ \mathcal{T}^{(0)}_{\hbox{\tiny GL}}, \mathcal{F}^{(0)}_{\hbox{\tiny GL}} \big\}\simeq
\big\{ 1.39767231,\, 0.00094334 \big\}. 
\end{equation}
In the bottom panel of  Fig.~\ref{Fig:thermoGR}, we further see that for large and intermediate temperatures, the localized black holes (black disks) dominate over the uniform string (horizontal gray line given by \eqref{ThermoGRunif2}). However, as we decrease the temperature, one reaches a critical value, $\mathcal{T}^{(0)}_{\hbox{\tiny $H$}}= \mathcal{T}^{(0)}_c$, where a first order phase transition also occurs at the canonical ensemble, i.e. where the localized solution has $\Delta \mathcal{F}^{(0)} =0$ for finite $\mathcal{T}^{(0)}_c$. Namely, for $\mathcal{T}^{(0)}_{\hbox{\tiny $H$}} < \mathcal{T}^{(0)}_c$, it is the uniform black string that now has lower free energy.\footnote{\label{foot} Note that the equivalence of statistical ensembles (in the thermodynamic limit) assumes that the entropy increases when the temperature grows, i.e. that the specific heat of the system is positive. However, typically, this is not the case for black object systems whose temperature is inversely proportional to the energy and have negative specific heat. Thus, the fact that localized black holes dominates the microcanononical ensemble for small energies does not necessarily impliy that they should also dominate the canonical ensemble for large temperatures. This also means that, although the previous scenario turns out to  hold in the present system, the critical point where the first order phase transition occurs in the two ensembles are not in one-to-one correspondence, i.e. the $\mathcal{T}^{(0)}_c$ in \eqref{SG:canonicalFirstOrTransition} does not correspond to $\mathcal{E}^{(0)}_c$ in \eqref{SG:microcanonicalFirstOrTransition} when we use the thermodynamic expressions \eqref{ThermoGRunif} for the uniform string.} 
In the bottom panel of Fig.~\ref{Fig:thermoGR}, we also display the yellow dashed curve that describes the perturbative result \eqref{thermoLocGRpert2} for the localized black hole as found in \cite{Harmark:2003yz}. Again, this time not surprisingly, we see that this curve is basically on top of the black exact numerical curve. The perturbative analysis predicts that in the first order phase transition in the canonical ensemble should occur at     
\begin{equation} \label{SG:canonicalFirstOrTransition}
\big\{ \mathcal{T}^{(0)}_c, \mathcal{F}^{(0)}_c \big\} \simeq \big\{ 1.37929643, 0.00103491 \big\}, 
\end{equation}
while the perturbative analysis \eqref{thermoLocGRpert2} predicts $\big\{ \mathcal{T}^{(0)}_c, \mathcal{F}^{(0)}_c \big\}  \big|_{\rm pert} \simeq\big\{ 1.37996978, 0.00103138 \big\} $. Again, this corresponds to a very small  relative error w.r.t. \eqref{SG:canonicalFirstOrTransition}:  just $\sim0.3\%$. So, the perturbative result \eqref{thermoLocGRpert2} turns out to be an excellent approximation for the whole range of temperatures where the localized black hole exists which is impressive.

 Note that in the phase diagrams of Fig.~\ref{Fig:thermoGR}, when the (red) non-uniform string branch meets the (black) localized black hole family, the horizon topology changes from $S^8\times S^1$ into $S^9$ through a conical  topology-changing transition, i.e. via a double-cone geometry (akin to the conifold transitions that appear in Calabi-Yau spaces) \cite{Kol:2002xz,Kol:2003ja,Emparan:2019obu,Emparan:2024mbp}. One of the branches must also have a regular cusp (unless the topology-changing transition occurs exactly at this cusp). We find that this conical topology-changing transition (and cusp) occurs somewhere in the window\footnote{\label{foot:detailsLastKol}The last localized black hole we present has $\big\{ \mathcal{E}^{(0)}, \mathcal{S}^{(0)}_{\hbox{\tiny $H$}} \big\}= \big\{ 0.01047193,\, 0.00684939 \big\}$, i.e. $\Delta\mathcal{S}^{(0)}_{\hbox{\tiny $H$}} \approx -0.00002055$ and  $\big\{ \mathcal{T}^{(0)}_{\hbox{\tiny $H$}}, \mathcal{F}^{(0)} \big\}= \big\{ 1.35713550,\,  0.00117638 \big\} $, i.e.  $\Delta\mathcal{F}^{(0)} \approx 0.00001721$. On the other hand,  the last non-uniform string we present has $\big\{ \mathcal{E}^{(0)}, \mathcal{S}^{(0)}_{\hbox{\tiny $H$}} \big\}= \big\{ 0.01065378,\,  0.00698355 \big\}$, i.e. $\Delta\mathcal{S}^{(0)}_{\hbox{\tiny $H$}} \approx -0.00002292$ and  $\big\{ \mathcal{T}^{(0)}_{\hbox{\tiny $H$}}, \mathcal{F}^{(0)} \big\}= \big\{ 1.35409244,\,  0.00119743 \big\} $, i.e.  $\Delta\mathcal{F}^{(0)} \approx 0.00001990$. Strictly speaking, it could happen that the topology-changing occurs at a value of  $\mathcal{E}^{(0)}_{\hbox{\tiny K}}$ slightly above the maximum in  \eqref{SG:ConicalTransition} if the non-uniform and/or uniform branches extend a bit further before merging. For our purposes, it is not enlightening to stretch the numerical codes to pinpoint the value of $\mathcal{E}^{(0)}_{\hbox{\tiny K}}$ with more accuracy.} 
 \begin{align} \label{SG:ConicalTransition}
& 0.01047193  \lesssim  \mathcal{E}^{(0)}_{\hbox{\tiny K}}  \lesssim  
 0.01065378, \qquad  \nonumber \\
&  1.377650 \gtrsim  \mathcal{T}^{(0)}_{\hbox{\tiny K}} \gtrsim 1.35409244,
\end{align}
where the left bound is found computing the maximum value of $\mathcal{E}^{(0)}$ (minimum value of $\mathcal{T}^{(0)}_{\hbox{\tiny $H$}}$) for which localized black holes solutions exist (black disks), and the right bound follows from finding the maximum value of $\mathcal{E}^{(0)}$ (minimum value of $\mathcal{T}^{(0)}_{\hbox{\tiny $H$}}$) for which non-uniform black strings (red squares) exist. 

For reference, it should be noticed that the phase diagrams of Fig.~\ref{Fig:thermoGR} for $d=11$ are qualitatively very similar to those of Einstein gravity with  ${\mathbb R}^{(1,d-2)}\times S^1$ asymptotics in lower dimensions $d$, starting at $d=5$, where we also have (non-)uniform black strings and localized (Kaluza-Klein) black holes competing and dominating in similar manners/regions of parameter space \cite{Gubser:2001ac, Harmark:2002tr, Kol:2002xz,Wiseman:2002zc, Kol:2003ja, Harmark:2003dg, Harmark:2003yz, Kudoh:2003ki,Sorkin:2003ka,Kol:2003if,Sorkin:2004qq, Gorbonos:2004uc, Kudoh:2004hs,Gorbonos:2005px,Dias:2007hg, Harmark:2007md,Wiseman:2011by, Figueras:2012xj, Kalisch:2016fkm, Dias:2017uyv,Dias:2017coo,Kalisch:2017bin,Ammon:2018sin,Cardona:2018shd} (see reviews \cite{Kol:2004ww,Harmark:2005pp,Horowitz:2011cq}). The only (minor exception, for our purposes) is that the double-cone merger in the microcanonical diagram occurs below the uniform string curve for $5\leq d\leq 11$ while for $d\geq 12$ it occurs above \cite{Sorkin:2004qq,Figueras:2012xj}. For completeness, for $d \geq 14$, both the localized and non-uniform branches are fully above the uniform curve in the micro-canonical phase diagram (in particular, there is no longer a cusp in the $ \mathcal{S}^{(0)}_{\hbox{\tiny $H$}}(\mathcal{E}^{(0)})$ phase diagram) \cite{Sorkin:2004qq,Figueras:2012xj,Emparan:2018bmi,Emparan:2020inr}. 

Before concluding this section, we emphasize two non-trivial checks of our numerics.
Firstly, we find that our numerical non-uniform and localized vacuum solutions obey the gravitational first law \eqref{SGStaticFirstlawSmarr} with an error that is smaller than $0.01\%$.
Secondly, one finds that the second order phase transition $-$ \eqref{SG:microcanonicalGL} and \eqref{SG:canonicalGL} $-$ between uniform and non-uniform black strings (green diamond in Fig.~\ref{Fig:thermoGR}) coincides with the point where the onset of the Gregory-Laflamme instability occurs. Physically, this had to be the case but note that the two points are obtained using two completely independent  setups/formulations and numerical codes. Indeed, the onset of the Gregory-Laflamme instability was obtained solving a linear boundary value problem in section~\ref{sec:GL},  while the bifurcation point of the non-uniform string from the uniform string was obtained solving a non-linear boundary value problem in section~\ref{sec:SetupNonUnif}.  

\subsection{BFSS thermodynamics \& phase diagrams \label{sec:Results-BFSS}}

We are in position to discuss now the phase diagrams of the dual BFSS theory. Given the analysis of section~\ref{sec:MapEinsteinBFSS} and, in particular, of section~\ref{sec:BFSSthermo}, this is now a straightforward exercise.
Namely, we simply need to take the Einstein quantities plotted in the gravitational phase diagrams of Fig.~\ref{Fig:thermoGR} and apply the supergravity/BFSS map \eqref{QFTthermoMap} to get the dual BFSS thermodynamics and phase diagrams. 

\begin{figure}[tb]
    \centering
     \includegraphics[width=0.65\textwidth]{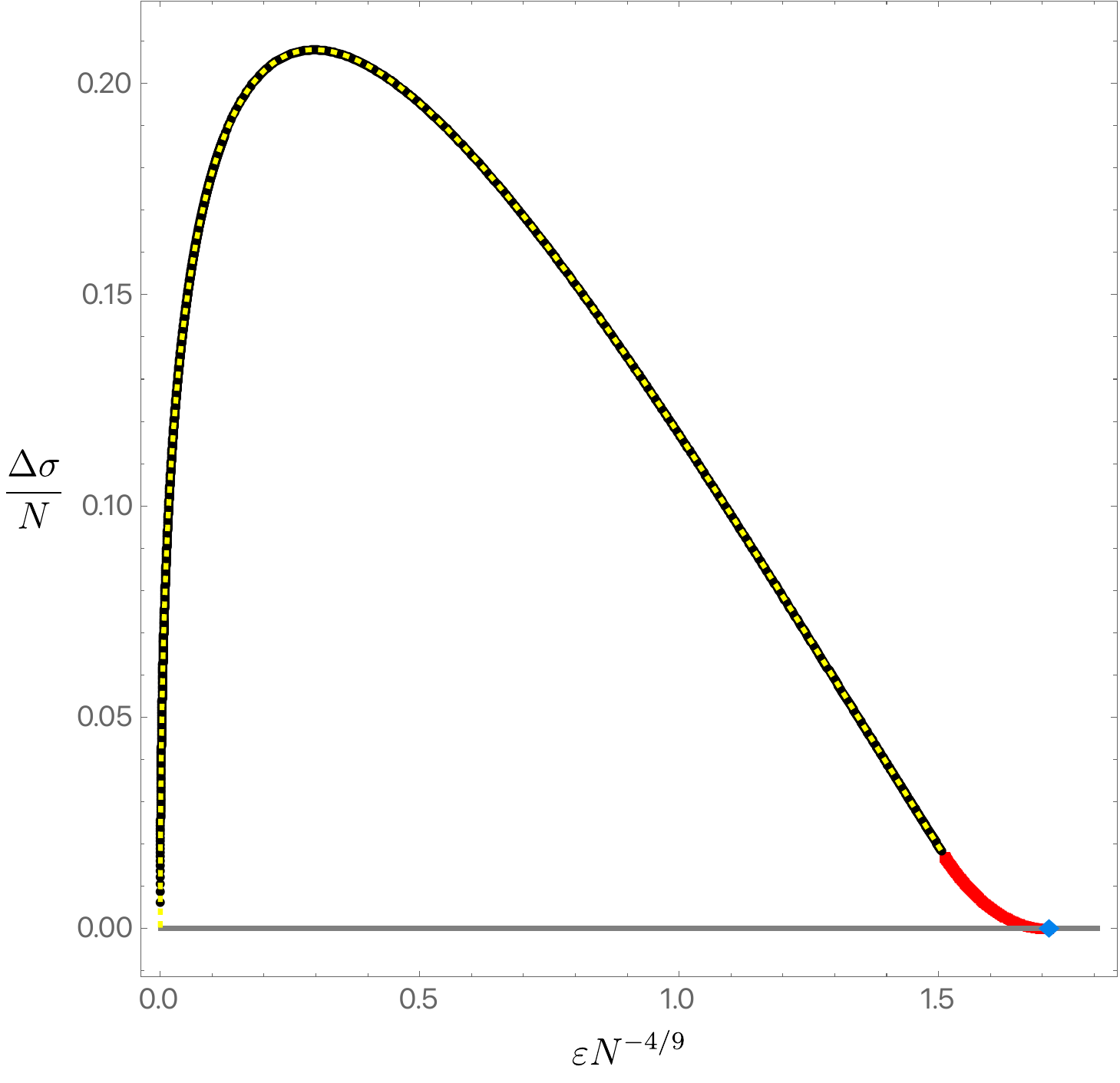} 
\includegraphics[width=0.65\textwidth]{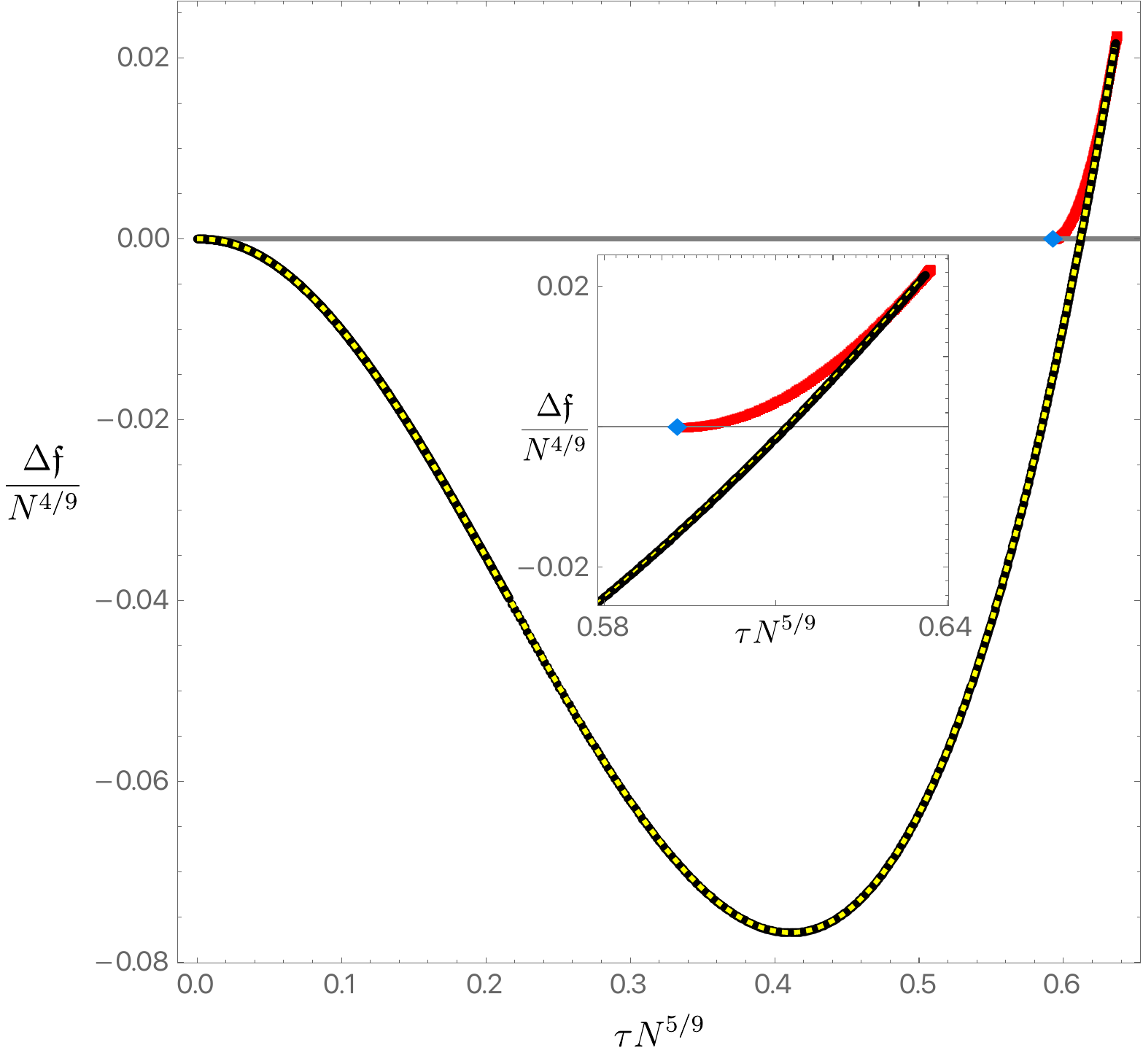}
    \caption{$\Delta \mathfrak{f}$, the free energy difference between a given thermal phase and the uniform phase at the same $\tau$, as a function of $\tau$. The black disks represent exact numerical data for the localized BFSS phase, while the red squares denote the non-uniform BFSS phase. 
    }
    \label{fig:BFSSphaseDiagrams}
\end{figure}

This time let us start with the discussion of the canonical phase diagram, shown in the bottom panel of Fig.~\ref{fig:BFSSphaseDiagrams}. For a given dimensionless temperature $\tau N^{5/9}$, the BFSS phase with smallest free energy density $\mathfrak{f}/N^{4/9}$ is the one that dominates the  canonical ensemble. Consistent with the color code used in Fig.~\ref{Fig:thermoGR}, the uniform BFSS phase  $-$ as given by \eqref{ThermoSYMunif} $-$ is the gray line, the black disks show exact numerical data for the localized BFSS phase and the red squares describe the exact numerical data for the non-uniform phase. As before, to make the presentation more clear, instead of $\mathfrak{f}(\tau)$ we plot the difference  $\Delta \mathfrak{f}(\tau)$ between the free energy density of a given BFSS solution and  $\mathfrak{f}(\tau)$ of the uniform BFSS phase with the same temperature. From the  bottom panel of Fig.~\ref{fig:BFSSphaseDiagrams} one concludes that the uniform BFSS dominates the canonical ensemble for small temperatures all the way up to $\tau=\tau_c$. At  the critical temperature  $\tau=\tau_c$ there is a first order phase transition and, for $\tau>\tau_c$, it is the uniform BFSS phase that has lower free energy. 
This is one of our main results. The BFSS canonical phase diagram aligns with the diagram conjectured by \cite{Itzhaki:1998dd}, but here we provide a precise prediction for the critical temperature (and free energy density) where the first-order phase transition occurs, namely
\begin{equation} \label{BFSS:canonicalFirstOrTransition}
\big\{ \tau_c N^{5/9} ,\mathfrak{f}_c N^{-4/9}\big\} \simeq \big\{ 0.61196576, -1.04043637 \big\}. 
\end{equation}
In  Fig.~\ref{fig:BFSSphaseDiagrams} we also display the dashed yellow curve which describes the perturbative prediction \eqref{ThermoSYMpertLoc2} for the localized BFSS phase. This curve is remarkably on top  of the exact numerical black curve not only for temperatures close to zero (where the approximation should be strictly valid) but along all the window of existence of the localized phase. In particular, the perturbative expression \eqref{ThermoSYMpertLoc2} predicts that the first order phase transition should occur at $\big\{ \tau_c N^{5/9} ,\mathfrak{f}_c N^{-4/9}\big\} 
 \big|_{\rm pert} \simeq\big\{ 0.61185189, -1.03989438 \big\} $. This corresponds to a relative error w.r.t. \eqref{BFSS:canonicalFirstOrTransition} of just $\sim 0.05\%$, i.e. \eqref{ThermoSYMpertLoc2} is indeed and excellent approximation for the whole range of temperatures where the localized BFSS exists.
 
 In Fig.~\ref{fig:BFSSphaseDiagrams} we also see that the (red) non-uniform BFSS phase never dominates the canonical phase diagram. It exists in the temperature range $\tau_{\hbox{\tiny GL}} \leq \tau< \tau_{\hbox{\tiny K}}$, where $\tau_{\hbox{\tiny GL}} $ is the temperature of the point where the non-uniform BFSS phase bifurcates (or merges) with the uniform BFSS phase in a second order phase transition; this is the BFSS dual of the gravitational Gregory-Laflamme onset point. On the other hand, $\tau_{\hbox{\tiny K}}$ is the temperature of the point where the BFSS non-uniform phase merges with the localized phase (very close to, if not exactly at, the cusp of the figure). Concretely we find that:
 \begin{align} \label{BFSS:GL}
& \big\{ \varepsilon_{\hbox{\tiny GL}} N^{-4/9} ,\sigma_{\hbox{\tiny GL}}/N\big\} \simeq \big\{ 1.71316176, 4.49546383 \big\}, \nonumber\\ 
& \big\{ \tau_{\hbox{\tiny GL}} N^{5/9} ,\mathfrak{f}_{\hbox{\tiny GL}} N^{-4/9}\big\} \simeq \big\{ 0.59280163, -0.95175652 \big\},
\end{align}
and\footnote{The last localized BFSS phase we present has $\big\{  \varepsilon_{\hbox{\tiny K}} N^{-4/9},\sigma_{\hbox{\tiny K}}/N \big\}= \big\{ 1.505439701,\,  4.15547171 \big\}$, i.e. $\Delta \sigma_{\hbox{\tiny K}}/N \approx 0.01845076 $ and  $\big\{   \tau_{\hbox{\tiny K}} N^{5/9}  , \mathfrak{f}_{\hbox{\tiny K}} N^{-4/9} \big\}= \big\{ 0.63614943,\, -1.13806127 \big\} $, i.e.  $\Delta\mathfrak{f}_{\hbox{\tiny K}} N^{-4/9} \approx 0.02163747 $. On the other hand,  the last non-uniform string we present has $\big\{  \varepsilon_{\hbox{\tiny K}} N^{-4/9}, ,\sigma_{\hbox{\tiny K}}/N \big\}= \big\{ 1.51261098,\, 4.16673593 \big\}  $, i.e. $\Delta ,\sigma_{\hbox{\tiny K}}/N \approx 0.01705693 $ and  $\big\{  \tau_{\hbox{\tiny K}} N^{5/9}, \mathfrak{f}_{\hbox{\tiny K}} N^{-4/9} \big\}= \big\{ 0.63708156 ,\, -1.14193965 \big\} $, i.e.  $\Delta\mathfrak{f}_{\hbox{\tiny K}} N^{-4/9} \approx 0.02252331 $. Strictly speaking, it could happen that the topology-changing occurs at a value of  $  \tau_{\hbox{\tiny K}} N^{5/9}$ slightly above the maximum in  \eqref{BFSS:ConicalTransition} if the non-uniform and/or uniform branches extend a bit further before merging. }
 \begin{align} \label{BFSS:ConicalTransition}
&   1.505439701  \lesssim \varepsilon_{\hbox{\tiny K}} N^{-4/9}  \lesssim  1.51261098, \nonumber\\ 
&  0.63614943  \lesssim  \tau_{\hbox{\tiny K}} N^{5/9}   \lesssim 0.63708156, 
\end{align}
where the left bound is found computing the maximum value of $\varepsilon$ (maximum value of $\tau$) for which  localized BFSS (black) solutions exist, and the right bound follows from finding the minimum value of $\varepsilon$ (maximum value of $\tau$) for which non-uniform BFSS (red) solutions exist. 

It is important to note that the values \eqref{BFSS:GL}
for $\varepsilon_{\hbox{\tiny GL}}$ and $\varepsilon_{\hbox{\tiny GL}}$, obtained using the nonlinear code, do match the values \eqref{GL:energy} and \eqref{GL:temp}, obtained using the linear code of section~\ref{sec:GL}. This is a non-trivial check of our computations since these two codes are completely independent and have different nature/formulation. Indeed, the nonlinear code constructs non-uniform phases starting at the point \eqref{BFSS:GL} where they merge with uniform phase, while the linear code of section~\ref{sec:GL} searched for the Gregory-Laflamme instability onset \eqref{GL:energy}-\eqref{GL:temp} of the uniform phase. The physical interpretation of the system indicates that the two points should be the same and the independent codes indeed find this is the case.

\begin{figure}[tb]
    \centering
     \includegraphics[width=0.47 \textwidth]{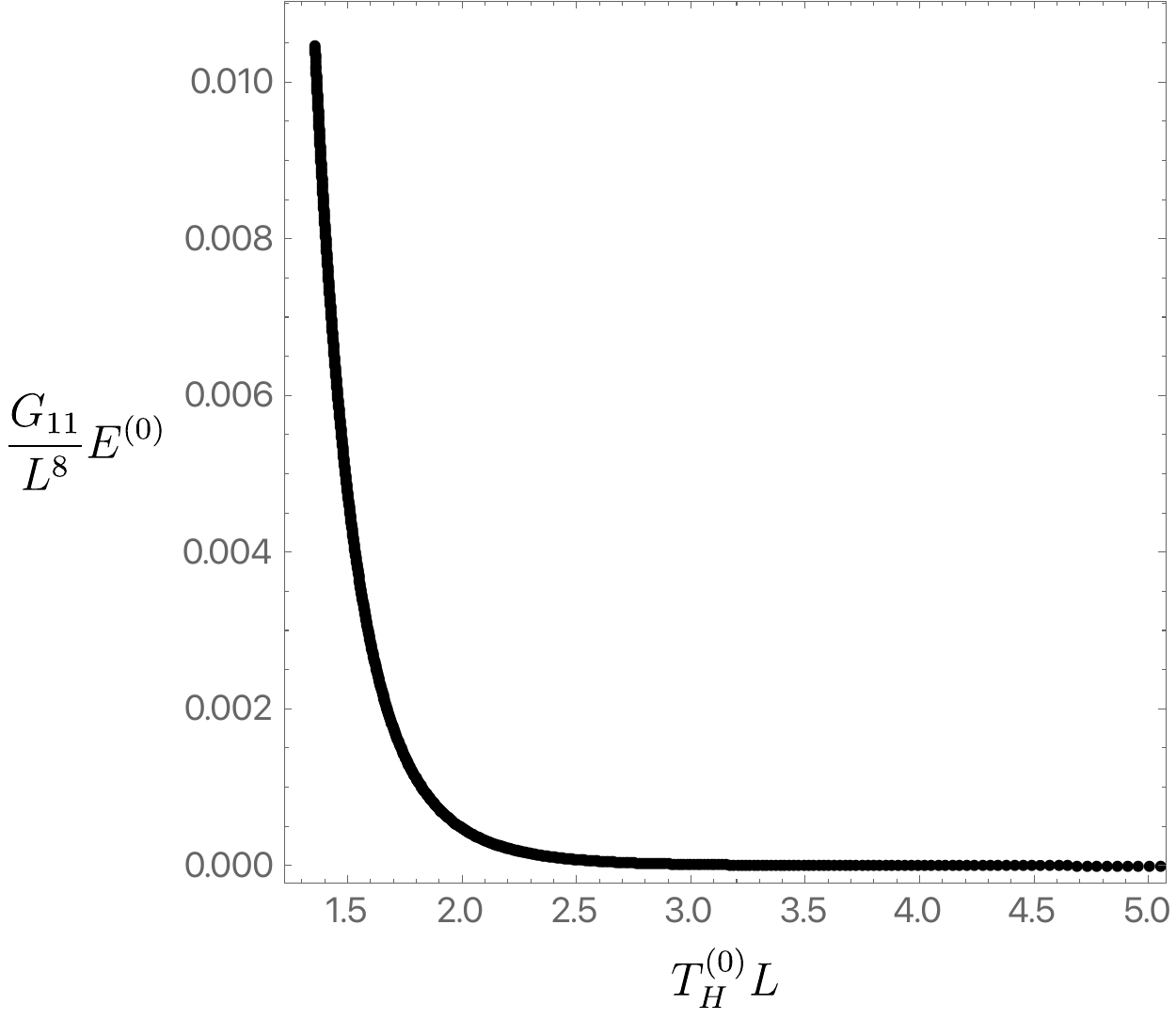}  \hspace{0.5cm}
\includegraphics[width=0.45\textwidth]{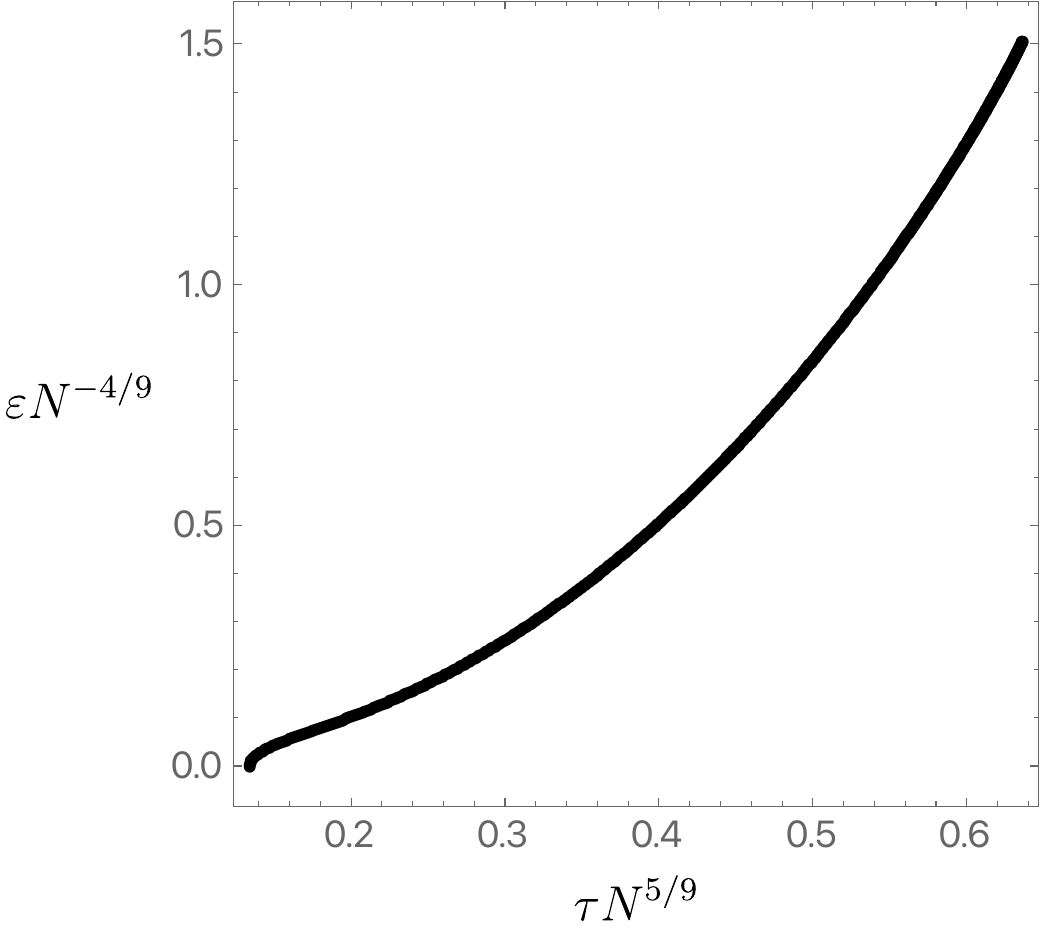}
    \caption{{\bf Top panel:} Localized black holes of 11-dimensional supergravity have negative specific heat, i.e. $\mathcal{E}^{(0)}$ is a decreasing function of $\mathcal{T}^{(0)}_{\hbox{\tiny $H$}}$. 
    {\bf Bottom panel:} Localized thermal states of the BFSS theory have positive specific heat, i.e. their energy density $\varepsilon$ is a increasing function of the temperature $\tau$. 
 }
    \label{fig:cvLoc}
\end{figure}

We now discuss the BFSS phase diagram in the microcanonical ensemble, that we left for last because it turns out to reveal a much more unexpected interesting structure. In the top panel of Fig.~\ref{fig:BFSSphaseDiagrams},  we plot $\Delta \sigma/N$, i.e. the entropy density difference between a given BFSS phase and the uniform phase at fixed energy density, as a function of the energy density $\varepsilon N^{-4/9}$. We see that for a large range of energies, namely for $0<\varepsilon< \varepsilon_{\hbox{\tiny K}}$ where the non-uniform phase does not exist, the (black) localized BFSS phase dominates over the uniform BFSS phase. However, and this comes as a surprise, for $\varepsilon_{\hbox{\tiny K}} <\varepsilon< \varepsilon_{\hbox{\tiny GL}}$ where the localized phase does not exist, the (red)  non-uniform BFSS phase dominates the microcanonical ensemble over the uniform BFSS phase.  
A topology-changing transition occurs at $\varepsilon = \varepsilon_{\hbox{\tiny K}}$, which is the holographic dual of the gravitational topology-changing transition \eqref{SG:ConicalTransition} described above  \cite{Kol:2002xz,Kol:2003ja,Emparan:2019obu,Emparan:2024mbp}. One finds that $\varepsilon_{\hbox{\tiny K}}$ is within the bound displayed in \eqref{BFSS:ConicalTransition}  where the lower bound is found computing the maximum value of $\varepsilon$ for which (black) localized BFSS solutions exist, and the upper bound follows from finding the minimum value of $\varepsilon$ for which (red) non-uniform BFSS solutions exist. Improved numerics would pin down $\varepsilon_{\hbox{\tiny K}}$ even further but is not required for our purposes. On the other hand, we find the value of $\varepsilon_{\hbox{\tiny GL}}$ to be the one displayed in \eqref{BFSS:GL}. This simply follows from applying the supergravity/BFSS map~\eqref{QFTthermoMap} to the gravitational Gregory-Laflamme instability onset \eqref{SG:microcanonicalGL} of the uniform phase.
The existence of an energy window where the non-uniform phase dominates the microcanonical ensemble is surprising and constitutes the main findings of our study. Note that, in the microcanonical ensemble, the phase transition that occurs at $\varepsilon=\varepsilon_{\hbox{\tiny GL}}$ is second order and for  $\varepsilon>\varepsilon_{\hbox{\tiny GL}}$, the uniform BFSS phase doinates the ensemble. 
We have also computed the specific heat associated with all the dominant BFSS phases in the canonical ensemble and found it to be always positive. At first glance, this might seem at odds with expectations for localized black holes where $\mathcal{E}^{(0)}$ is a decreasing function of $\mathcal{T}^{(0)}_{\hbox{\tiny $H$}}$. However, the supergravity/BFSS map \eqref{QFTthermoMap} ensures that the specific heat $\frac{1}{N}\frac{\partial \varepsilon}{\partial \tau}$ remains positive even for the BFSS dual of the localized black hole phase. This is explicitly shown in Fig.~\ref{fig:cvLoc}.

In principle, and this would be a remarkable holography precision test for the D0 brane/BFSS duality, it should be possible to derive the first order transition data \eqref{BFSS:canonicalFirstOrTransition} from a field theory calculation. Similarly,  it should also be possible to reproduce $\varepsilon_{\hbox{\tiny K}}$ and $\varepsilon_{\hbox{\tiny GL}}$ from a field theory calculation in the microcanonical ensemble. This might be a formidable challenge but we believe that our precise identification of these values on the supergravity description should motivate and guide such a search on the BFSS side  using computer lattice simulations, bootstrap methods, machine learning, and other methods that are being developed \cite{Hanada:2007ti, Catterall:2007fp, Anagnostopoulos:2007fw, Catterall:2008yz, Hanada:2008gy, Hanada:2008ez, Catterall:2009xn, Hanada:2009ne, Hanada:2013rga, Filev:2015hia, Hanada:2016pwv, Balthazar:2016utu, Han:2019wue, Bergner:2021goh, Rinaldi:2021jbg, Koch:2021yeb, Pateloudis:2022ijr, Hanada:2023rlk, Mathaba:2023non, Bodendorfer:2024egw} (see review~\cite{Lin:2025iir}).

\section{Final discussions \label{sec:conc}}

We have analyzed the low-energy limit of BFSS theory using its holographic supergravity dual. In particular, we determined the thermodynamic properties of the non-uniform and localized phases and compared them with the well-known uniform phase, both in the canonical and microcanonical ensembles. The existence of the non-uniform and localized BFSS phases arises from the Gregory-Laflamme instability of the uniform supergravity solution, which we explicitly demonstrated.  

Our perturbative results (for the localized phase) and numerical results (for both localized and non-uniform phases) in the canonical ensemble corroborate, and make precise, the longstanding conjecture of \cite{Itzhaki:1998dd}: at sufficiently low temperatures, the dominant phase is the localized BFSS thermal phase, dual to a black hole whose spatial horizon cross section has $S^9$ topology. We pinpoint the first-order phase transition between the localized and uniform BFSS phases at $\tau N^{5/9} \approx 0.6118$.  

Interestingly, the microcanonical ensemble reveals a richer structure. For sufficiently small energies, $0 < \varepsilon N^{-4/9} \lesssim 1.51$, the localized BFSS phase dominates. However, there exists an energy \emph{window}, $1.51 \lesssim \varepsilon N^{-4/9} \lesssim 1.7132$, in which a non-uniform phase-dual to a supergravity black string with $S^8 \times S^1$ topology-becomes dominant, breaking translational invariance along the M-theory circle. The second-order phase transition from the non-uniform to the uniform BFSS phase occurs at $\varepsilon N^{-4/9} \approx 1.7132$, a value that we compute with high precision.

The absence of a first-order phase transition in the BFSS microcanonical ensemble $-$ typically seen in transitions between localized black holes and uniform phases for $d \leq 13$ \cite{Wiseman:2002ti,Sorkin:2003ka,Kudoh:2003ki,Kudoh:2004hs,Headrick:2009pv,Kalisch:2017bin,Dias:2017uyv,Ammon:2018sin,Cardona:2018shd} with standard Kaluza-Klein asymptotics $-$ is somewhat surprising. A possible interpretation is the following\footnote{This was suggested to us by Juan Maldacena in a private communication.}. In conventional thermodynamic systems, first-order phase transitions in the microcanonical ensemble can give rise to mixed (or coexistent) phases, where distinct regions of each phase coexist spatially. In the present case, however, the boundary theory is quantum mechanical, which precludes spatial separation of phases. Instead, the system exhibits an intermediate phase $-$ the non-uniform black string $-$  that smoothly interpolates between the localized and uniform phases.

Localized BFSS phases dual to supergravity solutions with two or more black holes on the M-circle, like the ones we consider, are expected to exist \cite{Horowitz:2002dc,Harmark:2003eg,Harmark:2003yz,Dias:2007hg}. In fact, such solutions should exist even with different local masses and specific positions along the M-circle that satisfy equilibrium conditions \cite{Dias:2007hg}. However, these configurations are typically dynamically unstable, tending to merge into a single localized solution and therefore possessing lower entropy. There exists, in principle, an infinite set of such multi-localized solutions. Similarly, one expects an infinite family of non-uniform solutions featuring multiple lumpy regions (local maxima of the transverse radius along the circle) and pinching regions (local minima of the transverse radius).
More generally, the Carrollian map~\eqref{SG:CarrollianTransf}, together with the associated thermodynamical supergravity/BFSS map~\eqref{QFTthermoMap}, allows any black object solution with standard Kaluza-Klein asymptotics to be associated with one having D0-brane (i.e., $pp$-wave) asymptotics~\eqref{asympMetricBFSSdual}, and thus with a corresponding BFSS state. Given the plethora of higher-dimensional black hole solutions \cite{Emparan:2008eg,Dias:2010eu,Dias:2022mde,Dias:2022str,Dias:2023nbj} and the nontrivial nature of the map~\eqref{QFTthermoMap}, one must also entertain the possibility of additional, potentially unexpected phases that could dominate a generalized ensemble including rotations and/or supersymmetric mass deformations in the BMN sense \cite{Berenstein:2002jq,Costa:2014wya}. 

Considering the broader Dp-brane/SYM$_{p+1}$ dualities, a universal feature appears to emerge: in cases that have been studied, the localized SYM phase always dominates the microcanonical ensemble at sufficiently small energies. This behavior holds for the present D0-brane/SYM$_{1+0}$ duality, as well as for the D3-brane/SYM$_{3+1}$ (AdS$_5$/CFT$_4$) \cite{Dias:2015pda,Dias:2016eto}, the D1-brane/SYM$_{1+1}$ on a circle \cite{Aharony:2004ig,Aharony:2005ew,Catterall:2008yz,Dias:2017uyv,Catterall:2023tmr}, and the D1-D5 CFT$_2$/AdS$_3$ \cite{Avery:2010qw,Eberhardt:2019ywk,Bena:2024gmp,Dias:2025csz} dualities.  

In the canonical ensemble, however, the situation is more varied. In some cases (such as D0-brane/SYM$_{1+0}$ and D1-brane/SYM$_{1+1}$) the localized SYM phase dominates at low temperatures, while in others (D3-brane/SYM$_{3+1}$ and D1-D5 CFT$_2$/AdS$_3$) it never becomes dominant. Remarkably, the D0-brane/SYM$_{1+0}$ duality appears unique in that the non-uniform BFSS phase dominates the microcanonical ensemble over a finite energy window-a feature not observed in the other dualities listed above (perhaps for the reason described above).  

Focusing specifically on the D0-brane/BFSS duality, it would constitute a remarkable test of precision holography to recover the BFSS phase diagrams of Fig.~\ref{fig:BFSSphaseDiagrams}, as well as the associated critical energies and temperatures of the phase transitions, directly from a field theory analysis. Advancing this program will likely require further development of lattice, bootstrap, machine learning, and related techniques \cite{Hanada:2007ti, Catterall:2007fp, Anagnostopoulos:2007fw, Catterall:2008yz, Hanada:2008gy, Hanada:2008ez, Catterall:2009xn, Hanada:2009ne, Hanada:2013rga, Filev:2015hia, Hanada:2016pwv, Balthazar:2016utu, Han:2019wue, Bergner:2021goh, Rinaldi:2021jbg, Koch:2021yeb, Pateloudis:2022ijr, Hanada:2023rlk, Mathaba:2023non, Bodendorfer:2024egw} (see also review~\cite{Lin:2025iir}).  

A first step in this direction is to understand, using only BFSS degrees of freedom, the nature of the non-uniform and localized phases. Insights may be drawn from the D1-brane/SYM$_{1+1}$ on a circle duality, where the (non-)uniform and localized phases are comparatively well understood both in supergravity \cite{Dias:2017uyv} and in field theory \cite{Aharony:2004ig,Aharony:2005ew,Catterall:2008yz,Catterall:2023tmr}. In that system, useful observables for the confinement/deconfinement and uniform/localized transitions include the expectation values of Polyakov or Wilson loops along the Euclidean time and spatial circles, as well as the associated eigenvalue distributions \cite{Gross:1980he,Susskind:1997dr,Barbon:1998cr,Li:1998jy,Fidkowski:2004fc,Aharony:2004ig,Aharony:2005ew,Catterall:2010fx,Hanada:2016qbz}. At strong coupling, the eigenvalue distributions describe the positions of the collection of D0-branes in the type IIA system. For example, the Wilson loop along the spatial circle yields a uniform eigenvalue distribution for the uniform phase, a non-uniform but gapless distribution for the non-uniform phase, and a localized distribution with a gap for the localized phase \cite{Aharony:2004ig,Aharony:2005ew,Catterall:2010fx,Hanada:2016qbz}. Insights from this system may prove valuable for understanding the analogous phases in the BFSS system.

Although there has been extensive work in high-energy physics on understanding the canonical ensemble of various field theories $-$ often employing techniques such as localization \cite{Bobev:2023ggk} $-$ the microcanonical ensemble has only recently received significant attention \cite{Dias:2022eyq,Marolf:2022jra,Kim:2023sig}. Studying the microcanonical ensemble is considerably more challenging because it requires detailed knowledge of the underlying field theory microstates. This difficulty is a central reason why a complete resolution of the black hole information paradox remains out of reach. While recent progress has enabled the computation of the entropy of Hawking radiation \cite{Penington:2019npb,Almheiri:2019psf,Almheiri:2019yqk,Almheiri:2019hni,Almheiri:2019psy}, the precise quantum state of the radiation is still inaccessible \cite{Almheiri:2020cfm}. In this work, we uncover a rich microcanonical ensemble structure for BFSS quantum mechanics at low energies. Ideally, one would like to identify all relevant phases directly in terms of the BFSS degrees of freedom, though this endeavor inevitably requires a detailed understanding of the microscopic states themselves.

\begin{acknowledgments}

O.J.C.D. acknowledges financial support from the  STFC ``Particle Physics Grants Panel (PPGP) 2018" Grant No.~ST/T000775/1. J.~E.~S. has been partially supported by STFC consolidated grant ST/X000664/1 and by Hughes Hall College. The authors also acknowledge the use of the IRIDIS High Performance Computing Facility, and associated support services at the University of Southampton, in the completion of this work.
\end{acknowledgments}

\appendix
\section{Equations of motion for type II and 11-dimensional supergravities}
\label{sec:EOM}
For completeness, in this appendix we give the equations of motion that have, as solutions, the p-branes of type II supergravity and their near-horizon geometries.  Non(-uniform) and localized solutions also must solve these equations of motion (if they exist within the regime of parameters where type II supergravity is a good approximation to type II string theory). In the main text we are only interested on the $p=0$ branes of type IIA supergravity.

In the string frame, the equations of motion of type II action \eqref{action} are
\begin{eqnarray}\label{IIeomStringFr}
&& R_{ab}=-2\nabla_a\nabla_b \phi +\frac{1}{4}  \frac{1}{(p+2)!}\,e^{2\phi}{\biggl [} 2(p+2)\, F_a^{\,\: c_1\cdots c_{p+1}}F_{b c_1\cdots c_{p+1}} - g_{ab} \,F_{c_1\cdots c_{p+2}}F^{c_1\cdots c_{p+2}}{\biggr ]}, \nonumber \\
&& \nabla_c F^{c a_1 \cdots a_{p+1}}=0\,,\nonumber \\
&& \nabla_c\nabla^c \phi- 2\partial_c \phi \partial^c \phi +\frac{1}{4} \frac{p-3}{(p+2)!}\, e^{2\phi}  F_{c_1\cdots c_{p+2}}F^{c_1\cdots c_{p+2}}=0\,,
\end{eqnarray}
where we have introduced the RR field strength $F_{(p+2)}=\mathrm dA_{(p+1)}$.

In the Einstein frame, type II action (which includes the particular $p=0$ case of the IIA action \eqref{action}) reads
\begin{equation}
\label{actionE}
S_{II}^{\hbox{\tiny(E)}} = \frac{1}{16\pi G_{10}} \int \mathrm{d}^{10} x \sqrt{-\widetilde{g} } \, \Big( \widetilde{R}
- \frac{1}{2} \partial_\mu \phi \partial^\mu \phi - \frac{1}{2 (p+2)!} g_s^{\frac{1}{2}(p+1)} e^{\frac{1}{2}(3-p)\phi} (\mathrm dA_{(p+1)})^2 \Big),
\end{equation}
where Newton's constant is expressed in terms of the string length $\ell_s$ and string coupling $g_s$ as $16 \pi  G_{10}\equiv (2 \pi )^7 \ell_s^8  \,g_s^2$.
From this action we obtain the equations of motion in the Einstein frame
\begin{subequations}
\begin{multline}
 \widetilde{R}_{ab}-\frac{1}{2}\,\widetilde{R} \,\widetilde{g}_{ab}=\frac{1}{2}\left(  \partial_a \phi \partial_b \phi - \frac{1}{2}\, \widetilde{g}_{ab}\,\partial_c \phi \partial^c \phi \right) \\
+\frac{1}{2}\,g_s^{\frac{1}{2}(p+1)} e^{\frac{1}{2}(3-p)\phi}\left[ \frac{1}{(p+1)!}\, F_a^{\,\: c_1\cdots c_{p+1}}F_{b c_1\cdots c_{p+1}} -  \frac{1}{2 (p+2)!} \,\widetilde{g}_{ab}\,F_{c_1\cdots c_{p+2}}F^{c_1\cdots c_{p+2}}\right]
\end{multline}
\begin{equation}
\widetilde{\nabla}_c \left( e^{\frac{1}{2}(3-p)\phi} F^{c a_1 \cdots a_{p+1}}\right)=0\,,
\end{equation}
\begin{equation}
\widetilde{\nabla}_c\widetilde{\nabla}^c \phi- \frac{3-p}{4(p+2)!}\, g_s^{\frac{1}{2}(p+1)} e^{\frac{1}{2}(3-p)\phi}  F_{c_1\cdots c_{p+2}}F^{c_1\cdots c_{p+2}}=0\,,
\end{equation}
\label{IIeom}
\end{subequations}
where $\widetilde{\nabla}$ is the Levi-Civita connection of $\tilde{g}$.

Recall that the string and Einstein frame metrics are related by the transformations $g_{ab}=\widetilde{g}_{ab} e^{\frac{1}{2}\left(\phi-\phi _{\infty}\right)}$, while the dilaton and gauge field are the same in both frames.
 In the main text, we present the type II solutions \eqref{nonExtDp} and \eqref{NHnonExtDp} in the string frame.

The bosonic fields of  11-dimensional supergravity are the metric field $g$ and a 3-form gauge potential $A_{(3)}$ with associated field strength  $G_{(4)}=\mathrm{d}A_{(3)}$. The associated action is \cite{Cremmer:1978km}
\begin{equation}\label{actionMsugra}
S_{11} = \frac{1}{2 \kappa_{11}^2} \int \bigg( \star \mathcal{R} - \frac{1}{2} \, G_{(4)} \wedge \star\, G_{(4)} + \frac16 \, G_{(4)} \wedge G_{(4)} \wedge A_{(3)} \bigg)\,,
\end{equation}
where $\mathcal{R}$ is the Ricci volume form, $\mathcal{R}=R\,\star \mathbb{I}=R\,\mathrm{vol}_{11}$.
The (trace reversed, i.e. after using a contraction with the inverse metric to get rid of the Ricci scalar) equations of motion of 11-dimensional supergravity are 
\begin{align}\label{eqs:Msugra}
&R_{ab} = \frac{1}{12}\left[{G_{(4)}}_{acde} {G_{(4)}}_b{}^{cde} - \frac{1}{12} g_{ab} \,{G_{(4)}}_{cdef} {G_{(4)}}^{cdef}\right]\,, \nonumber\\
&\mathrm{d} \star  G_{(4)} = \frac{1}{2} G_{(4)} \wedge G_{(4)}.
\end{align}
where we contracted the equation of motion that follows from \eqref{actionMsugra} with the inverse metric to get the Ricci scalar $R$ to get the trace reversed equation of motion for the graviton \eqref{eqs:Msugra}.
All the solution discussed in the main text, namely the (non-)uniform and localized solutions, have $G_{(4)}=0$ and thus \eqref{actionMsugra} and   \eqref{eqs:Msugra} reduce effectively to the Einstein-Hilbert action and associated equations of general relativity in 11 dimensions.


\bibliography{refsLocalizedBFSS}{}
\bibliographystyle{JHEP}

\end{document}